%% file: main.tex
\documentclass[11pt]{article}
\usepackage[utf8]{inputenc} % allow utf-8 input
\usepackage[T1]{fontenc}    % use 8-bit T1 fonts
\usepackage{hyperref}       % hyperlinks
\usepackage{url}            % simple URL typesetting
\usepackage{booktabs}       % professional-quality tables
\usepackage{amsfonts}       % blackboard math symbols
\usepackage{nicefrac}       % compact symbols for 1/2, etc.
\usepackage{microtype}      % microtypography

\usepackage{algorithm}
\usepackage{algorithmic}

\usepackage{amsthm}
\usepackage{thmtools, thm-restate}
\usepackage{fullpage}
\usepackage{xcolor}
\usepackage{graphicx}
\graphicspath{{images/}{../images/}}
\usepackage{subfiles}
\usepackage[numbers]{natbib}

\newcommand{\ALG}{\text{ALG}}
\newcommand{\OPT}{\text{OPT}}
\newcommand{\Ex}{\mathbb{E}}
\newtheorem{theorem}{Theorem}
\newtheorem{lemma}{Lemma}
\newtheorem{definition}{Definition}
\newtheorem{prop}{Proposition}

\newtheorem{cor}{Corollary}
\newtheorem{remark}{Remark}
\newcommand{\nameOfProblem}{submodular $k$-secretary problem}
\newcommand{\nameOfProblemSL}{submodular $k$-secretary problem with shortlists}
\newcommand{\nameOfProblemMatroid}{submodular matroid secretary problem}
\newcommand{\nameOfProblemMatroidSL}{submodular matroid secretary problem with shortlists}

\newcommand{\streamingProblem}{maximizaing submodular functions in random order streaming model}
\newcommand{\nameSecSL}{secretary problem with shortlists}

\newcommand{\memory}{\tilde{O}(\frac{k}{\epsilon^2})}
\newcommand{\matroidmemory}{O(\frac{k}{\epsilon})}
\newcommand{\shortlist}{\tilde{O}\left(\frac{k}{\epsilon^2}\right)}
\newcommand{\query}{\tilde{O}(\frac{n}{\epsilon})}
\newcommand{\matroidshortlist}{\tilde{O}\left(\frac{k}{\epsilon}\right)}

\newcommand{\mcomment}[1]{}%{\textcolor{red}{#1}}

\newcommand{\rcomment}[1]{}%{\textcolor{red}{#1}}

\newcommand{\matcomment}[1]{}%{\textcolor{red}{#1}}

\newcommand{\toRemove}[1]{}%{\textcolor{cyan}{[Text in cyan to be Removed]#1}}

\newcommand{\etaMacro}{$c\frac{\log(1/\epsilon)}{\epsilon^2}  {\frac{1}{\epsilon^6} \log(1/\epsilon)  \choose {\frac{1}{\epsilon^4} \log(1/\epsilon)}}$~}
\usepackage{mathtools}

\newcommand{\settinga}{ {\mbox{$k \ge \alpha\beta$},} \settingb}
\newcommand{\settingb}{\mbox{$\beta\ge \frac{1}{\delta'}$}, $\alpha\ge 8(1/\delta') \log(1/\delta')$}

\title{Improved Submodular Secretary Problem with Shortlists\\
}

\author{%
 Mohammad Shadravan
 \thanks{Yale University, mohammad.shadravan@yale.edu} \\
}

\begin{document}

\maketitle

\begin{abstract}

\subfile{abstract}

\end{abstract}

%\ccomment{TODO:  Fix $1+ k^2/n$ in statement of streaming Theorem.  }
%\ccomment{TODO:  Fix the bounds in Theorem 1.  Make consistent.}
%\ccomment{TODO: Get rid of infinite domain.}
%\scomment{TODO: get the name of problem consistent using macro \mcomment{done}, related work\mcomment{done}, add bib\mcomment{done}, Rigorous proof for the $1-\delta$ probability of every item in $H_w$
%\mcomment{done}
%,fix $L=3\ln(1/\epsilon)$ \mcomment{done}, value oracle model \mcomment{done}, streaming update time terminology \mcomment{done}}

\section{Introduction}
 
\input{newIntroduction}

%\subsection{Organization}
%The rest of the paper is organized as follows. Section~\ref{sec:alg} describes our main algorithm (Algorithm \ref{alg:main}) for the \nameOfProblemSL, and demonstrates that its shortlist size is bounded by $\eta_\epsilon(k)=O(k)$. In Section \ref{sec:analysis}, we analyze the competitive ratio of this algorithm to prove Theorem~\ref{opttheorem}. In Section \ref{sec:streaming}, we provide an alternate implementation of Algorithm \ref{alg:main} that uses a memory buffer of size at most $\eta_\epsilon(k)$, in order to prove Theorem \ref{thm:streaming}. Finally, in Section \ref{sec:hardness}, we provide a proof of our impossibility result stated in Theorem \ref{hardness}. The proof of Theorem~\ref{thm:msub} along with the relevant algorithm appears in the appendix.

\section{Preliminaries}
\label{sec:analysis}

\input{preliminaries}

\section{Cardinality Constraints} %: improving dependecny on $1/\epsilon$}
\label{sec:cardinality}
\input{cardinality}

\section{Matroid Constraints}
\label{sec:matroid}
\input{algorithms}

\toRemove{
\section{Bounding the size of shortlist}
\label{sec:size}
\input{sizeBound}

}
\toRemove{
\subsection{Bounding the size of shortlist for Algorithm \ref{alg:SIIImax}}
\input{maxanalysis.tex}

}

%\section{Bounding the competitive ratio %(Proof of Theorem \ref{opttheorem})
%}

\subsection{Analysis of the algorithms (Matroids)} %(Proof of Theorem~\ref{opttheorem})}
\label{sec:analysis}

\input{analysis}

\section{ Matchoid Constraints}
\label{sec:matchoid}
\input{mmatroids}

%\section{Streaming %(Proof of Theorem \ref{thm:streaming})
%}
%\label{sec:streaming}
%\input{twocol/Supp/Suppstreaming}
%\subfile{streaming}

\toRemove{
\section{Broader Impact}
This work has applications in online mechanism design and submodularity in machine learning. In particular in data summarization and other submodular optimization problems in the streaming setting. Furthermore it has applications in analysing neural networks on the fly.
}

%\section{Impossibility Result (Proof of Theorem \ref{hardness})}
%\label{sec:hardness}
 
%\subfile{impossible}

%\section{Conclusion}
 
%\subfile{conclusion}

%\begin{thebibliography}{99}
%\subfile{bib}
%\end{thebibliography}

\bibliographystyle{plainnat}
\bibliography{mybib}


%\section{Supplementary}
%\label{Supp}
%\subfile{Supp/main}

\newpage
%\section{Preliminaries}
%\label{sec:analysis}
%\input{twocol/Supp/Suppmain}

\appendix

%\section{Preliminaries}
%\label{sec:analysis}
%\input{twocol/Supp/preliminaries}
%\input{Supp/preliminaries}

%\section{Bounding the competitive ratio %(Proof of Theorem \ref{opttheorem})
%}

\section{Missing parts in Cardinality Constraints}
\label{sec:analysis}
\input{Supp/CardSupp}

%\section{Missing Proofs in Matroid Constraints (Section 4)}
%\label{sec:analysis}
%%\input{twocol/Supp/Suppanlysis}
%\input{Supp/Suppanlysis}

\section{ Missing Proofs in the $p$-matchoid Constraints Section}
\label{sec:matchoidApp}
%\subfile{Supp/mmatroids}
%\subfile{Supp/}
%\input{twocol/Supp/Suppmmatroids}

\input{Supp/Suppmmatroids}

\mcomment{
\section{Missing Proofs in the Streaming Section %(Theorem 4) %(Proof of Theorem \ref{thm:streaming})
}
\label{sec:streamingApp}

\input{twocol/Supp/Suppstreaming}

\toRemove{
\appendix
\section{Do \emph{not} have an appendix here}
\textbf{\emph{Do not put content after the references.}}
Put anything that you might normally include after the references in a separate
supplementary file.
We recommend that you build supplementary material in a separate document.
If you must create one PDF and cut it up, please be careful to use a tool that
doesn't alter the margins, and that doesn't aggressively rewrite the PDF file.
pdftk usually works fine. 
\textbf{Please do not use Apple's preview to cut off supplementary material.} In
previous years it has altered margins, and created headaches at the camera-ready
stage. 
%%%%%%%%%%%%%%%%%%%%%%%%%%%%%%%%%%%%%%%%%%%%%%%%%%%%%%%%%%%%%%%%%%%%%%%%%%%%%%%
%%%%%%%%%%%%%%%%%%%%%%%%%%%%%%%%%%%%%%%%%%%%%%%%%%%%%%%%%%%%%%%%%%%%%%%%%%%%%%%
}
}
\end{document}

%% file: abstract.tex
First, for the \nameOfProblemSL\ \cite{us}, we present a near optimal $1-1/e-\epsilon$ approximation using shortlist of size %$O(k poly(1/\epsilon))$. 
$\shortlist$.
In particular, we improve the size of shortlist used in~\cite{us} from $O(k 2^{poly(1/\epsilon)})$ to %$O(k poly(1/\epsilon))$. 
$\shortlist$.
As a result, we present a fast $1-1/e-\epsilon$ approximation algorithm for random-order streaming of monotone submodular functions under cardinality constraints, using memory %$O(k poly(1/\epsilon))$. 
$\memory$.
%It exponentially improves the running time and memory of~\cite{us} in terms of $1/\epsilon$. 
Most importantly, the query time and running time of our algorithm is $\tilde{O}(n/\epsilon)$ ($n$ is the size of input).

Next we generalize the problem to matroid constraints, which we refer to as submodular matroid secretary problem with shortlists. It is a variant of the \textit{matroid secretary problem}~\cite{feldman2014simple}, in which the algorithm is allowed to have a shortlist. %The main question is how to achieve a constant competitive algorithm using a shortlist as small as possible.
We design an algorithm that achieves a $\frac{1}{2}(1-1/e^2 -\epsilon)$ competitive ratio 
for any $\epsilon>0$, 
using a shortlist of size
$\matroidshortlist$.
%$O(k  poly(\frac{1}{\epsilon}))$. 
This is especially surprising considering that the best known competitive ratio for the matroid secretary problem is $O(\log \log k)$~\cite{feldman2014simple,Lachish14}, where $k=rk(\mathcal{M})$. 
Moreover, we generalize our results to the case of $p$-matchoid constraints and give a $\frac{1}{p+1}(1-1/e^{p+1}-\epsilon )$ approximation using shortlist of size $\matroidshortlist$.
%$O(k poly(\frac{1}{\epsilon}))$, \mcomment{where $k:=$}
It asymptotically approaches the best known offline guarantee $\frac{1}{p+1}$~\cite{nemhauser1978analysis}. 
Furthermore, we show that our algorithms can be implemented in the streaming setting with the same approximation guarantees.

\newpage

%% file: newIntroduction.tex
%in the name of God

In the \textit{Secretary problem}, $n$ items arrive in random order.
The goal is to select the item with the highest value. All the selections are  made in an online manner. Once we observe one item we need to irrevocably decide whether or not to select that item%There is a simple strategy to achieve a $1/e$ competitive algorithm for this problem which is the best possible
%There is a simple strategy to achieve a $1/e$ competitive algorithm for this problem which is the best possible
~\citet{Dynkin:SovMath:1963}. 
Many variants and generalizations of the secretary problem have been studied in the literature, see e.g., \cite{Ajtai:2001,WILSON1991325,vanderbei1980optimal, wilson1991optimal, kleinberg, Babaioff:2008}.  
\citet{kleinberg, Babaioff:2008} introduced a 
{multiple choice secretary problem}, where the goal is to select $k$ items in a randomly ordered input so as to maximize  the { sum} of their values;
They provide an algorithm with competitive ratio that asymptotically approaches the optimal solution.
This problem has been further generalized to the case of submodular functions
\cite{Bateni,Gupta:2010}, in which
the value of the selected items is evaluated by a monotone submodular function. The algorithm can select at most $k$ items $a_1 \cdots, a_k$, in an online manner, from a randomly ordered sequence of $n$ items. The goal is to maximize $f(\{a_1,\cdots, a_k\})$.
The algorithm has a value oracle access to the function.
~\citet{kesselheim}, achieve a $1/e$-competitive competitive algorithm for this problem.
 The problem of maximizing a monotone submodular function under cardinality constraint is NP-hard. The best approximation algorithm possible is a $1-1/e$-approximation algorithm % by the greedy algorithm
 ~\cite{nemhauser1978best}. Furthermore, No online algorithm with the same guarantee is known for this problem.

~\citet{us}, introduced a model called  \textit{shortlist} model which is a relaxation of the online model. They study  if a  $(1-1/e-\epsilon)$ approximation is possible under this new model. 
In this model, the algorithm is allowed to keep a subset of items and upon receiving one new item add it to the shortlist or discard it. At the end, the output of the algorithm should be a subset of this shortlist. Optimistically, the goal is to keep this shortlist as small as possible, while achieving near optimal guarantees. \cite{us} present a $1-1/e-\epsilon-O(1/k)$ approximation for this problem using shortlist of size $O(k2^{poly(1/\epsilon)})$. Although the dependency on $k$ is linear but the dependency on $1/\epsilon$ is exponential. Therefore it is far from being practical.

The shortlist model has connections to 
another related problem,  {\it  \streamingProblem}~studied in~\cite{norouzi}. In this problem, items from a set $\cal U$ arrive in online manner and in random order,  the algorithm aims to select a subset $S \subseteq {\cal U}, |S|\le k$ in order to maximize $f(S)$. The streaming algorithm is allowed to maintain a {\it buffer} of size $\eta(k)\ge k$. 
However, this streaming problem is distinct from the \nameOfProblemSL.\ An algorithm in one model can not directly be converted to an algorithm in the other model.
Howerer~ \citet{us} show that their algorithms, can be implemented to use the same $\eta(k)=O(k2^{poly(1/\epsilon)})$ memory buffer for the random order streaming model.

 Recently streaming algorithms for maximizing a submodular function has been studied in a series of work.
 ~\citet{Badanidiyuru2014StreamingFly}, provide the first  one-pass streaming algorithm for maximizing a monotone submodular function subject to a $k$-cardinality constraint.
 They achieve ($1/2-\epsilon$)-approximation streaming algorithm, 
with a memory of size $O(\frac{1}{\epsilon}k\log k)$.
Recently,~\citet{kazemi2019submodular} improved  the memory buffer to $O(k/\epsilon)$.

\citet{norouzi} show that under some natural assumption
no $1/2+o(1)$ approximation ratio can be achieved
by any algorithm for streaming submodular maximization using $o(n)$ memory.
%that only queries the value of the submodular function on feasible sets (sets of size at most $k$) while using $o(n)$ memory.
They studied the random order streaming model in order to go beyond the upperbound for the adversarial order inputs. They present a $1/2+8\times 10^{-14}$ approximation using a memory buffer of size $O(k\log k)$.
 \citet{us} 
 substantially improve their result to  $1-1/e-\epsilon-O(1/k)$
 , by showing that their algorithm for the shortlist model can be converted into a random order streaming model. Furthermore, they improve the required memory buffer (in terms of $k$) to only $O_{\epsilon}(k)$. But one disadvantage of their algorithm is that their dependency on $1/\epsilon$ is exponential. 
 In this paper, we improve their algorithm and give a near optimal algorithm using memory $O(kpoly(1/\epsilon))$.

%In the streaming setting, \citet{Chakrabarti2015} provided a single pass streaming algorithm for monotone submodular function maximization under $k$-cardinality constraint, that achieves a $0.25$ approximation under adversarial ordering of input. Further, their algorithm requires $O(1)$ function evaluations per arriving item and $O(k)$ memory.
%The currently best known approximation  under  adversarial order streaming model is by~\citet{Badanidiyuru2014StreamingFly}, who achieve a $1/2-\epsilon$ approximation with a memory of size $O(\frac{1}{\epsilon}k\log k)$. 
In addition to the simple cardinality constraint, more general constraints
have been studied in the literature.
%For more general constraints than the simple cardinality constraint,
%~\citet{Chakrabarti2015} give a $1/4p$ approximation algorithm for streaming monotone submodular functions maximization subject to the intersection of $p$ matroid constraints.  
\citet{Chekuri}
give a $1/4p$ approximation algorithm for streaming monotone submodular functions maximization subject to to $p$-matchoid constraints.  
The $p$-matchoid constraints generalize many basic combinatorial constraints such as the cardinality constraint,
the intersection of $p$ matroids, and matchings in graphs and hyper-graphs. 
%A formal definition of a $p$-matchoid is given in Section~\ref{sec:matchoid}. 
Recently,~\citet{feldman2018streaming} designed a more efficient algorithm with lower number of function evaluations achieving  the same approximation $1/4p$.
%Recently \citet{shadravan2020submodular}, generalize the result of~\cite{us} to the matroid and matchoid constraints. They provide a random order streaming algorithm with approximation $1/2(1-1/e^2-\epsilon)-O(1/k)$ and $\frac{1}{(p+1)}(1-1/e^{p+1}-\epsilon-O(1/k))$, for matroid constraints and matchoid constraint respectively. 
%Their algorithm requires $O(k2^{poly(1/\epsilon)})$ memory.
%We slightly improve their approximation guarantee. But we exponentially improve their memory buffer  and running time. 
We show that our algorithms can be implemented in the streaming setting using $O(k poly(\frac{1}{\epsilon}))$ memory.
For any constant $\epsilon>0$, our algorithms achieve a $1-1/e-\epsilon$, a $\frac{1}{2}(1-1/e^2-\epsilon)$, and a $\frac{1}{p+1}(1-1/e^{p+1}-\epsilon)$ approximation for random-order streaming of submodular functions, under cardinality, matroid, and $p$-matchoid constraints, respectively. 
Furthermore, the greedy algorithm yields a ratio of $1/(p + 1)$ for $p$-independent systems~\cite{nemhauser1978analysis}.  These ratios for greedy are tight for all $p$~\cite{hazan2006complexity}.
Therefore our results for $p$-matchoid constraints is asymptotically tight.

\paragraph{\textbf{The shortlist model}.}
%\paragraph{The \nameSecSL.}
%\paragraph{The \nameOfProblemSL.}
In~\cite{us}, a relaxation of the secretary problem  is introduced
%.  What if  the online algorithm for secretary problem
where the algorithm is allowed to  select a {\it shortlist} of items.
%that is larger than the number of items that ultimately need to be selected. 
After seeing the entire input, the algorithm can choose from the bigger set of items in the shortlist. 
%That is, in a {\nameOfProblem}, the algorithm is allowed to choose more than $k$ items as part of a shortlist. Then, after seeing the entire input, the algorithm can choose a subset of size $k$ from the bigger set of items in the shortlist. 
%when seeing items online, and ultimately maximizes a function of $k$ elements, 
%we are allowed to choose more than $k$ elements, and then after seeing the entire input, choose a  a subset of size $k$ from our bigger set. 
This model is closely related to the random order streaming model. A comprehensive comparison between these two models can be found in~\cite{us}.
The main result of~\cite{us} is an  algorithm  for \nameOfProblemSL\ that, for any constant $\epsilon>0$, achieves a competitive ratio of $1-\frac{1}{e} - \epsilon-O(\frac{1}{k})$ with shortlist of size $ O_{\epsilon}(k)$. 

\subsection{Problem definition} \label{sec:matroidDef}
%Motivated by the improvements achieved for the competitive ratio of  \nameOfProblem\ in the \textit{shortlist} model~\cite{us}, we define submodular matroid secretary problem with shortlists.

%we ask if similar improvements can be achieved by relaxing the \nameOfProblemMatroid~to have a shortlist.
%We now give a more formal definition. 
We are given matroid $\mathcal{M}=(\mathcal{N},\mathcal{I})$, with $rk(\mathcal{M})=k$.
Items from a set ${\mathcal{ U}} = \{a_1, a_2, \ldots, a_n\}$  %(pool of items) 
arrive in a uniformly random order over $n$ sequential rounds. The set ${\mathcal U}$ is apriori fixed but unknown to the algorithm, and the total number of items $n$ is known to the algorithm. In each round, the algorithm irrevocably decides whether to add the arriving item to a { shortlist} $A$ or not. %The goal is to return,  at the end of $n$ rounds, a shortlist $A$ of size at most $\eta(k) \ge k$ to maximize the  `value' of best $k$ subset of this shortlist. More formally, 
%We are given a monotone submodular function $f:2^{\cal U} \rightarrow \mathbb{R}$, in a value oracle model. %Let the set $A \subseteq {\cal U}, |A|\le \eta(k)$ denote the shortlist of items selected by the online algorithm,  then 
The algorithm's value at the end of $n$ rounds is given by 
%\begin{equation*}
%\ALG = \Ex[\max_{S\subseteq A, S\in \mathcal{I}} f(S)
%\end{equation*}
\[\ALG = \Ex[\max_{S\subseteq A, S\in \mathcal{I}} f(S)]\] 
where $f(\cdot)$ is a monotone submodular function. The algorithm has value oracle access to this function.
The optimal offline utility is given by
\[\OPT:=f(S^*), \text{ where } S^*=\arg \max_{S \subseteq [n], S\in \mathcal{I}} f(S).\]
We say that an algorithm for this problem achieves a competitive ratio  $c$ using shortlist of size $\eta(k)$, if at the end of $n$ rounds, $|A|\le \eta(k)$ and $\frac{\ALG}{\OPT}\ge c$.

Given the shortlist $A$, since the problem of computing the solution $\arg \max_{S\subseteq A, S\in \mathcal{I}} f(S)$ can itself be computationally intensive, our algorithm will also track and output a subset $A^* \subseteq A$, and $A^* \in \mathcal{I}$. %We will lower bound  the competitive ratio by bounding $\frac{f(A^*)}{f(S^*)}$.

The problem definition for $p$-matchoid constraint is similar, but $S$ needs to be an independent set in all the matroids $\mathcal{M}_i$, for $i\in [q]$.

%\paragraph{\textbf{Our results.}}

\subsection{ Related Work}
In the \textit{matroid secretary problem}, the elements of a matroid $\mathcal{M}$ arrive in random order. Once we observe an item we need to irrevocably decide 
whether or not to accept it. The set of selected elements should form an independent set of the matroid.
The goal is to maximize the total sum of the values assigned to these elements.
It has applications in welfare maximizing online mechanism design for domains in which
the sets of simultaneously satisfiable agents form a matroid~\cite{babaioff2007matroids}.

The existence of a constant competitive algorithm is a long-standing open problem.
It has been shown that for some special cases of the matroid secretary problem, $O(1)$-compettetive algorithms exists. But for general case the problem is still open.
~\citet{Lachish14} provides  the first $\Omega(1/\log \log(k))$-competitive algorithm (the hidden constant is $2^{-2^{34}}$).
~\citet{feldman2014simple} give a simpler  order-oblivious $1/(2560(\log\log
(4k)+ 5))$-competitive algorithm.
%There are some $O(1)$-competitive algorithms 
\toRemove{
In particular, if a set of $n$ weights is assigned uniformly at
random to the elements of the ground set, then a $5.7187$-competitive algorithm can be obtained
for any matroid~\cite{soto2013matroid}. Furthermore, even
with adversarial arrival order of the elements,
a $16(1-1/e)$-competitive algorithm can be obtained  as long as the weight assignment is still done at random.
}
%\paragraph{Preemption Model}
For the preemption model, which is relaxation of the online model that we can substitute one item,
~\citet{Buchbinder:2014}  present a randomized $0.0893$-competitive
algorithm for cardinality constraints using $O(k)$ memory.

%chekuri et al. obtain a
%(1 − )/(2 + e)-competitive algorithm for this case using O(k log k/2)-space

\subsection{Our Results}

%We design an algorithm that achieves a $\frac{1}{2}(1-1/e^2-\epsilon-O(1/k))$ competitive ratio for any constant $\epsilon>0$, using a shortlist of size $O(k)$ for the matroid secretary problem with shortlists. 
%This is especially surprising considering that the best known competitive ratio for the matroid secretary problem is $O(\log \log k)$. We are also able to get a constant competitive algorithm using shortlist of size  at most $k$ and also a constant competitive algorithm in the preemption model.

\begin{restatable*}{theorem}{cardThm} \label{cardtheorem}
For any constant $\epsilon>0$, there exists an  algorithm (Algorithm \ref{alg:card}) for the \nameOfProblemSL\ that achieves a competitive ratio of $1-\frac{1}{e} -\epsilon \mcomment{ -O(\frac{1}{k})}$, with shortlist of size
$\shortlist$.
%$\eta_\epsilon(k)$.  
The running time of this  algorithm is $\query$. % $O_{\epsilon}(n)$.
\end{restatable*}

%Throughout the paper  $\eta_\epsilon(k)=O(kpoly(1/\epsilon)$) and the hidden constant in $O_{\epsilon}(.)$ notation is $O(poly(1/\epsilon))$.

%O\left(\frac{k}{\epsilon}\log^2(\frac{1}{\epsilon})\right)$.
\mcomment{$\eta_\epsilon(k)=$\etaMacro$k$ for some constant $c$.}
%$\eta_\epsilon(k)=O(2^{poly(1/\epsilon)}k)$.
This is an exponential speed-up of the algorithm presented for \nameOfProblem\  in the previous work~\cite{us}. Furthermore, the algorithm can be readily parallelized among as many as $O(1/\epsilon)$  processors.

\begin{restatable*}{theorem}{matroidThm} \label{opttheoremMatroid}
For any constant $\epsilon>0$, there exists an  algorithm (Algorithm \ref{alg:matroid}) for the \nameOfProblemMatroidSL\ that achieves a competitive ratio of $\frac{1}{2}(1-\frac{1}{e^2} -\epsilon \mcomment{-O(\frac{1}{k})})$, with shortlist of size $\matroidshortlist$. %$\eta_\epsilon(k)$.  
The running time of this  algorithm is $O(nk)$.
\end{restatable*}

This result is especially surprising considering that the best known competitive ratio for the matroid secretary problem is $\Omega(1/\log \log k)$. It implies a constant competitive algorithm using shortlist of size  at most $k$ and also a constant competitive algorithm in the preemption model.

{

\toRemove{
\begin{restatable*}{theorem}{thmpreemption} \label{thm:preemption}
For the matroid secretary problem in the preemption model, and matroid secretary problem 
that uses shortlist of size at most $\eta(k)=k$,
there is an algorithm 
%the algorithm~\ref{alg:main} 
%(with $\alpha=\beta=1$) 
that achieves a constant competitive ratio. %$\frac{1}{2} (1-1/e) (1-1/e^2-\epsilon)$.
\end{restatable*}
}
}

Furthermore,  for a more general constraint, namely $p$-matchoid constraints we prove the following. %(defined in section~\ref{sec:matchoid}):
%we prove:

\begin{restatable*}{theorem}{matchoidThm} \label{opttheoremmatchoid}
For any constant $\epsilon>0$, there exists an  algorithm for the submodular secretary problem with $p$-matchoid constraints that achieves a competitive ratio of $\frac{1}{p+1}(1-\frac{1}{e^{p+1}} -\epsilon \mcomment{-O(\frac{1}{k})})$, with shortlist of size %$\eta_\epsilon(k)$. 
$\matroidshortlist$.
%Here,  $\eta_\epsilon(k)=O(2^{poly(1/\epsilon)}k)$. 
The running time of this  algorithm is $O(n\kappa^{p})$, where $\kappa= \max_{i\in[q]} rk(\mathcal{M}_i)$.
\end{restatable*}

The  proposed  algorithm  also  has  implications  for  another  important  problem
of submodular function maximization under random order streaming model.
\begin{prop}
For random-order streaming model, we can achieve the same approximation guarantees as  stated in the above three lemmata, using memory of size $\memory$, $\matroidmemory$ and $\matroidmemory$ respectively.
\end{prop}
\begin{proof}
 The only difference is that instead of using online max algorithm ( in line 6 of the algorithm~\ref{alg:card}) we use an offline version of the algorithm, which only keep track of the max element using $O(1)$ memory. As a result we achieve a $O(\log(1/\epsilon))$ factor improvement in terms of memory used in all the three theorems stated above.
\end{proof}
% and $p$-matchoid constraints. 
%It will be discussed in
%See Chapter~\ref{sec:streaming}.
\mcomment{State the lemmata here.}

%% file: preliminaries.tex
 The following properties of submodular functions are well known (e.g., see~\cite{Buchbinder:2014,Feige:2011,Feldman:2015}). %\ref{bookorpaper}).

\begin{definition}
Given a monotone submodular function $f$, and subsets $A,B$ in the domain of $f$, we use $\Delta_f(A|B)$ to denote $f(A\cup B)-f(B)$. 
\end{definition}

\begin{lemma}\label{marginalsum}
Given a monotone submodular function $f$, and subsets $A,B$ in the domain of $f$, we use $\Delta_f(A|B)$ to denote $f(A\cup B)-f(B)$. 
For any set $A$ and $B$, $\Delta_f(A|B) \le \sum_{a\in A\setminus B} \Delta_f(a|B)$
\end{lemma}

\begin{lemma}\label{sample}
Denote by $A(p)$ a random subset of $A$ where each element has probability at least $p$ to appear in $A$ (not necessarily independently). Then $E[f(A(p))] \ge (1-p) f(\emptyset) + (p)f(A)$
\end{lemma}

\begin{lemma}[Chernoff bound for Bernoulli r.v.]
\label{lem:Chernoff}
Let $X = \sum_{i=1}^N X_i$, where $X_i = 1$ with probability $p_i$ and $X_i = 0$ with
probability $1 - p_i$, and all $X_i$ are independent. Let $\mu = \Ex(X) = \sum_{i=1}^N p_i$. Then, 
\[P(X \geq (1 + \delta)\mu) \le e^{{-\delta^2\mu}/{(2+\delta)}} \]
for all $\delta>0$, and
\[P(X \leq (1 - \delta)\mu) \le e^{-\delta^2\mu/2}\]
for all $\delta\in(0,1)$. 
\end{lemma}

\begin{definition}
(\textbf{Matroids}). A matroid is a finite set system $\mathcal{M} = (\mathcal{N} , \mathcal{I})$, where $\mathcal{N}$ is a set and $\mathcal{I}\subseteq 2^{\mathcal{N}}$
is a family of subsets such that: (i) $\emptyset \in I$, (ii) If $A \subseteq B \subseteq N$ , and $B \in I$, then $A \in I$,
(iii) If $A, B \in I$ and $|A| < |B|$, then there is an element $b \in B\setminus A$ such that $A + b \in I$. In a matroid $\mathcal{M} = (\mathcal{N} , \mathcal{I})$, $N$ is called the ground set and the members of $\mathcal{I}$ are called independent sets of the matroid. The bases of $\mathcal{M}$ share a common cardinality, called the
rank of $\mathcal{M}$ ( denote it by $rk(\mathcal{M})$).
\end{definition}

\begin{definition} \label{def:matchoid}
(\textbf{Matchoids}). Let $\mathcal{M}_1 = (\mathcal{N}_1, \mathcal{I}_1), \cdots ,\mathcal{M}_q = (\mathcal{N}_q, \mathcal{I}_q)$ be $q$ matroids over overlapping groundsets. Let $\mathcal{N} = \mathcal{N}_1\cup  \cdots \cup \mathcal{N}_q$ and $\mathcal{I} = \{S\subseteq \mathcal{N} : S\cap \mathcal{N} \in \mathcal{I}_{\ell},
 \forall \ell\}$. The finite set system
$\mathcal{M}_p = (\mathcal{N} , \mathcal{I})$ is a $p$-matchoid if for every element $e \in \mathcal{N}$ , $e$ is a member of
at most $p$ matroids.
%$\mathcal{N}$ for at most $p$ indices $\ell \in [q]$.
\end{definition}

\begin{lemma} \label{extension}
For any matroid $\mathcal{M}$, with $rk(\mathcal{M})=k$. Every independent set $I \in \mathcal{I}$, with $|I|<k$ can be extended to a base $I'\supset I$, with $|I'|=k$.
\end{lemma}

\begin{lemma} 
\label{lem:Brualdi} (\textbf{Brualdi}~\cite{brualdi1969comments} )
If $A, B$ are any two bases of matroid $M$ then there exists a
bijection $\pi$ from $A$ to $B$, fixing $A\cap B$, such that $A - x + \pi(x) \in M$ for all $x \in A$.
\end{lemma}

In~\cite{us}, a $(n, m)$-ball-bin random set is defined  as follows.
A set of random variables $X_1, \cdots, X_m$ defined in the following way. Throw $n$ balls into
$m$ bins uniformly at random. Then set $X_j$ to be the number of balls in the $j$-th bin. They call the
resulting $X_j$’s a $(n, m)$-ball-bin random set. They use these variables to define $(\alpha,\beta)$-windows as follows.

\begin{definition}[$(\alpha,\beta)$ windows~\cite{us}] \label{def:windows}
Let $X_1,\ldots,X_{k\beta}$ be a $(n,k\beta)$-ball-bin random set.
Divide the indices $\{1,\ldots, n\}$ into $k\beta$ slots, where the $j$-th slot, $s_j$, consists of $X_j$ consecutive indices in the natural way, that is, slot $1$ contains the first $X_1$ indices, slot $2$ contains the next $X_2$, etc.
 %, $X_i$ would be the number of balls in the $i$-th bin.
 Next, we define $k/\alpha$ windows, where window $i$ consists of $\alpha \beta$ consecutive slots, in the same manner as we assigned slots.
\end{definition}

\begin{definition} \label{def:config}
For each item $e$ in the input, define $Y_e\in [k\beta]$ as the random variable indicating the slot in which $e$ appears. We call vector $Y\in [k\beta]^n$ a \textit{configuration}.
\end{definition}

To reduce notation, when clear from context, we will use $s$ and $w$ to also indicate
the set of items in the slot $s$ and window $w$ respectively.
\mcomment{is it necessary?}
Additionally, for any slots $s,s'$, %in window $w$, 
 we use notation $s \succ s'$ denotes $s$ appears after $s'$.
%all the slots $s$ in the sequence of slots in window $w$ that appear after slot $s'$ \mcomment{(not including $s'$)}.

%\subsection{Some useful properties of $(\alpha, \beta)$ windows}

\mcomment{
Some useful properties of $(\alpha, \beta)$ windows is proved in~\cite{us}. It is used in %Definition~\ref{def:windows} and used 
in Algorithm \ref{alg:main}. 
We summarize them in %this section.
the appendix. In particular we need the following:
}

%% file: cardinality.tex
%Besm allah alrahman alrahim
In this section, we focus on the cardinality constraints, namely submodular $k$-secretary problem with shortlists. 
\citet{us}, give a near optimal approximation algorithms for this problem using shortlist of size $O_{\epsilon}(k)$, where the hidden constant is $O(2^{poly(1/\epsilon)})$. 
Although the running time of their algorithm is linear in $n$, but the large hidden constant that exponentially depends on $1/\epsilon$ makes this algorithm far from practical. 
In this section we propose a fast algorithm that improves the dependency on $1/\epsilon$. We achieve an improved approximation ratio $1-1/e-\epsilon \mcomment{-O(1/k)}$ using shortlist of size $\shortlist$.
%But the the dependency of our algorithm on $1/\epsilon $ is $O(\frac{1}{\epsilon} \log (1/\epsilon))$.

\cardThm

We make some changes to the algorithm and analysis of~\citet{us}.
The main modification is in the way the algorithm selects elements inside a window. %Similar to~\cite{us}, 
The building block of the algorithm are a more advanced version of $(\alpha,\beta)$-windows defined in~\cite{us}, we call it stochastic windows (refer to Definition~\ref{def:windows}).
In contrast the algorithm does  not need to choose the best $\alpha$-subsequence $\tau^*$, and return the $\gamma(\tau^*)$ defined on that subsequence among  ${\alpha \beta \choose \alpha}$ many subsequences. This number of selections in a window is the reason for having a hidden constant in the $O_{\epsilon}(k)$ that exponentially depends on $1/\epsilon$. We alleviate the selection method in a window by keeping track of $\alpha$ subsets. We reduce the total number of selected items in a window to $\sqrt{L}\alpha \beta \log (1/\epsilon)$ and totally to $k\sqrt{L} \beta \log(1/\epsilon)=O\left(\frac{k}{\epsilon^2}\right)$. %(In the last section we even further improve this bound to make it independent of $\alpha$).
%Consequently, the hidden constant in $O_{\epsilon}(k)$ shortlist would be $ (1/\epsilon) \log(1/\epsilon)^2 =\tilde{O}(1/\epsilon)$.
Furthermore we improve the total number of queries and the running time to $\tilde{O}(n/\epsilon)$.
%in the final section.

\subsection{Algorithm description (cardinality constraint)}

The algorithm divides the input into $(\alpha,\beta)$-windows, $1,\cdots, W$.
%The output of the algorithm is $S$. 
We denote by $S$ the  solution that the algorithm keeps in each iteration.
It is initially an  empty set, and it will be incremented by adding a subset of  %$\alpha$
items 
in each window. Let's denote by $S_{1,\cdots, w}$, the set $S$ by the end of window $w$.
Additionally, the algorithm keeps track of all the selected items in a set $R$, that we call it a \textit{shortlist}. 
It is initially an empty set, and it grows  by adding each element that algorithm selects. The shortlist is the set of items that might be selected later on by the algorithm and can be added to the set $S$. Any other item that is not selected as part of the shortlist will be discarded immediately.  Let's denote by $R_{1,\cdots, w}$, the shortlist defined on the first $w$ windows.
Throughout the paper, if  the subscript of $S$ and $R$ is not stated explicitly, we mean $S_{1,\cdots,w-1}$ and $R_{1,\cdots, w-1}$ respectively. \mcomment{is it necessary}

In each window $w$, the algorithm keeps track of $L$ sets  $H_1,\cdots, H_{L}$, with $H_i$ being either $\emptyset$ or $|H_i|=i$.
In each slot $s$ in window $w$, the algorithm tends to select elements $m_i$ for $i\in U_s$ ($U_s$ is a range defined in Definition~\ref{def:q}). Each $m_i$  corresponds to set $H_i$. The element $m_i$ is the element with maximum marginal gain with respect to $S\cup H_i$.
\begin{equation*} \label{eq:maxmi}
  m_i\leftarrow\underset{x\in s\cup R}{\arg\max }  \Delta(x|S\cup H_i)
\end{equation*}

\begin{remark}
Each maximum element can be found in an online manner by the online max algorithm (Algorithm 1 in~\cite{us}) using shortlist of size $O(\log (1/\epsilon))$.
\end{remark}

At the end of slot $s$, for $i\in U_s$,
we add $ m_i$ to the shortlist $R$.
Moreover  \mcomment{Define +}, if $\Delta(H_{i+1}|S) < \Delta(H_{i}+ m_i|S)$,
we update 
%\begin{equation*} \label{eq:Hi}
$H_{i+1} \leftarrow H_{i}+ m_i$.
%\end{equation*}
%where $S$ is the solution selected in the previous windows.

\begin{definition}[Range for each slot] \label{def:q}
Define $q:=1-(1-\frac{1}{k\beta})^{k}$ for slot $s$. %(the probability is the same  for all slots).
Also define %$U_s:=(qs-\delta_s, qs+\delta_s)$
\begin{align}
U_s:=(\ell_{s},u_{s})=(q\cdot s-\delta_s, q\cdot s+\delta_s),
\end{align}
where $\delta_s:=4\sqrt{qs\log(1/\epsilon)} \le 4\sqrt{\alpha \log(1/\epsilon)}$.

\end{definition}

\begin{algorithm*}[h!]
  \caption{~\bf{ {\bf Cardinality-Constraint}}}
  \label{alg:cardconst} 
  \label{alg:tmp} 
\begin{algorithmic}[1]
\STATE Inputs:  submodular function $f$, window $w$, parameter $\epsilon \in (0,1]$, and set $S$, $R$ and $L$. \mcomment{$\alpha, \beta$} 

\STATE Initialize %$L\leftarrow \alpha(1+\sqrt{\alpha \log(1/\epsilon)})$, 
$H_{\ell} \leftarrow \emptyset, \forall 0\le \ell \le L$

\FOR {every slot $s$ in window $w$\mcomment{, $j=1,\ldots, \alpha\beta$} }
  %\STATE Concurrently:
  %\FOR{  $1 \le \ell \le L$, concurrently }
  
  \FOR{  $\ell \in U_s$}
  
  %subsequences of previous slots $\tau\subseteq \{s_1, \ldots, s_{j-1}\}$ of length $|\tau|<\alpha$ \label{li:subb}\\
  %\hspace{0.44in} in window $w$, 
  \STATE $R' \leftarrow Sample(R,1/(k\beta))$ \COMMENT{sample a set of size $|R|/(k\beta)$ from $R$}
  \STATE call the \textit{online max algorithm} (Algorithm 1 in~\cite{us} ) to compute, with probability $\epsilon/2$:  \\
  $m_{\ell} \leftarrow\underset{x\in s\cup R'}{\arg\max }  \Delta(x|S\cup H_{\ell-1}).$

  \mcomment{
  %\ref{alg:matroidmax} 
  with the following inputs: 
  %\begin{center}
  \begin{itemize}%[leftmargin=0.7in]
  \item   number of items $N=|s_j|+1$, $\delta=\frac{\epsilon}{2}$, and
  \item item values $I=(a_0, a_1, \ldots, a_{N-1})$, with 
      \begin{eqnarray*} 
  a_0 & := & \max_{x\in R} \Delta(x|S\cup H_i  ) \\
     a_\ell & := & \Delta(s_j(\ell)| S \cup H_i ),  \forall 0<\ell\le N-1
     \end{eqnarray*}
 where $s_j(\ell)$ denotes the $\ell^{th}$ item in the slot $s_j$. 
  \end{itemize}
}
\STATE $M_{\ell} \leftarrow$ 
The {shortlist} %and maximum element 
returned by
%output of 
the above \textit{online max algorithm} 
{for slot $s$ and set $H_{\ell-1}$.} 

%Add \hspace{0.44in} all items except the dummy item $0$ to the shortlist $A$. 
%That is, \label{li:sube} $$A\leftarrow A\cup  (A(j)\cap s_j)$$

% Add $m_i$ to the shortlist $R$. % for $1\le i \le \alpha$. 

\IF{$\Delta(H_{\ell}| S)   < \Delta(H_{\ell-1} + m_{\ell} | S)$ %and $(1+\epsilon) \times H_{\ell-1} < \Delta(H_{\ell-1} + m_{\ell} | S)$  
}
\STATE  $H_{\ell} \leftarrow H_{\ell-1} +m_{\ell}$
\STATE  $R \leftarrow R+ ( \{m_{\ell}\} \cap M_{\ell})$
\ENDIF
\ENDFOR
\ENDFOR
 
\STATE return $H_{L}$ 
%return $A$, $A^*$. %$A=T_1 \cup \cdots  \cup T_{W}$
\end{algorithmic}
\end{algorithm*}

\begin{algorithm*}[h!]
  \caption{~\bf{ {\bf Submodular}  Secretary with Shortlists}}
  \label{alg:card} 
  \label{alg:tmp} 
\begin{algorithmic}[1]
\STATE Inputs: number of items $n$, submodular function $f$, parameter $\epsilon \in (0,1]$. 
\STATE Initialize: $S \leftarrow \emptyset, R
\leftarrow \emptyset$, 
constants $\alpha \ge 1, \beta \ge 1$, $L \leftarrow u_{\alpha \beta}$.
%which depend on the constant $\epsilon$.
\STATE Divide indices $\{1,\ldots, n\}$ into $(\alpha, \beta)$ windows. 
\FOR {window $w= 1, \ldots, W=k/\alpha$} 
\STATE $S_w \leftarrow$ Cardinality-Constraint$(S,w,R, L)$
\STATE $S \leftarrow S \cup S_w$
\ENDFOR
\STATE $S' \leftarrow \text{a sample of size } k \text{ from } S$
\STATE return $S' \cap R$
\mcomment{\STATE return $A$, $A^*$. }
\end{algorithmic}
\end{algorithm*}

\subsection{Analysis of the algorithm: cardinality constraint}

In this section, we prove Theorem~\ref{cardtheorem}.
First of all, we can bound the size of shortlist:

%Now we give an overview of the approximation guarantee of the algorithm.

\textbf{Proof overview.}
%Similar to~\cite{us}, 
We first lower bound $\Ex[f(S_{1,\cdots, W})]$,
\mcomment{define OPT, W}
and then we can lower bound $\Ex[f(S\cap R)]$.
In particular, \mcomment{in proposition ?} we prove competitive ratio $1-1/e-\epsilon \mcomment{-O(1/k)}$ for Algorithm~\ref{alg:card} by choosing large enough parameters $\alpha, \beta$  that are depending on $1/\epsilon$. %\nameOfProblemSL. 
Similar to~\cite{us}, a crucial idea is to show that given the \textit{history} of the selection made by the algorithm in  windows $1,\cdots, w-1$, the probability that any of the $k$ items in the optimal solution $S^*$ \mcomment{define $S^*$}  appears either in $w$ or in the shortlist $R$ is at least $\frac{\alpha}{k}$.
Additionally, the elements of $S^*$ are distributed independently and uniformly at random in the $\alpha \beta$ slots of $w$.
Since we have modified the algorithm, the structure of the elements that get selected in windows $1,\cdots, w-1$ (the history of the algorithm) is slightly different from the structure of selected elements in Algorithm 2 in~\cite{us}, namely $T_{1,\cdots,w-1}$ \mcomment{Define T in our paper before this section} (refer to Definition 3 in~\cite{us}). However,  we are still able to prove the aforementioned property. The main reason is that under new selection criteria in our Algorithm~\ref{alg:card}, removing one item that is not selected by the algorithm would not change the output of the $\arg\max$ in line 5 of Algorithm~\ref{alg:card}. Hence if we remove one of the items not selected by the algorithm, all the subsets $H_i$ in a window remain unchanged and consequently %$S_w$ and 
$S$ remains unchanged. Therefore, we can still prove similar  properties proven for $(\alpha,\beta)$-windows in~\cite{us}.

\mcomment{
By choosing $\alpha$ and $\beta$ large enough, it is easy to see that
there are roughly $\alpha$ elements of $S^*$ in a window $w$, and they appear in different slots w.h.p.
The algorithm in~\cite{us}, tries to identify the slots in $w$ containing elements of $S^*$. In particular, it is looking for an $\alpha$-subsequence of slots $\tilde{\tau}_w\mcomment{=\{s_1,\cdots,s_{\alpha}\}}$ containing elements of $S^* \cap w$.
Because the algorithm is not aware of the optimal solution and the fact that elements arrive in an online manner, it is not possible to {predict those slots}.
The idea in~\cite{us}, is to try all $\alpha$-subsequences $\tau$ of slots in $w$ and choose the one with highest marginal gain
(tries as many as 
${\alpha \beta \choose \alpha}$, \mcomment{$\alpha$-}subsequences).
 Moreover, knowing the slots in $w$ containing the $S^*$ is not enough to know which element in those slots are part of $S^*$. Thus in each slot they choose the element with the highest   marginal gain with respect to the elements selected in the previous slots (refer to the definition of $\gamma$ function in~\cite{us}, eq.~(1)). Consequently, they argue that the  marginal gain of the selected item in each  slot of $\tilde{\tau}_w$ is more than the marginal gain of the element of $S^*$ in the same slot. 
 }

 In our analysis, we use a novel variant of a window that we call it stochastic window. It is defined only for the purpose of analysis.   
 In regular window that consists of subsequent slots in one instance of randomly ordered input (or a configuration, refer to Definition~\ref{def:config}). But in the stochastic window after each slot we can switch 
 to another configuration that is consistent with current history $T$. The expected gain of the algorithm can be equivalently analyzed by combining   stochastic windows. 
 In the analysis, we lower bound the expected marginal gain of the algorithm in slots that we call them  \textit{active slots} ( Definition~\ref{def:active}).
 %A randomly selected element in an 
 Active slots contain an element that can be equally any element of $\OPT$. Thus we can lower bound the marginal gain in such slots.
 
 The new structure of stochastic windows helps us to  simplify the analysis by  eliminating dependency of elements appearing in a slot and the elements appeared in the previous slots that are not part of history $T$,  more precisely the total number of active slots seen previously (refer to Lemma~\ref{lemma:zsj}). 
  Note that the algorithm is not aware of the position of active slots. In the algorithm we keep track of sets $H_{\ell}$ ($0\le \ell \le L$) each with size of $\ell$. Intuitively, the selection made in the $j$-th active slot in a stochastic window increases marginal difference of level $j$ and $j-1$.
   Finally, we write recursive formulas to compute the expected marginal gain of the algorithm (eq.~\eqref{eq:wrec}).

\mcomment{ 
 Here, in this paper we argue that it is not necessary to consider all $\alpha$-subsequnces in window $w$. 
Consider $\alpha$-subsequence $\tilde{\tau}_w=\{s_1,\cdots, s_t\}$, containing $S^*\cap w$. \mcomment{by s and w we mean elements in each} The algorithm in~\cite{us} greedily chooses the set $\gamma(\tilde{\tau}_w)=\{i_1,\cdots, i_t\}$ containing one element from each slot $i_j$. Then, they lower bound the marginal gain of $i_j$ with respect to previously selected elements $S \cup \{i_1,\cdots, i_{j-1}\}$.
More precisely, using the aforementioned crucial property stated above, \mcomment{refer to lemma in this paper}
in their Lemma 11,  they show that for all $j=1,\ldots, t$, 
{\small 
\begin{equation}\label{eq:optdiff}
\Ex[\Delta_f(i_j|S\mcomment{_{1,\ldots, w-1}} \cup \{i_1, \ldots, i_{j-1}\})|T_{1,\ldots, w-1}, i_1, \ldots, i_{j-1}] \ge \frac{1}{k}\left((1-\frac{\alpha}{k})f(S^*)-f(S\mcomment{_{1,\ldots, w-1}} \cup \{i_1, \ldots, i_{j-1}\})\right)\ . 
\end{equation}
}
Now consider  subseuquence $\tilde{\tau}_w=\{s_1,\cdots, s_t\}$. Modify the method in~\cite{us} for selecting corresponding elements in this subsequence, namely $\gamma(\tau) \mcomment{=\{i_1,\cdots, i_t\}}$ \mcomment{(refer to eq.~(1) in~\cite{us})} as follows.
Suppose that after selecting $i_j$ in slot $s_j$, the algorithm substitutes the current $j$ elements selected so far from slots $s_1,\cdots, s_j$
\mcomment{as part of $\gamma(\tilde{\tau}_w)$ }
with
a subset $C_j$ of $w$ \mcomment{ of size $j$}, whose marginal gain w.r.t. $S$ is larger (the elements of $C_j$ do not need to be from slots in $\tilde{\tau}_w$, they should be from the shortlist).
\mcomment{more than the marginal gain of the first $j$ elements of $\gamma(\tilde{\tau}_w)$.} %namely $\{i_1,\cdots, i_j\}$.
%Let's call this subset $C_j$.
Thus, for all $j=1,\cdots, t$, 
\begin{equation}\label{eq:cprop}
\Delta(C_j|S) \ge
\Delta(C_{j-1} + i_j |S).
%\Delta(\{i_1,\cdots, i_j\}|S).
\end{equation}

where \mcomment{argmax latex}
\[
i_j:= \underset{x\in s_j\cup R}{\arg\max } {\Delta(x|S\cup C_{j-1})}.
\]
Given that the subset $C_j$ selected by the algorithm is part of the shortlist, and conditional on the \textit{history} of the selection made by algorithm, i.e., $T$, we can prove an equation similar to eq.~\eqref{eq:optdiff} for subsets $C_j$. 
{\small 
\begin{align*}
\Ex[\Delta_f(i_j|S\mcomment{_{1,\ldots, w-1}} \cup C_{j-1})|T]  \ge  \frac{1}{k}\left(f(S^*)-f(S\mcomment{_{1,\ldots, w-1}} \cup C_{j-1})\right)\ . 
\end{align*}
}
Consequently, \mcomment{similar to Lemma 12 in~\cite{us}} we can show
{\small 
\begin{align}\label{eq:oneovereC}
&\Ex[ f(S^*) - f(S_{1,\ldots, w-1} \cup  C_j) | T] \\ &\le 
\Ex[ f(S^*) - f(S_{1,\ldots, w-1} \cup  C_{j-1} \cup \{i_j\}) | T] \\ &\le 
%e^{-j/k}
\frac{1}{e}
\left( f(S^*)-f(S_{1,\ldots, w-1} \cup C_{j-1})\right)\ .
\end{align}
}

}
 %Instead, in our algorithm we keep track of $\alpha$ subsets $H_1, \cdots, H_{\alpha}$. 

\mcomment{
 Now the question is how should the algorithm create such sets $C_j$ in a \mcomment{memory efficient way and } online manner. 
 Note that the algorithm is not aware of $\tilde{\tau}_w$, the slots containing $S^*\cap w$.
  In order to do that, the algorithm keeps track of $\alpha$ sets $H_1,\cdots, H_{\alpha}$.
 One main observation is the following:
 \begin{lemma}
 Suppose $\tilde{\tau}_w$ is the subsequence containing $S^*\cap w$. For $0\le j < \alpha$, define $C_j$ to be the latest set $H_j$ before observing slot $s_{j+1}$, and $C_{\alpha}$ to be the set $H_{\alpha}$ at the end of window $w$. Then, $\{C_j\}_{j=1}^{\alpha}$ satisfy equation~\eqref{eq:cprop} and therefore~\eqref{eq:oneovereC}. \mcomment{$j=1 \text{ to } t$}
 \end{lemma}

The intuition behind the proof is inductive.  Suppose at the time the algorithm arrives in the slot $s_j$, we know $H_{j-1}$ satisfies eq.~\eqref{eq:cprop} and we set $C_{j-1}=H_{j-1}$. Then the algorithm find the element with maximum marginal gain w.r.t. $S\cup H_{j}$, namely $i_j$. Now from line 8 of the algorithm\mcomment{ref}, either
\[
\Delta(H_j|S ) \ge \Delta(H_{j-1}+i_j|S)
\]
or we update $H_j=H_{j-1}+i_j$. In either case and later on in the algorithm we will have
\[
\Delta(H_j|S ) \ge \Delta(C_{j-1}+i_j|S)
\]
Thus this property holds true for the latest $H_j$ before observing slot $s_{j+1}$. Thus by setting $C_{j}=H_{j}$, before observing $s_{j+1}$, the eq.~\eqref{eq:cprop} holds true for $C_{j+1}$ too.
Note that the algorithm is not aware of the position of slots $s_j$, and therefore the sets $\{C_j\}_{j=1}^{\alpha}$. That is why the algorithm keeps track of all $\alpha$ subsets $H_i$, each corresponding to subsets of size $i=1,\cdots, \alpha$.
}
 
%Intuitively, if the algorithm knows that some slot $s$ in $w$  contains an element of $S^*$, it still does not know how many other element of $S^*\cap w$ appear before $s$. Thus it could be the first, the second, $\cdots$, or the $\alpha$-th \mcomment{t=?} element of  $S^*\cap w$. Suppose it is the $j$-th element. The algorithm should find $C_j$. 

%\subsection{Bounding $\Ex[f(S_{1,\cdots, W})]$ \mcomment{is $W$  defined?}}
\subsubsection{Analysis of the Algorithm: Bounding $\Ex[f(s)]$ }

%\scomment{Begin:Trying out something------------------------}

%\begin{lemma}
%\label{lem:indepi}
%For any window $w$, slot $s,s'$ in window %$w, \ldots, W$, and $i\in S^*$, 
%$$\mathbb{P}(Y_i=s | T_{1,\ldots, w-1})=\mathbb{P}(Y_i=s' | T_{1,\ldots, w-1})$$
%\end{lemma}
%\begin{proof}
%By applying Bayes rule to Lemma \ref{afterw}.
%\end{proof}

In this section, %we use the observations from the previous sections to 
we lower bound the expected value of $f(S)$ (Theorem~\ref{cardtheorem}). First we define a few notations that will be used in the proof.
%to show the existence of a random subsequence of slots $\tau_w$ of window $w$ such that we can lower bound %$f(S_{1,\ldots, w-1}\cup S_w)- f(S_{1,\ldots, w-1})$ in terms of $\OPT -f(S_{1,\ldots, w-1})$. This will be used to lower bound 
%the expected increment $\Delta_f(S_w|S) = f(S\cup S_w) -f(S)$ for a window $w$ over all possible configurations.

%Similar to~\cite{us}, define $H_{\ell}^s$ to be the set $H_{\ell}$ in the Algorithm~\ref{alg:card} at the end of slot $s$.

\begin{definition}[Hierarchy of Selections]
Let's denote by $H_{\ell}^s$, the set $H_{\ell}$  defined in the Algorithm~\ref{alg:card} at the end of slot $s$ when the algorithm runs on a sequence of slots.
%( or a (stochastic) window $w$ ). 
%Define $T_w:=\bigcup_{i=1}^{\alpha} H_i$.
\mcomment{
More precisely, define $H_{\ell}^s$ recursively as:
\[
H_{0}^{s} = \emptyset
\]
\[
H_{\ell}^{s} := H_{\ell-1}^{s} + \underset{}{}
\]}
\end{definition}

Note that the above definition is general in the sense that the sequence of slots are not necessarily a randomly ordered  instance of the input ${\mathcal{ U}}$.
(especially, it also works for stochastic windows defined below)

%define $T$
%Define $T(w,s)$ as  
Then define the following random variable $T(w,s)$ that will track all the  selections made by the algorithm up to slot $s$ in a sequence of slots ( it could be a window or a stochastic window defined in Definition~\ref{def:stocwind}).
\begin{definition} [History]\label{def:T}
For slot $s$ in (stochastic) window $w$ define 
\begin{align*}
T(w,s):= \{(H_{\ell}^{s'},\ell, s')|s\succ s', 1\le \ell \le L\}, and \\
Supp(T(w,s)):=\bigcup_{1\le \ell \le L, s\succ s'} H_{\ell}^{s'},
\end{align*}
If the configuration is not clear from the context, we make the notation explicit by $T(w,s)(Y)$ for configuration $Y$ (refer to Lemma~\ref{config}).
We use shorthand $R(w,s)$ to denote $Supp(T(w,s))$. Furthermore, we denote by $T_{1,\cdots, w-1}$  the subset of history defined before  $w$ and by $T_w$ the subset defined on $w$.
\end{definition}

Now, we define a \textit{stochastic $(\alpha,\beta)$-window} $\bar{w}$.
A window as defined in Definition~\ref{def:windows} refers to  subsequent elements in an instance of the input arrival (or a configuration refer to Definition~\ref{def:config}). However, we define a \textit{stochastic window} $\bar{w}$ as defined below. The main difference is that elements of different slots belong to different configurations, under one condition that the configuration of a slot must satisfy the history of algorithm if it is run on previous slots. More precisely

\begin{definition}[Stochastic $(\alpha,\beta)$-Window]\label{def:stocwind}
 Define a \textit{stochastic window} $\bar{w}$ containing $\alpha\beta$ subsequent slots  $s_1, \cdots, s_{\alpha\beta}$ s.t. elements of  each slot  are coming from same configuration (Definition~\ref{def:config}). 
More precisely, 
the elements of slot $s_i$ in $\bar w$  are elements of slot $s_i$ in configuration  $Y_i$,  $1\le i \le \alpha \beta$.
%$Y_1,\cdots, Y_{\alpha \beta}$.
%These configurations must satisfy the property that $T(\bar w,s_i)(Y_{i+1}) = T(\bar w,s_{i})(Y_{i})$ for $1\le i \le \alpha \beta$.
Suppose $Y_1,\cdots, Y_{i}$ are selected, then  $Y_{i+1}$ is being selected uniformly at random from all the configurations satisfying 
$T(\bar w,s_i)(Y_{i+1}) = T(\bar w,s_{i})(Y_{i})$ (The initial case $T(\bar w, s_0)(Y_0)=T_{1,\cdots, w-1}$ is the history of algorithm in previous windows).

%For slot $s$, we choose elements in $s$ given $T(w,s)$, i.e., for item $e\in I$, $\Pr(e\in s)=\Pr(e\in s|T(w,s))$. 
\end{definition}

\toRemove{
\begin{definition}[Stochastic $(\alpha,\beta)$-Window]\label{def:stocwind}
 Define a \textit{stochastic window} $\bar{w}$ defined on slots $s_1,\cdots, s_{\alpha\beta}$ s.t. elements of  each slot  are coming from one configuration. 
More precisely, 
the elements of slot $s_i$ in $\bar w$  are elements of slot $s_i$ in configuration  $Y_i$,  $1\le i \le \alpha \beta$.
%$Y_1,\cdots, Y_{\alpha \beta}$.
%These configurations must satisfy the property that $T(\bar w,s_i)(Y_{i+1}) = T(\bar w,s_{i})(Y_{i})$ for $1\le i \le \alpha \beta$.
Suppose $Y_1,\cdots, Y_i$ are selected, then  $Y_{i+1}$ is selected uniformly at random from all the configurations satisfying 
$T(\bar w,s_i)(Y_{i+1}) = T(\bar w,s_{i})(Y_{i})$ (The initial case $T(\bar w, s_0)(Y_0)=T_{1,\cdots, w-1}$ is the history of algorithm in previous windows).

%For slot $s$, we choose elements in $s$ given $T(w,s)$, i.e., for item $e\in I$, $\Pr(e\in s)=\Pr(e\in s|T(w,s))$. 
\end{definition}
}

An implication of the above definition is that there is a universal $T=T(\bar w,s_{\alpha\beta})$  for the entire stochastic window $\bar{w}$, such that $Y_i$ must be compatible with $T(\bar w,s_i)$ for $1\le i\le \alpha \beta$.
Also note that there might not exist a sequence of items with the same elements as in a stochastic window. Since there might be several copies of an element in a stochastic window (in different slots). 

%First for a window $w=s_1,\cdots, s_{\alpha\beta}$ define the following random process: X
The following lemma implies that instead of finding the expected marginal gain of the algorithm in a window $w$  w.r.t. previously selected elements $S$, we can find the expected marginal gain in a stochastic window
defined on slots with the same indices.
$\bar w$ w.r.t. $S$. %stochastic windows 

\begin{lemma}
%Furthermore, 
Suppose $S$ is the  solution  kept by the algorithm over slots $1, \cdots, s$, in an instance of randomly-ordered input ${\mathcal{ U}}$. 
The expected marginal gain of the element selected by the algorithm in $w$ with respect to $S$, i.e., $S_w$ can be written as
\begin{align*}
    \Ex_{w}[\Delta_f(S_w|S) | T_{1,\cdots, w-1}] = %\Ex_{T_w}[\Delta_f(S_w|S) | T_{1,\cdots, w-1}] =
    \Ex_{\bar w}[\Delta_f(S_{\bar w}|S) | T_{1,\cdots, w-1}] \ .
    %\sum_{1 \le i \le \alpha \beta} \Delta_f()
\end{align*}
Here, $S_{\bar w}$ is the the selection made by the algorithm in $\bar w$.
\end{lemma}
\begin{proof}
%The first equality is trivial. 
We can write
$$
\Ex_{w}[\Delta_f(S_w|S) | T_{1,\cdots, w-1}] = \Ex_{T_w}[\Delta_f(S_w|S) | T_{1,\cdots, w-1}] =  \Ex_{T_{\bar w}}[\Delta_f(S_{\bar w}|S) | T_{1,\cdots, w-1}] \ .
$$
The first equality is based on definition of $T_w$. The second one is because of the definition of $\bar w$.
For the second equality, we need to make sure that for any given $T$,
$\Pr(T_w=T) = \Pr(T_{\bar w}=T)$.
%It is because 

We prove it by induction on the index of slot $s$ in $\bar w$. 
We denote the configuration corresponding to $w$, by $Y$. 
For $s_0$, by definition $T(\bar w, s_0)(Y_0)=T_{1,\cdots, w-1}= T(w,s_0)$. By induction hypothesis, for slot $s_{i}$ and any
 given $T$, we have
$\Pr(T(\bar w,s_i)(Y_i)=T) = \Pr(T(w,s_i)(Y)=T)$. Now for slot $s_{i+1}$ and given 
$T'=T+U$, where $T\subseteq T'$ is the subset of $T'$ up to slot $s_i$, then
\begin{align*}
\Pr(T(\bar w,s_{i+1})(Y_{i+1})=T') &=\Pr(T(\bar w,s_{i})(Y_i)=T) \times \Pr\left(T(\bar w,s_{i+1})(Y_{i+1})
=T' | T(\bar w,s_{i})(Y_i)=T \right) \\
&= \Pr(T(w,s_{i})(Y)=T) \times \Pr\left(T(\bar w,s_{i+1})(Y_{i+1})
=T' | T(\bar w,s_{i})(Y_i)=T \right)
\end{align*}
Now 
\begin{align*}
\Pr\left(T(\bar w,s_{i+1})(Y_{i+1})
=T' | T(\bar w,s_{i})(Y_i)=T \right) &= \frac{\text{number of configurations satisfying } T'}{\text{number of configurations satisfying } T} \\
&=\Pr\left(T( w,s_{i+1})(Y)
=T' | T( w,s_{i})(Y)=T \right) \ .
\end{align*}

\end{proof}

\mcomment{high level idea of definition and next lemma}

%\begin{definition}
%We use shorthand $R(\bar w,s)$ to denote $Supp(T(\bar w,s))$.
%\end{definition}
\begin{prop}
For the final set $S$ returned by the algorithm over slots $1,\cdots, k\beta$, we can decompose 
the expected value of $f(S)$ over $k/\alpha$ stochastic windows as follows.
$$
\Ex[f(S)]= \sum_{i=1}^{k/\alpha} \Ex_{\bar w_i} \Delta_f(S_{\bar w_i}|S) \ ,
$$
where $\bar w_i$ is stochastic window creating the  $i \alpha \beta$-th, $\cdots, $ $(i+1)\alpha \beta -1$-th slots.
\end{prop}
The rest of this section lower bounds the expected marginal gain of the elements algorithm selects in a stochastic window.

%Now we create a random subsequence $\tau_w$ of slots in stochastic window $\bar{w}$ as follows:

\begin{definition}[active slots]  \label{def:active}
%$Z_s$ and $\tau_{\bar w}$]
\label{def:tauw}
For every slot $s$ in the stochastic window $\bar w$, create set $Z_s \subseteq S^*$ as follows:
%Create sets of items $Z_s, \forall s\in w$  as follows: 
add every item from $i\in S^*\cap s$ independently with probability $\frac{1}{k \beta p_{i,s}}$ to $Z_s$, where 
$p_{i,s}:=\Pr(i \in s \cup Supp(T(\bar w,s)) | T(\bar w,s))$.
%(where $p_{i,s}$ is defined in eq.~\ref{eq:pijBound}).
Then, for every item $i\in S^*\cap Supp(T(\bar w,s))$, with probability $\frac{1}{k\beta}$, add $i$ to $Z_s$. 
 %for all $s\in w$.
 %\mcomment{for a randomly chosen slot $s$ in $w$.} 
 Furthermore, define subsequence $\tau_{\bar w}=(s_1, \ldots, s_t)$ as the sequence of slots in $\bar w$ with $Z_s\ne \emptyset$, we call them active slots.
\end{definition}

Now we define a notation $t(\bar w, s)$ representing the total number of active slots seen before a slot $s$. If it is greater than or equal $u_s$ we use $u_s-1$ instead.
\begin{definition}
Define $t(\bar w,s):= \sum_{s'\preceq s} 1(Z_{s'}\ne \emptyset) $ and  $r(\bar w,s) = \min \{ \sum_{s'\preceq s} 1(Z_{s'}\ne \emptyset) , u_s-1 \} $, where $u_s$ is defined in Definition~\ref{def:q}.
\end{definition}

\mcomment{
\begin{definition}[$Z_s$ and $\tau_w$]
\label{def:tauw}
 Create sets of items $Z_s, \forall s\in w$  as follows: for every slot $s$, add every item from $i\in S^*\cap s$ independently with probability $\frac{1}{k \beta p_{i,s}}$ to $Z_s$. Then, for every item $i\in S^*\cap Supp(T_{1,\cdots, w-1})$, with probability $\alpha/k$, add $i$ to $Z_s$ for all $s\in w$.
 %\mcomment{for a randomly chosen slot $s$ in $w$.} 
 Define subsequence $\tau_w$ as the sequence of 
slots with $Z_s\ne \emptyset$. 
\end{definition}
}

\begin{lemma}
\label{lem:Zs}
For any slot $s$ in a stochastic window $\bar w$, 
given  $T(\bar w,s)$, 
all $i, i' \in S^*, i\ne i'$ will appear in $Z_s$ independently with probability $\frac{1}{k\beta}$, i.e.,  the random variables  $\mathbf{1}(i\in Z_s | T(\bar w,s))$ are i.i.d. for all $i \in S^*$, and 
\[
\Pr(i \in Z_s | T(\bar w,s)) = \Pr(i'\in Z_s | T(\bar w,s)) = \frac{1}{k\beta} \ .
\]
%Also, given $T$, for every $i \in S^*$, the probability to appear in $Z_s$ is equal for all slots $s$ in window $w$.Further, each $i\in S^*$ occurs in  $Z_s$ of at most one slot $s$.
\end{lemma}
\mcomment{
\begin{lemma}
\label{lem:Zs}
Given any $T_{1,\ldots, w-1}=T$, for any slot $s$ in window $w$, all $i, i' \in S^*, i\ne i'$ will appear in $Z_s$ independently with probability $\frac{1}{k\beta}$. Also, given $T$, for every $i \in S^*$, the probability to appear in $Z_s$ is equal for all slots $s$ in window $w$. Further, each $i\in S^*$ occurs in  $Z_s$ of at most one slot $s$.
\end{lemma}
}
{
\begin{proof}
The proof is similar to Lemma 10 in~\cite{us}, and it is a direct implication of Lemma~\ref{lem:ijindep} in the Appendix.
%which is for the new construction of $T(\bar w,s)$. \mcomment{full proofin appendix}
\end{proof}
}

\mcomment{
%\begin{proof}
First consider $i\in S^*\cap Supp(T)$. Then, $\Pr(i\in Z_s|T) = \frac{\alpha}{k}\times \frac{1}{\alpha \beta} = \frac{1}{k\beta}$ by construction. Also, the event $i\in Z_s|T$ is independent from $i'\in Z_s|T$ for any $i'\in S^*$ as $i$ is independently assigned to a $Z_s$ in construction. Further, every $\in S^*\cap T$ is assigned with equal probability to every slot in $s$.
}
\mcomment{
Now, consider $i\in S^*, i\notin Supp(T)$. Then, for all slots $s$ in window $w$,
$$\Pr(i \in Z_s|T) =\Pr(Y_i=s | T) \frac{1}{p_{i,s}k\beta} =  p_{i,s}\times \frac{1}{p_{i,s}k\beta} = \frac{1}{k\beta},$$
where $p_{i,s}$ is defined in \eqref{eq:pijBound}. We used that 
$p_{i,s}=\Pr(Y_i=s|T)$ for $i\notin Supp(T)$. 
Independence of events  $i\in Z_s|T$ for items in $S^*\backslash Supp(T)$ follows from Lemma \ref{lem:ijindep}, which ensures $Y_i=s|T$ and  $Y_j=s|T$ are independent for $i\ne j$; and from independent selection among items with $Y_i=s$ into $Z_s$. 
}
\mcomment{
The fact that every $i\in S^*$ occurs in at most one $Z_s$ follows from construction: $i$ is assigned to $Z_s$ of only one slot if $i\in Supp(T)$; and for $i\notin Supp(T)$, it can only appear in $Z_s$ if $i$ appears in slot $s$.
%\end{proof}
}

% We will show that the subsequence $(a_1, \ldots, a_{\alpha \beta})$ has close to $\alpha$ non-zero  items, and will lower bound the increment provided by the greedy subsequence $S_w=\gamma(\tau^*)$ over $f(S_{1,\ldots, w-1})$ by the increment provided by this subsequence. 

%\begin{lemma}
%For any slot $s$ in $w$, given $T_{1,\ldots, w-1}=T$, variables $\mathbf{1}(i\in Z_s|T_{1,\ldots, w-1}=T)$ for $i\in S^*$ are i.i.d. with probability $\frac{1}{k\beta}$ to be true. 
%\end{lemma}
%\begin{proof}
%\scomment{Should be straightforward by previous lemmas}
%\end{proof}

\mcomment{high level idea of this lemma}
From the above lemma we can see that for any active slot a randomly selected element in $Z_s$ is equally likely to be any element of $S^*$.
\begin{prop}\label{lemma:zs}
For $i, i' \in S^*$, {and slot $s$ in $\bar w$},  %$\tilde{\tau}_w=(s_1,\cdots, s_t)$,
\begin{align*} 
\label{eq:pk}
&\Pr(i\in Z_{s}| T(\bar w,s), Z_s\ne \emptyset)
= \Pr(i'\in Z_{s}|T(\bar w,s), Z_s \ne \emptyset)  
\ge \frac{1}{k} \ .
\end{align*}
\end{prop}

An important advantage of stochastic windows is the following lemma. 
\begin{lemma}\label{lemma:zsj}
For $i, i' \in S^*$, {and slot $s$ in $\bar w$}, and $1\le j \le t$, %and $\tilde{\tau}_w=(s_1,\cdots, s_t)$,
\begin{align*} 
%\label{eq:pk}
&\Pr(i\in Z_{s}| T(\bar w,s), s=s_j)
= \Pr(i'\in Z_{s}|T(\bar w,s), s=s_j)  
\ge \frac{1}{k} \ .
\end{align*}
\end{lemma}
\begin{proof}
The proof is from previous lemma and the fact that $Y_s$ only depends on $T(\bar w,s)$ and is independent of the rest of the elements in $Y_1, \cdots ,Y_{s-1}$ and therefore it is independent of $t(\bar w, s-1)$. More precisely,
\rcomment{
\begin{align*}
\Pr(i\in Z_s| T(\bar w,s), s=s_j) &= \Pr(i\in Z_s| T(\bar w,s), Z_S\ne \emptyset, \exists Y_1,\cdots, Y_{s-1} \text{ satisfying } T(\bar w,s) \text{ s.t. } s=s_j ) \\
&= \Pr(i\in Z_s|T(\bar w,s), Z_s\ne \emptyset) \times \mathbb{I}(\exists Y_1,\cdots, Y_{s-1} \text{ satisfying } T(\bar w,s) \text{ s.t. } s=s_j)
\end{align*}
}
\begin{align*}
\Pr(i\in Z_s| T(\bar w,s), s=s_j) &= \Pr(i\in Z_s| T(\bar w,s), Z_s\ne \emptyset, t(\bar w,s-1)=j-1 ) \\
&= \Pr(i\in Z_s|T(\bar w,s), Z_s\ne \emptyset) 
\end{align*}
%The last line is because the event $(\exists Y_1,\cdots, Y_{s-1} \text{ satisfying } T(\bar w,s) \text{ s.t. } s=s_j)$ is independent of $Z_s$.
The last line is because $Z_s$ and $t(\bar w,s-1)$ are independent.
\end{proof}

\mcomment{
\begin{lemma}\label{lemma:zs}
For $i, i' \in S^*$, and $\tilde{\tau}_w=(s_1,\cdots, s_t)$,
\begin{align} 
\label{eq:pk}
&\Pr(i\in Z_{s_j}| T(w,s_j))
= \Pr(i'\in Z_{s_j}|T(w,s_j))  
\ge \frac{1}{k} \ .
\end{align}
\end{lemma}
}

\mcomment{
\begin{lemma}\label{lemma:zs}
For all $i, i' \in S^*\backslash \{Z_{s_1} \cup \ldots \cup Z_{s_{j-1}}\}$,
\begin{align} 
\label{eq:pk}
&\Pr(i\in Z_{s_j}| Z_{s_1}, \ldots , Z_{s_{j-1}}, T_{1,\cdots, w-1})
= \Pr(i'\in Z_{s_j}| Z_{s_1}  \ldots  Z_{s_{j-1}},T_{1,\cdots, w-1})  
\ge \frac{1}{k} \ .
\end{align}
\end{lemma}
}

\begin{definition} \label{def:ml}
For stochastic window $\bar w$,
define $m_{\ell}^s$ to be $m_{\ell}$ as defined in Algorithm~\ref{alg:cardconst} running over $\bar w$, at the end of slot $s$, which is
%as defined in eq.~\eqref{eq:maxmi}.
\begin{equation*}
m_{\ell}^s:=\underset{x\in s\cup R(\bar w,s)}{\arg\max}{\Delta(x|S\cup H_{\ell-1}^{s-1})}, 
\end{equation*}

\rcomment{
Also for the sequence \mcomment{could it be a set} $\tilde{\tau}_{\bar w}=(s_1,\cdots, s_t)$ defined in Definition~\ref{def:tauw}, define sequence $\mcomment{\mu_w} \mu_w=(i_1,\cdots, i_{\alpha'})$, for $\alpha'=\min(t,\alpha)$,
where
\begin{equation}
i_j:=m_{j}^{s_j},
\end{equation}
}
\rcomment{
Moreover, for $1\le j \le \alpha'$ define 
\begin{equation}
C_{j}:=H_{j}^{ (s_{j+1}) -1} \ ,
\end{equation}
}
\rcomment{
if $j+1> t$, set $s_{j+1}:=\alpha \beta+1$. 
\mcomment{check again define $s_{t}=\infty$ }
}
\rcomment{
We also use the notation $i_{1,\cdots, j}=(i_1,\cdots, i_j)$, for $1\le j\le \alpha'$.
}
\end{definition}

\rcomment{
The following observations are immediate:
\begin{prop} \label{lem:inc}
For slot $s\in w$, and $1\le \ell \le \alpha$,
\begin{equation} \label{eq:hplusm}
    \Delta(H_{\ell}^s |S) \ge \Delta(H_{\ell-1}^{s-1} + m_{\ell}^s |S) %\mcomment{write it as \Delta}
\end{equation}
\mcomment{is this used?}
\begin{equation}
\Delta(H_{\ell}^{s} |S ) \ge \Delta(H_{\ell-1}^{s} |S),
\end{equation}
\begin{equation}
\Delta(H_{\ell}^{s+1} |S ) \ge \Delta(H_{\ell}^{s} |S),
\end{equation}
also
\begin{equation} \label{eq:CL}
\Delta(C_{\ell}|S) \ge \Delta(C_{\ell-1}|S),
\end{equation}
moreover, 
\begin{equation}
\Delta(S_w|S) \ge \Delta(C_{\alpha}|S).
\end{equation}
\end{prop}
}

\mcomment{ add high levela idea.
Suppose $\tilde{\tau}_w$ is as defined in definition~\ref{def:tauw}, and
In the following lemma we lower bound the marginal gain of $i_j$, the element 
a randomly picked element of optimal solution in slot $s_j$ with respect to previously selected items.
}

\begin{lemma}
\label{lem:margij}
For slot $s$ in stochastic window $\bar w$, given $T$ and $1\le \ell \le L$, {
$$%\underset{Z_s}
\underset{\bar w}
{\Ex}\left[\Delta(m_{\ell}^s|S \cup H_{\ell-1}^{s-1}) | T(\bar w,s)=T, Z_s\ne \emptyset \right] \ge \frac{1}{k} \left(f(S^*)-f(S \cup H_{\ell-1}^{s-1}
)\right)\ .$$}
\end{lemma}

\mcomment{
\begin{lemma}
\label{lem:margij}
Given the sequence $\tau_w=(s_1, \ldots, s_t)$ defined in Definition \ref{def:tauw}, and $\mu_w=\{i_1,\cdots, i_{\alpha'}\}$ defined in Definition~\ref{def:ml},
for $1\le j \le \alpha'=\min(t,\alpha)$, {\small
$$\Ex[\Delta_f(i_j|S_{1,\ldots, w-1} \cup C_{j-1} %\{i_1, \ldots, i_{j-1}\}
)|T_{1,\ldots, w-1}, Z_{s_1},\cdots Z_{s_{j-1}} \mcomment{ i_{1,\ldots, j-1}}] \ge \frac{1}{k}\left((1-\frac{\alpha}{k})f(S^*)-f(S_{1,\ldots, w-1} \cup C_{j-1}
%\{i_1, \ldots, i_{j-1}\}
)\right)\ .$$}
\end{lemma}
}

\begin{proof}

%with probaility $\frac{1}{k\beta}$. %Since $a_s$ is  a randomly chosen item from $Z_s$, this implies $a_s=i$ independently with probability $1/k\beta$ for all $i \in S^*\backslash Z_{1,\ldots, s-1}$.

From Definition~\ref{def:ml},
%of $\gamma(\cdot)$ (refer to \eqref{eq:gamma}), 
$m_{\ell}^s$ is chosen greedily to maximize the increment
\[
\underset{x\in s\cup R(\bar w,s)}{\arg\max} \Delta(x|S\cup H_{\ell-1}^{s-1}),
\]

%$\Delta_f(i|S_{1,\ldots, w-1} \cup i_{1,\ldots, s-1})$ over all 
So $m_{\ell}^s$ belongs to $s \cup R(\bar w,s)
%Supp(T_{1,\ldots, w-1}) 
\supseteq Z_{s}$. 
Therefore, we can lower bound the marginal gain of $m_{\ell}^s$ w.r.t. previously selected items $S \cup H_{\ell-1}^{s-1}$ by the marginal gain of a randomly picked item $i$ from $Z_{s}$ as follows.
\begin{eqnarray*} 
& & %\underset{Z_s}
{\Ex}[\Delta(m_{\ell}^s|S\mcomment{_{1,\ldots, w-1}} \cup H_{\ell-1}^{s-1})|T(\bar w,s), Z_s\ne \emptyset] \\
(\text{using Proposition }~\ref{lemma:zs}) 
%\eqref{eq:pk} \mcomment{and \ref{}}) 
& \ge & \frac{1}{k} \sum_{i\in S^*} %\Ex[ 
\Delta(i|S\mcomment{_{1,\ldots, w-1}} \cup H_{\ell-1}^{s-1}) %\ecomment{|T(w,s),Z_s\ne \emptyset}] 
\\
(\text{using  Lemma~\ref{marginalsum}, monotonicity of $f$} ) & \ge & \frac{1}{k} %\ecomment{\Ex} 
\left( f(S^*)-f(S\mcomment{_{1,\ldots, w-1}} \cup H_{\ell-1}^{s-1}) \right) \ . \\
%(\text{using  monotonicity of }f) & \ge & \frac{1}{k}  \Ex[f(S^*)-f(S\mcomment{_{1,\ldots, w-1}} \cup C_{j-1})]\\
%(\text{using  Lemma \ref{lem:Zs} and Lemma~\ref{sample}}) & \ge & \frac{1}{k} \left(\left(1-\frac{\Ex[|\cup_{s'\in w} Z_{s'}|]}{k}\right)f(S^*) -f(S_{1,\ldots, w-1} \cup i_{1,\ldots, s-1})\right))\\
%(\text{using  Lemma~\ref{lem:Zs} and Lemma~\ref{sample}})  & \geq & \frac{1}{k} \left(f(S^*) -f(S\mcomment{_{1,\ldots, w-1}} \cup C_{j-1})\right)
\end{eqnarray*}
%Here, we used the submodularity of $f$ and 
%The last inequality uses the observation from Lemma \ref{lem:Zs} that given $T$, every $i\in S^*$ appears in $\cup_{s'\in w} Z_{s'}$ independently with probability $\alpha/k$, so that every $i\in S^*$ appears in $S^*\backslash\cup_{s'\in w} Z_{s'}$ independently with probability $1-\frac{\alpha}{k}$; along with Lemma \ref{sample} for submodular function $f$. 
%so that $\Ex[|\cup_{s'\in w} Z_{s'}|] \le {\alpha}$.
\end{proof}

\begin{cor}
\label{cor:margij}
For slot $s$ in stochastic window $\bar w$,  %{suppose} the sequence $\tau_{\bar w}=(s_1, \ldots, s_t)$ {is as} defined in Definition \ref{def:tauw}, 
given $T$, and $1\le j \le L$, we have

$$\underset{\bar w}
{\Ex}\left[\Delta(m_{j}^s|S \cup H_{j-1}^{s-1}) | T(\bar w,s)=T, s=s_j \right] \ge
\frac{1}{k} [ f(S^*)-f(S \cup H_{j-1}^{s-1}
)] \ .$$
%\frac{1}{k} \Ex \left[f(S^*)-f(S \cup H_{j-1}^{s-1}
%)| s=s_j \right]\ .$$}
\end{cor}
\begin{proof}
Similar to previous lemma but in the step that uses Prop.~\ref{lemma:zs}, we use Lemma~\ref{lemma:zsj}.
%the fact that the events $s=s_j$ and $Z_s$ are independent. 
\end{proof}

\begin{definition}
For stochastic window $\bar w$, and $0\le j \le L$ define 
\begin{equation*}
%A_{s,j} = \Ex_{\bar w}[f(S\cup H_{j}^s)| s\succeq s_j]
A_{s,j} = \Ex_{\bar w}[f(S\cup H_{j}^s)| r(\bar w,s)=j]
\end{equation*}
\end{definition}

\begin{cor}
\label{cor:margcj}
For a stochastic window $\bar w$, %{suppose} the sequence $\tau_{\bar w}=(s_1, \ldots, s_t)$ {is as} defined in Definition \ref{def:tauw}, %and $\mu_{\bar w}=\{i_1,\cdots, i_{\alpha'}\}$ is as defined in Definition~\ref{def:ml},
for a fixed slot $s$, and $j\in U_s$, %$1\le j \le L$,
\begin{align*}
\underset{\bar w}{\Ex}\left[\Delta(m_j^s|S \cup H_{j-1}^{s-1} %\{i_1, \ldots, i_{j-1}\}
)|  s=s_j \right] 
&\ge \frac{1}{k} \left(  f(S^*)-A_{s-1,j-1}   \right)  \ .
\end{align*}

\rcomment{
\begin{align*}
\underset{\bar w}{\Ex}\left[\Delta(i_j|S \cup C_{j-1} %\{i_1, \ldots, i_{j-1}\}
)|  s=s_j \right]  &\ge
\frac{1}{k} \left( \underset{\bar w}{\Ex} \left[f(S^*)-f(S \cup C_{j-1})| s=s_j  \right] \right) \\ 
&=\frac{1}{k} \left( \underset{\bar w}{\Ex} \left[f(S^*)-f(S \cup C_{j-1})| r(\bar w,s-1)=j-1  \right]  \right) \\
&=\frac{1}{k} \left(  [f(S^*)-A_{s-1,j-1}   \right)  \ .
\end{align*}
}

\rcomment{
{\small 
$$
\underset{j,Z_s}{\Ex}\left[\Delta(i_j|S \cup C_{j-1} %\{i_1, \ldots, i_{j-1}\}
)|  s=s_j \right]  \ge \frac{1}{k} ( \underset{j}{\Ex} \left[f(S^*)-f(S \cup C_{j-1})|  s=s_j   \right]  ) \ .$$}
}

\toRemove{
{\small
$$\underset{T}{\Ex} \left[
\underset{j,Z_s}{\Ex}\left[\Delta(i_j|S \cup C_{j-1} %\{i_1, \ldots, i_{j-1}\}
)| T(\bar w,s), s=s_j \right] \right] \ge \frac{1}{k} \left( \underset{T}{\Ex} \left[ \underset{j}{\Ex} \left[f(S^*)-f(S \cup C_{j-1})| T(\bar w,s), s=s_j   \right] \right] \right) \ .$$}
}
%\Ex[\Delta(i_j|S_{1,\ldots, w-1} \cup C_{j-1} %\{i_1, \ldots, i_{j-1}\}
%)|T(w,s), s = s_j ] \ge \frac{1}{k} \Ex \left[f(S^*)-f(S_{1,\ldots, w-1} \cup C_{j-1})| T(w,s), s= s_{j}, \ecomment{ j\le r} \right]\ .$$}
\end{cor}
%\ecomment{(The randomness of the expectation is on randomness of $s, T(w,s), j$, $Z_i$'s and input distribution)}

\mcomment{handle $j=1$}
\begin{proof} \mcomment{check again}
%By substituting $C_{j-1}=H_{j-1}^{(s_{j})-1}$ and $i_j=m_{j}^{s_j}$ in 
From Corollary~\ref{cor:margij}, {for slot $s$} %and $1\le j \le L$,
%\ecomment{by conditioning on $s=s_j$} and using Lemma~\ref{lemma:zsj}
%the fact that $Z_{s_j} \ne \emptyset$,  
we have

\begin{align*}
%&\Ex[\Delta(i_j|S_{1,\ldots, w-1} \cup C_{j-1}
%)|T(w,s), s=s_j ] \\ \ge  
&\underset{\bar w}{\Ex} \left[\Delta(m_j^{s}|S \cup H_{j-1}^{s-1}
)|T(\bar w,s), s=s_j \right]  \ge 
\frac{1}{k} \left( 
f(S^*)-f(S \cup H_{j-1}^{s-1} )  \right)\ .
%\frac{1}{k} \left( {\Ex}
%\left[f(S^*)-f(S \cup H_{j-1}^s ){| s=s_j} \right] \right)\ .
\end{align*}

\rcomment{
\begin{align*}
%&\Ex[\Delta(i_j|S_{1,\ldots, w-1} \cup C_{j-1}
%)|T(w,s), s=s_j ] \\ \ge  
&\underset{Z_s}{\Ex} \left[\Delta(i_j|S \cup H_{j-1}^{(s_j)-1}
)|T(\bar w,s), s=s_j \right]  \ge 
\frac{1}{k} \left( {\Ex}
\left[f(S^*)-f(S \cup C_{j-1} ){| s=s_j} \right] \right)\ .
\end{align*}
}
By taking  expectation on
%$T(\bar w,s)$ 
$\bar w$ conditioned on $s=s_j$ 
%and conditioning on $s=s_j$ 
from both sides we get the first line
%($s$ is uniformly distributed in $w$)
%$T(w,s), s=s_{j}$, \ecomment{ $s, j\le r$ from both sides}  %and using Proposition~\ref{prop:Tsucc}
%implies the Corollary. 

\begin{align*}
\underset{\bar w}{\Ex}\left[\Delta(m_j^s|S \cup H_{j-1}^{s-1} %\{i_1, \ldots, i_{j-1}\}
)|  s=s_j \right]  &\ge
\frac{1}{k} \left( \underset{\bar w}{\Ex} \left[f(S^*)-f(S \cup H_{j-1}^{s-1})| s=s_j  \right] \right) \\ 
&=\frac{1}{k} \left( \underset{\bar w}{\Ex} \left[f(S^*)-f(S \cup H_{j-1}^{s-1})| r(\bar w,s-1)=j-1 , Z_s\ne \emptyset \right]  \right) \\
&=\frac{1}{k} \left( \underset{\bar w}{\Ex} \left[f(S^*)-f(S \cup H_{j-1}^{s-1})| r(\bar w,s-1)=j-1  \right]  \right) \\
&=\frac{1}{k} \left(  f(S^*)-A_{s-1,j-1}   \right)  \ .
\end{align*}

%\ecomment{(The first expectation is on r.v. $j$ and randomness of input distribution and $Z$). 
%Note that by conditioning on $s=s_j$, we can use Lemma~\ref{lemma:zsj}.
%Lemma~\ref{lemma:zs} holds true (since it will increase the $\Pr(i\in Z_s)$ for $i\in S^*\setminus Supp(T)$). 

%By taking  expectation on $T(\bar w,s)$ and conditioning on $s=s_j$ from both sides 
%($s$ is uniformly distributed in $w$)
%$T(w,s), s=s_{j}$, \ecomment{ $s, j\le r$ from both sides}  %and using Proposition~\ref{prop:Tsucc}
%implies the Corollary. 
The third line is because of independence of $Z_s$ and %$r(\bar w,s-1)$. 
$S\cup H_{j-1}^{s-1}$.
\end{proof}

\mcomment{Using standard techniques for the analysis of greedy algorithm, the following corollary of the previous lemma can be derived: given any $T_{1,\ldots, w-1}=T$:}

%As defined in Definition~\ref{def:q}, $\Pr(Z_s\ne \emptyset) = q$.
\begin{lemma}
For $q$ defined in Definition~\ref{def:q},we have
$\Pr(Z_s\ne \emptyset) = q$.
\end{lemma}
\begin{proof}
First note that from Lemma~\ref{lem:Zs}, for each element $e\in S^*$, $\Pr(e\in Z_s| T(\bar w,s)) = \frac{1}{k\beta}$.
Since $Z_s$ and $T(\bar w,s)$ are independent, $\Pr(Z_s\ne \emptyset)=(1-\frac{1}{k\beta})^k =q$.
\end{proof}

\rcomment{
\begin{definition}
Now suppose $q:= \Pr(Z_s\ne \emptyset)=1-(1-\frac{1}{k\beta})^{k}$ for slot $s$ (the probability is the same  for all slots).
Also define $U_s:=(qs-\delta_s, qs+\delta_s)$, where $\delta_s:=4\sqrt{qs\log(1/\epsilon)} \le 4\sqrt{\alpha \log(1/\epsilon)}$.
\end{definition}
}

\begin{lemma}
For slot $s$ in $\bar w$, w.p. at least $1-\epsilon$, we have 
$r(\bar w,s) \in U_s $.
\end{lemma}
\begin{proof}
Since random variables $\mathbf{1}(Z_s=\emptyset)$ ($\forall s\in \bar w$) are independent, % and $r(\bar w,s) = \sum_{s'\preceq s} 1(Z_{s'}\ne \emptyset) $. 
and $\Ex[r(\bar w,s)] = qs$,
 by Chernouf bound,
\[
\Pr\left(r(\bar w,s) \le (1-\delta) qs \right) \le e^{-\frac{\delta^2 qs}{2}}  ,
\] 
By setting $\delta=\frac{4}{\sqrt{qs}} \sqrt{\log(2/\epsilon)}$, we have 
\[
\Pr(r(\bar w,s) \notin U_s) \le \epsilon \ .
\] 
\end{proof}

\begin{lemma}
For each slot $s$ in $\bar w$, and $j< u_s$, 
\[
f(S\cup H_{s}^{j}) \ge f(S\cup H_{s}^{j-1}).
\]
\end{lemma}
\begin{proof}
The proof is by induction on $s$. Assuming the for all layers less than $u_{s-1}$ we have the above property, based on line 8 of the algorithm we can prove it for slot $s$ up to layer $u_{s}$.
\end{proof}

\toRemove{
\begin{prop}
%Suppose $q'= q \cdot \Pr(r(\bar w,s)\in U_s)\ge q (1-\epsilon)$, 
For %$0\le j \le u_s$,
$ j \in U_s$, 
from Cor.~\ref{cor:margcj},  for $q=\Pr(Z_s\ne \emptyset)$ we have
\begin{align*}
A_{s,j}&= q \times \left(A_{s-1,j-1} + \Ex_{\bar w}[\Delta(m_j^s|S\cup H_{j-1}^{s-1}) |s=s_j] \right)+ (1-q) \times A_{s-1,j} \nonumber\\
&\ge q \times \left(A_{s-1,j-1} + \frac{1}{k}\left(f(S^*) - A_{s-1,j-1} \right) \right)+ (1-q) \times A_{s-1,j} \ .
\end{align*}
Also for $j \le l_s$ we have $A_{s,j}\ge A_{s-1,j}$.
\end{prop}
}

\toRemove{
\begin{proof}
\begin{align*}
A_{s,j} &= \Ex_{\bar w}[f(S\cup H_{j}^s)| r(\bar w,s)=j] \\
&= \Pr(Z_s\ne \emptyset) \cdot  [ \Pr(r(\bar w,s)+1 \in U_s) \cdot \left(A_{s-1,j-1} + \Ex_{\bar w}[\Delta(i_{j}|S\cup H_{j-1}^{s-1}) |s=s_j]  \right) +\Pr(r(\bar w,s)+1 > u_s)  ) ]
\end{align*}
Note that from above lemma if slot $s$ is active and $r(\bar w,s-1)+1 \notin U_s$, i.e., when 
we have 
$f(S\cup H_{r(\bar w,s)}^{s}) \ge f(S\cup H_{r(\bar w,s)}^{s-1}) \ge f(S\cup H_{r(\bar w, s-1)}^{s-1})$.

redefine $r$ without min. above $u_s$  zero.
\end{proof}
}

\toRemove{
\begin{align}
A_{s+1,j+1}= q (A_{s,j} + \Delta(i_{j+1}|S\cup C_{j}) |s+1=s_j )+ (1-q) A_{s,j+1}.
\end{align}
}
\begin{definition}\label{def:B}
For a fixed slot $s$, define $$B_s:=%\Ex_{\bar w, j} \left[f\left(S\cup H_{j}^{s}\right)|r(\bar w,s)=j \right] = 
\Ex_{\bar w} \left[f\left(S\cup H_{r(\bar w,s)}^{s}\right)\right] = \Ex_{j \sim r(\bar w,s)} \left[ A_{s, j} \right] .$$
\end{definition}
%Then if the $r(\bar w,s)$ is less than $\alpha$, 
%\begin{align*} 
%B_{s} \ge q \times \left(B_{s-1} + \frac{1}{k}(f(S^*)-B_{s-1}) \right) + (1-q) \times B_{s-1}.
%\end{align*}
%where $B_0=f(S)$.
%Now $B_s = B_s \times (1-1/\alpha)$
%We will consider $s'=\alpha\beta (1-\epsilon)$, then w.p. $1-\epsilon$ for slot $s'$ and all the slots $s$ before $s'$ in $\bar w$ we have $r(\bar w,s) \le \alpha$. Thus we set
%$q':=q(1-\epsilon)$. 
\begin{lemma}
Suppose $\theta=  \Pr(r(\bar w,s)\in U_s)\ge  (1-\epsilon)$. Then we get the following equation,
\begin{align}\label{eq:wrec}
%&f(S\cup S_w) \ge B_{\alpha \beta} \nonumber \\
&B_{s} \ge q \times \left(B_{s-1} + \frac{1}{k}(\theta f(S^*)-B_{s-1}) \right) + (1-q) \times B_{s-1}, & \forall 1\le s \le \alpha \beta \nonumber \\
&B_0=f(S),
\end{align}
where $f(S\cup S_w) \ge B_{\alpha \beta}$.

\end{lemma}
\begin{proof}
%Suppose slot $s$ is the $j$-th active slot.
%Suppose $t=\sum_{s'\preceq s} 1(Z_{s'}\ne \emptyset)$.
%If $r(\bar w, s-1) < u_s-1$, 
%If $t\in U_s$,  %$t < u_s-1$, 
%then we  have $r(\bar w, s) = r(\bar w,s-1) + 1 < u_s$.
%So the algorithm makes the  update in line 8 of the algorithm for $\ell=j$. 
From Cor.~\ref{cor:margcj}, for $j\in U_s$,
\begin{align*}
\underset{\bar w}{\Ex}\left[\Delta(m_j^s|S \cup H_{j-1}^{s-1} %\{i_1, \ldots, i_{j-1}\}
)|  s=s_j \right] 
&\ge \frac{1}{k} \left(  f(S^*)-A_{s-1,j-1}   \right) .
\end{align*}
Suppose $t=\sum_{s'\preceq s} 1(Z_{s'}\ne \emptyset)$.
If $t\in U_s$  and $ s$ is active slot, then for $j=r(\bar w,s)$, we have $r(\bar w,s-1)= j-1$. Also if $s$ is not active then $r(\bar w,s)=r(\bar w,s-1)$. 
Thus from Cor.~\ref{cor:margcj},
\toRemove{
\begin{align*}
A_{s,j} \ge q \times \left(A_{s-1,j-1} + \frac{1}{k}\left(f(S^*) - A_{s-1,j-1} \right) \right)+ (1-q) \times A_{s-1,j-1} \ .
\end{align*}
}
\begin{align*}
A_{s,r(\bar w,s)} \ge q \times \left(A_{s-1,r(\bar w,s-1)} + \frac{1}{k}\left(f(S^*) - A_{s-1,r(\bar w,s-1)} \right) \right)+ (1-q) \times A_{s-1,r(\bar w,s-1)} \ .
\end{align*}

But if $t\notin U_s$, %$r(\bar w, s-1) = u_s-1$, 
%then we  have $r(\bar w, s) = u_s-1$. Thus 
we can only guarantee $A_{s,j}\ge A_{s-1,j}$. In fact, in this case we can use a weaker guarantee and use
\toRemove{
$$A_{s,j} \ge q \times (A_{s-1,j-1}-\frac{1}{k}A_{s-1,j-1}) + (1-q) \times A_{s-1,j}.$$
}
\begin{align*}
A_{s,r(\bar w,s)} \ge q \times \left(A_{s-1,r(\bar w,s-1)} - \frac{1}{k} A_{s-1,r(\bar w,s-1)}  \right)+ (1-q) \times A_{s-1,r(\bar w,s-1)} \ .
\end{align*}

With probability $\theta$, we know the first case happens, i.e.,  $t\in U_s$, thus

\toRemove{
\begin{align*}
B_s&:=\Ex_{\bar w} \left[f\left(S\cup H_{r(\bar w,s)}^{s}\right) \right] \\
&\ge q (1-1/k)\times \left(\Ex_{\bar w} \left[f\left(S\cup H_{r(\bar w,s-1)}^{s-1}\right)\right]  \right) \\
&+ (1-q) \times \Ex_{\bar w} \left[f\left(S\cup H_{r(\bar w,s-1)}^{s-1}\right)\right] + \frac{1}{k} \underbrace{\Ex_{\bar w} [f(S^*) |  r(\bar w,s) \in U_s \text{ and } r(\bar w,s-1) < u_s-1 ] }_{\ge\theta f(S^*)}.
\end{align*}
}
%Thus, $B_s= \Ex_{\bar w} \left[ A_{s, r(\bar w,s)} \right] \ge q \times  $

\begin{align*}
B_s&:=\Ex_{j\sim r(\bar w,s)} \left[A_{s,j} \right] 
\ge q (1-1/k)\times \left(\Ex_{j\sim r(\bar w,s-1)} \left[A_{s-1,j}\right]  \right) \\
&+ (1-q) \times \Ex_{j \sim r(\bar w, s-1)} \left[A_{s-1,j}\right] + \frac{q}{k} \times \underbrace{\Ex_{\bar w} [f(S^*) |  t \in U_s]}_{\ge\theta f(S^*)} \\
&\ge q(1-1/k) B_{s-1} + (1-q)B_{s-1} +\frac{q}{k} \theta f(S^*) \ .
\end{align*}
\end{proof}

\rcomment{
we should go upto $s'=\alpha \beta (1-\epsilon)$.?
}

\begin{prop}
Combining all $k/\alpha$ windows together we have a set $S$ of size at most $k(1+\frac{(4\sqrt{\alpha\log(1/\epsilon)})}{\alpha})$, such that
$$
\frac{\Ex[f(S)]}{\theta \cdot OPT} \ge 1-(1-q/k)^{k\beta} \ge 1-e^{-q\beta } \ge 1-e^{-1}-2\epsilon.
$$
Thus if we sample a set of size $k$ from $S$, its expected value is at lest $(1-1/\sqrt{\alpha})(1-e^{-1}-2\epsilon)\theta \cdot OPT $.
By setting $\alpha=1/\epsilon^2$, we get
$\Ex[f(S)]\ge(1-e^{-1}-3\epsilon) OPT$.
\end{prop}

\begin{remark}
In the above analysis, we have $\alpha + 4\sqrt{\alpha \log(1/\epsilon)} \le k$, thus $\epsilon \ge  \Omega(1/\sqrt{k})$. %Similar restrictions also exists in~\cite{us}.
\end{remark}

\begin{remark}
If we impose  cardinality constraint $k$ on the expected number of selections made by the algorithm, we can choose  any $\epsilon > 0$ .
\end{remark}

\begin{lemma} \label{lem:Rsize}
The size of the shortlist $R$ that Algorithm~\ref{alg:card} uses is at most $16k\sqrt{\alpha\log(1/\epsilon)} \beta \log (2/\epsilon)$. % = O(\log (1/\epsilon))$.
\end{lemma}

\begin{proof}
There are total of $\alpha \beta (k/\alpha) = k \beta$ slots. In each slot, we run $4\sqrt{\alpha\log(1/\epsilon)}$ \textit{online max} algorithms, each add elements of $M_i$ with size $4\log(2/\epsilon)$  to the shortlist $R$. Thus, the algorithm add $(k\beta)(4 \log(2/\epsilon)\alpha)$ items to the shortlist $R$.
\end{proof}

\begin{lemma}
The running time and query complexity of the algorithm are $\tilde{O}(n/\epsilon)$.
\end{lemma}
\begin{proof}
For each new item algorithms make $|U_s|=O(\sqrt{\alpha \log(1/\epsilon)})$ queries. Additionally we examine one $|R|/k\beta$ many elements of shortlist in each slot. Thus the total number of queries over the input is $O(n\sqrt{\alpha \log(1/\epsilon)})=\tilde{O}(n/\epsilon)$. 
\end{proof}

\begin{prop}
We can get the same result for secretary with shortlist model by invoking the \textit{online max} algorithm in~\cite{us}, and the size of shortlist will be $\log(1/\epsilon)$ times the size of memory. 
\end{prop}

\mcomment{
Next, we compare $\gamma(\tau_w)$ to $S_w=\gamma(\tau^*)$ . Here, $\tau^*$ was defined has the `best' greedy subsequence of length $\alpha$ (refer to \eqref{eq:Sw} and \eqref{eq:taustar}). To compare it with $\tau_w$, we need a bound on size of $\tau_w$. 
\begin{lemma}
\label{lem:lengthtau}
For any real $\delta\in (0,1)$, 
and if $k \ge \alpha\beta$, $\alpha \ge 8\log(\beta)$ and $\beta \ge 8$, 
%$\beta \ge \frac{3}{\delta}$, $\alpha \ge 4 \log(3\beta/\delta)$, and $\frac{2\alpha}{k} \le \frac{\delta}{3}$, 
then given any $T_{1,\ldots, w-1}=T$,
$$(1-\delta)\left(1-\frac{4}{\beta}\right)\alpha \le |\tau_w| \le (1+\delta)\alpha,$$
with probability $1-\exp(-\frac{\delta^2\alpha}{8\beta})$.
\end{lemma}
}

\mcomment{
%\label{sec:streaming}
%\ccomment{Put here a short section explaining on how extend the algorithm to work for streaming.}
In this section, we show that Algorithm \ref{alg:main} can be implemented in a way that it uses a memory buffer of size at most $\eta(k)=O(k)$; and the number of objective function evaluations for each arriving item is $O(1+\frac{k^2}{n})$. This will allow us to obtain  Theorem \ref{thm:streaming} (restated below) as a corollary of Theorem \ref{opttheorem}.
\thmStreaming*
In the current description of Algorithm \ref{alg:main}, there are several steps in which the algorithm potentially needs to store $O(n)$ previously seen items in order to compute the relevant quantities. 
%We discuss how to modify those steps to obtain an equivalent implementation but with memory buffer requirement of only $O(k)$. 
First, in Step \ref{li:subb}, in order to be able to compute $\gamma(\tau)$ for all less than $\alpha$ length subsequences $\tau$ of slots $s_1, \ldots, s_{j-1}$, the algorithm should have stored all the items that arrived in the slots $s_1, \ldots, s_{j-1}$. However, this memory requirement can be reduced by a small modification of the algorithm, so that at the end of iteration $j-1$, the algorithm has already computed  $\gamma(\tau)$ for all such $\tau$, and stored them to be used in iteration $j$. In fact, this can be implemented in a memory efficient manner, in the following way. For every subsequence $\tau$ of slots $s_1, \ldots, s_{j-1}$ of length $<\alpha$, consider prefix $\tau'=\tau\backslash s_{j-1}$. Assume $\gamma(\tau')$ is available from iteration $j-2$. 
If $\tau'=\tau$, then $\gamma(\tau)=\gamma(\tau')$. Otherwise, in Step 6 of iteration $j-1$, the algorithm must have considered the subsequence $\tau'$ while going through all subsequences of length less than $\alpha$ of slots $s_1, \ldots, s_{j-2}$. Now, modify the implementation of Step 6 so that  the algorithm also tracks the (true) maximum $M_{j-1}(\tau')$ of $a_0, a_1, \ldots, a_N$ for each $\tau'$. Then, $\gamma(\tau)$ can be obtained by extending $\gamma(\tau')$ by $M_{j-1}(\tau')$, i.e.,  $\gamma(\tau)=\{\gamma(\tau'), M_{j-1}(\tau')\}$. Thus, at the end of iteration $j-1$, $\gamma(\tau)$ would have been computed for all subsequences $\tau$ relevant for iteration $j$, and so on. In order to store these $\gamma(\tau)$ for every subsequence $\tau$ (of  at most $\alpha$ slots from $\alpha \beta$ slots), we require a memory buffer of size at most $\alpha^2{\alpha \beta \choose \alpha} = O(1)$.  
%In order to implement steps 6--9 for window $w$, and for every slot $s_j$, the algorithm needs a buffer to store $\tau, \gamma(\tau)$ for every subsequence  $\tau$ of length at most $\alpha$, of the $j-1$ slots in window $w$. This requires a memory buffer of size at most $\alpha \beta \times 2\alpha{\alpha \beta \choose \alpha}$. 
Secondly, across windows and slots, the algorithm keeps track of $R_w, S_w, w=1,\ldots, k/\alpha$ where $W=k/\alpha$. In the current description of Algorithm \ref{alg:main}, these sets are computed after seeing all the items in window $w$ in Step~\ref{li:Rw}. Thus, all the items arriving in that window would be needed to be stored in order to compute them, requiring $O(n)$ memory buffer. However, the alternate implementation discussed in the previous paragraph reduces this memory requirement to $O(k)$ as well. Using the above implementation, at the end of iteration $\alpha \beta$ for the last slot $s_{\alpha\beta}$ in window $w$, we would have computed and stored $\gamma(\tau)$ for all the subsequences  $\tau$ of length $\alpha$ of slots $s_1,\ldots, s_{\alpha\beta}$. 
$R_w$ is simply defined as union of all items in  $\gamma(\tau)$ over all such $\tau$ (refer to \eqref{eq:Rw}). And, $S_w = \gamma(\tau^*)$  for the best subsequence $\tau^*$ among these subsequences (refer to \eqref{eq:Sw}). 
Thus, computing $R_w$ and $S_w$ does not require any additional memory buffer. Storing $R_w$ and $S_w$ for all windows requires a buffer of size at most $\sum_w |R_w| + |S_w| = \frac{k}{\alpha} \times \alpha {\alpha \beta \choose \alpha}+ k = O(k).$
%Thus the subsequences considered in   to either add slot $s_j$. , by using the observation  that $R_w$ (as defined in \eqref{eq:Rw}) can be formed simply by adding all the union of all items in $A_j(\tau)$ The union of items these subsequences form $R_w$. Further, it needs to store $S_w$ for each window, which is of size $\alpha$. 
Therefore, the total buffer required to implement Algorithm \ref{alg:main} is of size $ O(k)$. 
%Finally, let's bound the number of objective function evaluations for each arriving item. Each arriving item is processed in Step 6, where objective function is evaluated twice for each $\tau$ to compute the corresponding $a_i$. Since there are atmost ${\alpha \beta \choose \alpha}$ subsequences $\tau$ for which this quantity is computed, the total number of objective function evaluations is  bounded by $2 {\alpha \beta \choose \alpha}=O(1)$. 
%This concludes the proof of Theorem \ref{thm:streaming}.
Finally, let's bound the number of objective function evaluations for each arriving item. Each arriving item is processed in Step 6, where objective function is evaluated twice for each $\tau$ to compute the corresponding $a_i$. Since there are atmost ${\alpha \beta \choose \alpha}$ subsequences $\tau$ for which this quantity is computed, the total number of times 
%the %Algorithm~\ref{alg:SIIImax} is called is 
%objective function evaluations is  
this computation is performed is bounded by $2 {\alpha \beta \choose \alpha}=O(1)$. However, for each $\tau$, we also compute $a_0$ in the beginning of the slot. Computing $a_0$ for each $\tau$ involves taking max over all items in $R_{1,\ldots, w-1}$, and requires $2|R_{1,\ldots,w-1}|\le 2k {\alpha \beta \choose \alpha}$ evaluations of the objective function. Due to this computation, in the worst-case, the update time for an item can be $ 2k {\alpha \beta \choose \alpha}^2 + 2 {\alpha \beta \choose \alpha}= O(k)$. However, since $a_0$ is computed {\it once} in the beginning of the slot for each $\tau$, the  total update time over all items is bounded by $2k {\alpha \beta \choose \alpha}^2 \times k\beta + {\alpha \beta \choose \alpha} \times n = O(k^2+n)$. Therefore the amortized update time for each item is $O(1+\frac{k^2}{n})$.
\scomment{replaced by above:Also each subroutine call on a slot $s_j$ goes over all elements in $s_j$ and also $R_{1,\cdots, w-1}$ to find the maximum element. 
%In the current format of the algorithm  
%The update time can be made to be at most $\frac{O(k^2)}{n}$ in this way. 
When we pass elements of $s_j\cup R_{1,\cdots, w-1}$ to the online subroutine~\ref{alg:SIIImax} one by one, for each element $e\in  s_j$ we can also pass $\frac{|R_{1,\cdots, w-1}|}{|s_j|}$ elements from $R_{1,\cdots, w-1}$. %Therefore, the expected update time for each element in $s_j$ would be  $\frac{O(k^2)}{n}$.
Therefore the amortized update time
would be $O(1+\frac{k^2}{n})$.
%in the beginning of each slot when we pass
Also note that the worst case update time can be $O(k)$.}
This concludes the proof of Theorem\ref{thm:streaming}.
}

%To conclude,  the following is obtained as a corollary of Theorem \ref{alg:main}.

%\noindent {\bf Theorem 2 (restated)}
%{\em For any constant $\epsilon>0$, there exists an algorithm for the \nameStreaming that uses a memory of size at most $\eta_\epsilon(k)=O(k)$, and achieves $1-\frac{1}{e} -\epsilon -O(\frac{1}{k})$ approximation to $\OPT$.
%Also it processes each arriving item with $O(1)$ evaluations
%of the objective function.}

\toRemove{
\subsubsection{Streaming Max-Coverage }
to be completed
}

{

\toRemove{
\subsection{Improving the number of queries }
In this section we show that we can reduce the total number of queries required in each round from $\alpha$ to $O(\sqrt{\alpha\log(1/\epsilon)})=\tilde{O}(1/\epsilon)$. It is done by a slight modification of the algorithm in which in the inner loop for slot $s$, instead of going over $0\le \ell \le \alpha$, we only consider $s/\beta-\sqrt{s/\beta \log(1/\epsilon)} \le \ell \le s/\beta+\sqrt{s/\beta\log(1/\epsilon)}$.
By using a Chernoff bound we can show that w.p. at least $1-\epsilon$, the total number of slots $s'\succeq s$ with $Z_{s'}\ne \emptyset$, is between
$s/\beta-\sqrt{s/\beta \log(1/\epsilon)} \le \ell \le s/\beta+\sqrt{s/\beta\log(1/\epsilon)}$. In other words, w.p. $(1-\epsilon)$, we have $s=s_{j+1}$  for some $j$ in the above range.
Therefore we miss each element of the findl solution w.p. $1-\epsilon$, which appears in the approximation factor by using Lemma~\ref{sample} (note that in this lemma we do not need sampling independently). 
As a result this also improves memory to $O(k\sqrt{\alpha}\beta)$.
}
}

\toRemove{
\subsection{Improving memory from $O(\frac{k}{\epsilon^2})$ to $k+\tilde{O}(\frac{1}{\epsilon^2})$}
}

%% file: algorithms.tex
In this section, we focus on the matroid constraints. We study the \nameOfProblemMatroidSL as defined in Section~\ref{sec:matroidDef}. 
\mcomment{add citation for arxiv preprint}
\mcomment{\cite{me} provide a $\frac{1}{2}(1-1/e^2-\epsilon)-O(1/k)$ approximation algorithms for this problem using shortlist of size $O_{\epsilon}(k)$, where the hidden constant is $O(2^{poly(1/\epsilon)})$. 
Although the running time of their algorithm is linear in $n$, but the large hidden constant that exponentially depends on $1/\epsilon$ makes this algorithm far from practical.} %In this section 
%We propose a new algorithm that improves the approximation ratio, and improves the dependency of size of the shortlist on $1/\epsilon$. 
Our algorithm achieves an approximation ratio $\frac{1}{2}(1-1/e^2-\epsilon)$ using shortlist of  size $O(k/\epsilon)$.
%But the the dependency on $1/\epsilon $ is $O(poly(1/\epsilon))$.
%size $O_{\epsilon} (k)$. 
Note that in this section $k:=rk(\mathcal{M})$ is the rank of the given matroid $\mathcal{M}$. 
We can use both stochastic and non-stochastic windows in the analysis of the algorithm in this section.

%\matroidThm

\mcomment{Discuss related work}

The algorithm is similar to the algorithm for the cardinality constraints described in Section~\ref{sec:cardinality}.
\matcomment{
The building block of our algorithm is again $(\alpha,\beta)$-windows (refer to~\ref{def:windows}). We divide the input into $(\alpha,\beta)$-windows.} 
We make some modifications to the Algorithm~\ref{alg:cardconst}, Algorithm~\ref{alg:card}
and the underlying procedure that it calls, i.e., the \textit{online max algorithm} (Algorithm 1 in~\cite{us}).
The main difficulty in designing algorithms for the \nameOfProblemMatroidSL\ in comparison with the simpler \nameOfProblemSL\ is that the algorithm needs to make sure  the set of elements that are going to be returned as the output of the algorithm is an independent set.
For the cardinality constraints, the algorithm could add up to $k$ items to the set of current solution $S$ without worrying about independence of the new set. Whereas for the matroid constraints we might need to remove some of the items from the current solution $S$,  in order to make it independent. The main difference of the algorithm in this section and Section~\ref{sec:cardinality} is the way that the new algorithm deals with these removals. In addition to  oracle access to the submodular function $f$, we assume access to an \textit{independence oracle}. The independence oracle can verify in $O(1)$ whether or not a set is an independent set of the given matroid. 
%Our algorithms are intuitive. 

First we define functions $g$ and $\theta$ in eq.~\eqref{eq:g}, \eqref{eq:theta}. 
The function $g(e,S)$ is counterpart of $\Delta(e,S)$  in matroid setting. In other words, $g$ maximizes the marginal gain of $S+e$, after removing possibly one element $e'$ (selected by $\theta$) to make $S+e-e'$ an independent set.
\mcomment{take care of all $g()<0$}.
%The, we slightly modify the \textit{online max algorithm}, and propose  Algorithm~\ref{alg:matroidmax}, 
The we use a slight modification of the \textit{online max algorithm},
for the following problem {( \textit{Secretary Problem with Replacement}):}
we are given an (independent) set of a matroid and we want to add one item to this set, from a pool of items that are arriving in an online manner,  and keep it an independent set by possibly removing some other item from the set. The goal is to maximize the $f$-value of the new set.
 
\matcomment{
Then we extend this idea from adding one item to adding multiple items to a given independent set. 
}

\matcomment{
More precisely, in each window $w$ (defined similar to the one in Section~\ref{sec:cardinality}), in addition to set $S_w$ that is going to be added to the current solution $S$, we also remove a corresponding set from $S$ to make it an independent set of the matroid. }
%The analysis 
A crucial lemma in the analysis of the algorithm is Brualdi lemma (refer to Lemma~\ref{lem:Brualdi}). This lemma gives a bijection between two bases of a matroid.
We employ the Brualdi Lemma in our Lemma~\ref{replaced}, in which we use the bijection provided by Brualdi Lemma to lower bound the $f$-value of the remaining set after removing one item by %Algorithm~\ref{alg:matroidmax}. 
online max algorithm.
Intuitively, we prove the marginal gain of the new set, after adding a new item $a$ and possibly removing some other item to make the set independent, is at least as much as when we remove the corresponding element of $a$ from the bjiection provided by Brualdi Lemma, namely $\pi(a)$. Then we argue that $\pi(a)$ is distributed almost uniformly among elements of the current solution $S$, thus by Lemma~\ref{sample} we can lower bound the $f$-value of the remaining set.

\mcomment{
We make some changes to the algorithm and analysis of~\citet{us}.
The main modification is in the way the algorithm selects elements inside a window. Similar to~\cite{us}, the building block of the algorithm are $(\alpha,\beta)$-windows defined in~\cite{us}.
But the algorithm does  not need to choose the best $\alpha$-subsequence $\tau^*$, and return the $\gamma(\tau^*)$ defined on that subsequence among  ${\alpha \beta \choose \alpha}$ many subsequences. This number of selections in a window is the reason for having a hidden constant in the $O_{\epsilon}(k)$ that exponentially depends on $1/\epsilon$. We alleviate the selection method in a window by keeping track of $\alpha$ subsets. We reduce the total number of selected items in a window to $\alpha^2 \beta \log (1/\epsilon)$. Consequently, the hidden constant in $O_{\epsilon}(k)$ shortlist would be $ (1/\epsilon) \log(1/\epsilon)^2 =\tilde{O}(1/\epsilon)$.
}

\subsection{Algorithm Description}
Before describing our main algorithm we design a subroutine for 
a problem that we call it \textit{secretary problem with replacement}:
we are given a matroid $\mathcal{M}=(\mathcal{N},\mathcal{I})$ and an independent set $S\in \mathcal{I}$.
A pool of items
$I=(a_1,\cdots, a_N)$ arriving sequentially in a uniformly random order,
find an element $e$ from $I$ that can be added to $S$ after removing 
possibly one element $e'$ from $S$ such that the set remains independent, i.e., $S+e-e' \in \mathcal{I}$.
The goal is to choose element $e$ and $e'$ in an online manner with maximum marginal increment $g(e,S)=f(S+e-e')-f(S)$.
%which contains an item with expected value close to the maximum
%marginal increment. 
More precisely define function $g$ as:
\begin{definition} \label{def:g}
For an independent set $S\in \mathcal{I}$, and $e\in \mathcal{N}$ define
\begin{equation}
\label{eq:g}
g(e,S):= f(S+e-\theta(e,S)) - f(S),
\end{equation}
\end{definition}
%Now define function 
where $\theta$ is defined as: 
\begin{equation}
\label{eq:theta}
\theta(e,S) :=
 \underset{e'\in S}{\arg\max}
 %\arg\max_{e'\in S} 
 \{ f(S+e-e')| S+e-e' \in \mathcal{I} \}.
\end{equation}
%Also define $g$ as 
%% $$g(e):=\max_{e': S+e-e' \in \mathcal{I}} F(S+e-e') - F(S)$$
% $$g(e,S):= f(S+e-\theta(e,S)) - f(S)$$
We will consider the variant in which we are allowed to have a shortlist, where the  algorithm can add items to 
a shortlist and choose one item from the shortlist at the end.
We employ the \textit{oneline max algorithm}, Algorithm 1, in~\cite{us} to find:
\[
m \leftarrow\underset{x\in U}{\arg\max }\  g(x,S).
\]

\mcomment{
For the \textit{secretary problem with replacement}, we give Algorithm~\ref{alg:matroidmax}, which is a simple modification of the \textit{online max algorithm} (Algorithm 1) in~\cite{us}.
}
\mcomment{remove secretary replacement algorithm}

\begin{lemma} [refer to Proposition 3 in~\cite{us}]
%Algorithm~\ref{alg:matroidmax}, 
The online max algorithm, 
returns  element  $e$ with maximum  $g(e,S)$ with probability $1-\delta$, thus it 
achieves a  $1-\delta$ competitive ratio for the \textit{secretary problem with replacement} (element with maximum $g$). %with shortlists of size logarithmic in $1/\delta$.
using shortlist of size  $O(\log 1/\delta)$.
\end{lemma}

\begin{algorithm*}[ht]
  \caption{~\bf{ {\bf Matroid-Constraint}}}
  \label{alg:matroidconst} 
\begin{algorithmic}[1]
\STATE Inputs:  submodular function $f$, parameter $\epsilon \in (0,1]$, and set $R$. \mcomment{$\alpha, \beta$} 
{
\STATE Initialize $SH \leftarrow \emptyset$ %\bar{H}_{\ell}=\emptyset, \forall 0\le \ell \le \alpha$
}

\FOR {every slot $s$ \matcomment{ in window $w$ } }
  %\STATE Concurrently:
 %\matcomment{ \FOR{  $1 \le \ell \le \alpha$, concurrently} }
  
  %subsequences of previous slots $\tau\subseteq \{s_1, \ldots, s_{j-1}\}$ of length $|\tau|<\alpha$ \label{li:subb}\\
  %\hspace{0.44in} in window $w$, 
  \STATE $R' \leftarrow Sample(R,1/(k\beta))$ \COMMENT{sample a set of size $|R|/(k\beta)$ from $R$}
  
  \STATE call the \textit{online max algorithm} (Algorithm 1 in~\cite{us} ) to compute, with probability $\epsilon/2$:  \\

$m_{s} \leftarrow\underset{x\in s\cup R'}{\arg\max }\  g(x,SH).$
\mcomment{S to SH}

%$m_{\ell} \leftarrow\underset{x\in s\cup R}{\arg\max }\  g(x,S + H_{\ell-1} - \bar{H}_{\ell-1}).$

\STATE $o_{s} :=  \theta(m_{s}, SH)$.
 
%\STATE $o_{\ell} :=  \theta(m_{\ell}, S + H_{\ell-1}- \bar{H}_{\ell-1})$.
  
\STATE $M_{s} \leftarrow$ 
The {shortlist} %and maximum element 
returned by
%output of 
the above \textit{online max algorithm} 
{for slot $s$ and set $SH$.} 

%Add \hspace{0.44in} all items except the dummy item $0$ to the shortlist $A$. 
%That is, \label{li:sube} $$A\leftarrow A\cup  (A(j)\cap s_j)$$

% Add $m_i$ to the shortlist $R$. % for $1\le i \le \alpha$. 

\IF{$ f(SH) < f(SH + m_{s}- o_{s} )$  \mcomment{define + - for sets} } 

%\IF{$ f(S + H_{\ell} - \bar{H}_{\ell}) < f(S + H_{\ell-1} - \bar{H}_{\ell-1} + m_{\ell}- o_{\ell} )$  \mcomment{define + - for sets} } 
%\STATE  $H_{\ell} \leftarrow H_{\ell-1}   +m_{\ell}$
%\STATE $\bar{H}_{\ell} \leftarrow \bar{H}_{\ell-1}+o_{\ell}$
\STATE $SH \leftarrow SH+m_{s}-o_{s}$
\STATE  $R \leftarrow R+ ( \{m_{s}\} \cap M_{s})$
\ENDIF
%\ENDFOR
\ENDFOR
 
\STATE return $SH$ %$H_{\alpha}, \bar{H}_{\alpha}$ 
%return $A$, $A^*$. %$A=T_1 \cup \cdots  \cup T_{W}$
\end{algorithmic}
\end{algorithm*}

\begin{algorithm*}[ht]
  \caption{~\bf{ {\bf Submodular} Matroid Secretary with Shortlists}}
  \label{alg:matroid} 
\begin{algorithmic}[1]
\STATE Inputs: number of items $n$, submodular function $f$, parameter $\epsilon \in (0,1]$. 
\STATE Initialize: $S \leftarrow \emptyset, R
\leftarrow \emptyset$, 
constants $ \beta \ge 1$ which depend on the constant $\epsilon$.
\STATE Divide indices $\{1,\ldots, n\}$ into  $(k, \beta)$ window. 
%\FOR {window $w= 1, \ldots, W=k/\alpha$} 
%\STATE $S_w, \bar{S}_w \leftarrow$ Matroid-Constraint$()$
%\STATE $S \leftarrow S + S_w -\bar{S}_w$
%\ENDFOR
\STATE $S\leftarrow $ Matroid-Constraint($R$)
\STATE return $S \cap R$
\end{algorithmic}
\end{algorithm*}

We divide the input into %sequential blocks that we refer to as $(\alpha, \beta)$ windows. (Refer to   Definition~\ref{def:windows}).
$k\beta$ slots.
%1 in~\cite{us} or the appendix.
%Here $k=rk(\mathcal{M})$. 
%\ccomment{Would an example or a figure help here to define slots and windows?}
%Thus, $q^{th}$ slot is composed of indices $\{\ell, \ldots, r\}$, where $\ell=X_1+...+X_{q-1}+1$ and  $r=X_1+...+X_q$. 
%Further, if the ordered the input is $\bar a_1, \dots, \bar a_n$ then we say that the items inside the slot $s_q$ are $\bar a_{\ell}, \bar a_{\ell+1}, \dots, \bar a_{r}$ 
%To reduce notation, when clear from context, we will use $s_q$ and $w$ to also indicate the {\it set of items} in the slot $s_q$ and window $w$ respectively.
%The idea is to divide the inputs into $k/\alpha$ blocks (i.e., windows) for a constant $\alpha\ge 1$ such that 
%\ccomment{Shouldn't we say what alpha is because the intuition depends on it}
Intuitively, for large enough %$\alpha$ and 
$\beta$,
%roughly $\alpha$ items from the optimal set $S^*$ are likely to lie in each of these windows, and further, 
it is unlikely that two items from $S^*$  appear in the same slot. 
%The algorithm selects $\alpha$ items from each window, with at most one item from each of $\alpha \beta$ slots in a window.
\matcomment{The algorithm can focus on identifying a constant number (roughly $\alpha$) of optimal items from each of these windows, with at most one item coming from each of the $\alpha \beta$ slots in a window. }
%Similar to~\cite{us}, the core of our algorithm is a subroutine that 
%Algorithm~\ref{alg:matroidmax} accomplishes this task in an online manner in each slot.
%The algorithm chooses an  $\alpha$-subsequence of the $\alpha \beta$ slots that maximize the marginal gain with respect to previously selected items.
\matcomment{
At the end of window, the algorithm chooses a subset of size $\alpha$, $S_w$, to be added to the current solution $S$. Furthermore, the algorithm removes a subset of $S$ of size at most $\alpha$ from $S$, correpsonding to $S_w$, in order to make it an indepdendent set of the matroid.
}
\mcomment{
At the end, the algorithm chooses the 'best' greedy $\alpha$-subsequence of  slots in a window.}
%But the difference is that 
For matroid constraints, in contrast with the cardinality constraints, adding items from a new window to the current solution $S$ could make it a non-independent set of matroid $\mathcal{M}$. In order to make the new set independent we have to remove some items from $S$. The removed item corresponding to $e$ will be $\theta(e,S)$ as defined in~\eqref{eq:theta}. We need to take care of all the removals for newly selected items in the window. 
%Therefore we have to slightly change the definitions in~\cite{us}. 
%Therefore we adapt new notations for the algorithm with matroid constraints:

%To implement this idea, we use a greedy selection method that considers all possible $\alpha$ sized subsequences of the $\alpha\beta$ slots in a window, and aims to identify the subsequence that maximizes the increment over the best items identified so far. Because of the removal of elements in the definition of $g$, we slightly modify the definition of function $\gamma$ compared to its definition in~\cite{us}.

\matcomment{
\textbf{Notations}. 
Throughout  the algorithm, we keep track of 
$S_{1,\cdots, w}$, the current solution from window $1,\cdots ,w$, and $R_{1,\cdots, w-1}$ the set of the elements that the algorithm keeps in window $1,\cdots, w-1$ as shortlist.
\mcomment{$A^*$ the elements of $S_{1,\cdots, w-1}$ that are selected by the algorithm~\ref{alg:matroidmax} (w.p. $1-\delta$).}
%we use shorthand $S_{1, \cdots ,w}$ to denote $S_1 \cup \cdots \cup S_w$.
Throughout this section, if  the subscript of $S$ and $R$ is not stated explicitly, we mean $S_{1,\cdots,w-1}$ and $R_{1,\cdots, w-1}$ respectively. \mcomment{is it necessary}
}

\matcomment{
\begin{definition} \label{def:Hbar}
 For a slot $s \in \{1,\cdots, \alpha \beta\}$ in window $w$.
Let's denote by $H_{\ell}^s$, the set $H_{\ell}$ as defined in the Algorithm~\ref{alg:card} at the end of slot $s$.  Similarly, define $\bar{H}_{\ell}^s$ to be the set $\bar{H}_{\ell}$ at the end of slot $s$. (also initialize  $H_{\ell}^0=\bar{H}_{\ell}^0=\emptyset$)
%Define $T_w:=\bigcup_{i=1}^{\alpha} H_i$.
\mcomment{
More precisely, define $H_{\ell}^s$ recursively as:
\[
H_{0}^{s} = \emptyset
\]
\[
H_{\ell}^{s} := H_{\ell-1}^{s} + \underset{}{}
\]
}
\mcomment{Define $\bar{H}_{\ell}$}
\end{definition}
}

%% file: sizeBound.tex
Given the algorithm description, it is not very difficult to show that %\scomment{in expectation or with high probability?} \mcomment{Neither. It is exact. the max alg should terminate after selecting $O(\log(1/\delta))$ elements. the gaurantee is w.h.p.} 
our algorithm uses a shortlist of size at most $O(k)$ if $\alpha, \beta$ are constants. 
The number of items selected by Algorithm \ref{alg:SIIImax} is at most $L$.
By choosing appropriate size $L$ for shortlist $A$,
%$L=\ln (1/\epsilon)+\ln(1/\delta)+\sqrt{\ln^2{1/\delta}+2m\ln(1/\delta)\ln(1/\epsilon)}$ 
w.h.p. we can gaurantee  that $A$ contains the maximum element $M$.
%A key observation here is that the expected number of items selected by Algorithm \ref{alg:SIIImax} 
%\ref{alg:tmp} 
%\mcomment{1?} 
%is bounded by $\log(1/\delta)$. 
\scomment{Change $\epsilon,\delta$ to get max with probability $1-\epsilon$.}
%\begin{theorem}
%\label{maxanalysis}
%Algorithm~\ref{alg:SIIImax}, with parameter $u=n\epsilon$,  
%and shortlist of size $L= \ln(1/\epsilon)+\ln(1/\delta)+\sqrt{\ln^2{1/\delta}+2\ln(1/\delta)\ln(1/\epsilon)}$
%selects the maximum element with 
%probability $(1-\epsilon-\delta)$, \end{theorem}

%This result can be further extended to high probability bounds using standard techniques:

%\scomment{add appropriate lemma about Algorithm 1.}
Now, we can deduce the desired bound on the size of shortlist returned by Algorithm \ref{alg:tmp}
\begin{prop}
The size of shortlist $A$ selected by Algorithm~\ref{alg:tmp} is $O(k)$.
%\scomment{proposition about size of shortlist returned by Algorithm \ref{alg:tmp}.}
\end{prop}
\begin{proof}
Note that for each window $w=1,\ldots, k/\alpha$, and for each of the $\alpha\beta$ slots in this window, the subroutine in Algorithm \ref{alg:tmpWindow} runs  Algorithm \ref{alg:SIIImax} for ${\alpha \beta \choose \alpha}$ times (for all $\alpha$ length subsequences) to add at most $\log(1/\delta)$ items each time to . Therefore, the above algorithm selects $ k \beta {\alpha \beta \choose \alpha} = O(k)$ items as part of the shortlist.
\end{proof}

%% file: maxanalysis.tex
%\subsection{Analysis of Algorithm \ref{alg:} for the basic secretary problem with shortlist}
\scomment{Can we add a lemma showing that expected number of items selected will be $\log(1/\delta)$ when $u=N\delta$}
\mcomment{same as previous comment}
\scomment{Also we need to discuss that when we say shortlist of size $\eta(k)$, is that expected number of items or high probability? If it is high probability, we need to define what it means to say shortlist of size $\eta(k)$. Does it mean $\eta(k)$ with prescribed probability $1-\epsilon'$?}
\scomment{the results here will move to the previous section, and proofs for high prob bound will go to appendix}

\begin{theorem}
$\mathbb{E}[|S|] = \ln n$.
\end{theorem}
\begin{proof}
%Suppose $F(S_i) = f(T)$, where $|T|=m$. If $a_i \notin T$ then it is not selected. 
%Because if $a_i\notin T$ and is selected then it should have positive $F$ marginal value, which means $F(S_i) = F(S_{i-1} \cup \{a_i\} ) > F(S_{i-1}) =  f(T) $, it is a contradiction. 
%Thus only elements in $T$ will be selected at position $\ell$.  
Suppose $S_i=\{a_1, \cdots, a_i\}$.
Consider all permutations of $S_i$, an element will be selected at position $i$ if it is equal to $\max_{a\in S_i} F(a)$, thus the probability that it gets selected is $1/i$. Therefore the expected number of selections will be at most $\sum_{i=1}^{n} \frac{1}{i}  = \ln n $
%The expected number of elements selected at position $\ell$ is at most $\frac{m}{\ell}$. 
\end{proof}

%In the rest we will make the following assumption:\\
%\textbf{Assumption. } %There is a unique $m$-tuple $OPT$ with maximum $f(OPT)$ in the input. 
%There is a unique maximum element.

\scomment{What is $\delta$ in the theorem below? The proof below seems to suggest it is high probability statement, not exact.}
\begin{theorem}\label{maxanalysis}
Algorithm~\ref{alg:SIIImax}, with parameter $u=n\epsilon$,  selects a set $S$ with 
$$|S|<\ln (1/\epsilon)+\ln(1/\delta)+\sqrt{\ln^2{1/\delta}+2m\ln(1/\delta)\ln(1/\epsilon)}$$ 
and $\mathbb{E}[\max_{a\in S} a]=(1-\epsilon-\delta)OPT$, where $OPT$ is the max element in the input.
\end{theorem}
\begin{proof}
We use Freedman's inequality.
 If $\{a_1,\cdots, a_i\}$ has a unique maximum, % subset of size $m$, 
define $Y_i$ to be a random variable indicating whether the algorithm has selected $a_i$  or not, where $Y_i=1-\frac{1}{i}$ if $a_i$ is selected and $Y_i=-\frac{1}{i}$ otherwise. 
 If it has not unique solution define $Y_i=0$. ($a_i$ will not be selected)
 Also define $\mathcal{F}_i=\{Y_{n},Y_{n-1}, \cdots, Y_{n-i+1}\}$.

Let $X_i=\sum_{j=n-i+1}^{n} Y_j$,
then $\{X_i\}$ is a martingle, 
because $E[X_{i+1}|\mathcal{F}_{i}] = X_i+E[Y_{n-i}|\mathcal{F}_i]$.
 If $\{a_1,\cdots, a_i\}$ has a unique maximum element, $E[Y_{n-i}|\mathcal{F}_i]=(1/i)(1-1/i)+(1-1/i)(-1/i)=0$,
 otherwise $E[Y_{n-i}|\mathcal{F}_i]=0$. So in both cases $E[X_{i+1}|\mathcal{F}_{i}] =X_i$.
%X_i+E[Y_{n-i}]=X_i$. Also $|X_{i+1}-X_i| < 1$.
%frac{m}{i} (1-\frac{m}{i}) +(1-\frac{m}{i})\frac{m}{i} 
As in the Freedman's inequality, let $L=\sum_{i=n\epsilon}^{n} Var(Y_i| F_{i-1})$. 
\begin{align*}
L  =  \sum_{i=n\epsilon}^{n} \frac{1}{i}  (1-\frac{1}{i})^2 + (1-\frac{1}{i}) (\frac{1}{i})^2 
< \sum_{i=n\epsilon}^{n} \frac{1}{i} =\ln (1/\epsilon)
\end{align*}
Therefore,
\[
Pr(X_{n-n\epsilon}\ge \alpha \text{ and }  L\le \ln (1/\epsilon) ) \le exp(-\frac{\alpha^2}{\ln (1/\epsilon)+ 2\alpha })   < \delta 
\]
Thus we get $\alpha > \ln(1/\delta)+\sqrt{\ln^2{1/\delta}+2\ln(1/\delta)\ln(1/\epsilon)}$.
%Thus we get $\alpha > \ln(1/\epsilon)(1+\sqrt{1+2m})$. 
Also $|S| = X_{n-n\epsilon} + \ln(1/\epsilon)$. Therefore 
\[
Pr(|S| \ge  \ln (1/\epsilon)+\ln(1/\delta)+\sqrt{\ln^2{1/\delta}+2\ln(1/\delta)\ln(1/\epsilon)}  )  \le \delta
\]
So with probability $(1-\delta)$, $|S| \le \ln (1/\epsilon)+\ln(1/\delta)+\sqrt{\ln^2{1/\delta}+2\ln(1/\delta)\ln(1/\epsilon)} $.  %\frac{m\ln c}{\sqrt{\epsilon}} + m\ln c $ .
%Suppose $OPT$ is the optimal solution, i.e $|OPT|=m$ and  $f(OPT)$ is maximum in subsets of $I$.
%Since $\max $ is a submodular function, 
Also $\mathbb{P}(OPT \in\{a_{n\epsilon},\cdots, a_n \}  ) =(1-\epsilon)$.
%$E[f(OPT\cap \{a_{n\epsilon},\cdots, a_n \} )] = (1-\epsilon)OPT$.
Therefore $E[\max_{a\in S} a] \ge (1-\epsilon)OPT-\delta OPT$. 
%By setting $1/c=\epsilon$, $|S|=m\ln (1/\epsilon)+ \sqrt{2m \log (1/\epsilon)} $ and $E[F(S)]=(1-2\epsilon)OPT$.
\end{proof}

\begin{theorem}
Any online algorithm needs to select at least $\frac{1}{2}\log(1/\epsilon)-\frac{1}{2}$ elements, in expectation, to select the maximum element  with probability at least $(1-\epsilon)$ in a random permutation. (we assume $n> 1/\epsilon$)
\end{theorem}
\begin{proof}
Let $I_i=\{a_1,\cdots, a_{n/2^{i-1}}\}$, $T_i=\{a_{n/2^i+1}, \cdots, a_{n/2^{i-1}}\}$, and $R_i=I_1\setminus I_i$, for $i=1, \cdots, \log(1/\epsilon)$. Suppose $M_i$ is the maximum element in $I_i$. Let $S$  be the set of selected elements by algorithm at the end of execution.
Suppose $\epsilon_i= E[M_i\notin S| M_i\in T_i]$,
then $E[|S\cap T_i|] \ge \frac{1}{2} (1-\epsilon_i)$.
Therefore $E[|S|] \ge \sum_{i=1}^{\log(1/\epsilon)} \frac{1}{2} (1-\epsilon_i) $. Also w.p. $\frac{1}{2^{i}}$, $M_1 \in T_i$, 
thus $\sum_{i=1}^{\log(1/\epsilon)} \frac{1}{2^{i}} \epsilon_i  \le \epsilon$. 
(Note that we use the fact $ E[M_i\notin S| M_i\in T_i \text{ and }  M_i=M_1] \le \epsilon_i$, i.e, if algorithm selects one element it will select 
it even if we increase its value and keep the rest untouched) 
Now $E[|S|]$ is minimized under above constraint if $\frac{1}{2^{\log(1/\epsilon)}}\epsilon_{\log(1/\epsilon)} = \epsilon$ and the rest are zero.
Hence $E[|S|] \ge \frac{1}{2}\log(1/\epsilon)-\frac{1}{2}$.
\end{proof}

%\begin{prop}
%For a m-submodular function $f$, any online algorithm needs to select at least $\frac{m}{2}\log(m/\epsilon)-\frac{m}{2}$ elements, in expectation, to select a set $S$, with $|S|\le m$  such that  $E[f(S)] \ge (1-\epsilon)OPT$, in a random permutation. 
%\end{prop}
%\begin{proof}
%Apply previous theorem on m separate 1 to n matching.
%\end{proof}

%% file: analysis.tex
\input{newProof.tex}

%% file: newProof.tex
In the next section, we will show that $\Ex[f(S\cap R)] \ge \frac{1}{2}(1-\frac{1}{e^2} - \epsilon \mcomment{-O(\frac{1}{k})}) f(S^*)$ to provide a bound on the competitive ratio of Algorithm~\ref{alg:matroid}, for \nameOfProblemMatroidSL.

\begin{definition}
Denote by $SH(s)$ the set $SH$ in the algorithm at the end of slot $s$.
\end{definition}

\matcomment{
\begin{definition}
For slot $s$ in window $w$, and $1 \le \ell \le \alpha$, define \mcomment{ $s\ge 1$ define for $s=0$}
\begin{equation}
SH{(w, \ell, s)} :=( S_{1,\cdots, w-1} \cup H_{\ell-1}^{s-1} ) \setminus \bar{H}_{\ell-1}^{s-1}\ .
\end{equation}
\end{definition}
}

\mcomment{rearrange definitions}

\mcomment{
For any subsequence $\tau=(s_1,\ldots, s_\ell)$ of the $\alpha\beta$ slots in window $w$, define the greedy subsequence $\gamma(\tau)$ of items as:
\begin{equation}
\label{eq:gamma}
\gamma(\tau):=\{i_1, \ldots, i_\ell\},
\end{equation}
where
\begin{equation}
\label{eq:ij}
i_j := \underset{i\in s_j \cup R_{1, \ldots, w-1}}{\arg\max}  g(i, S{(w-1,\tau, j-1)}).
\end{equation}
}

\toRemove{
\begin{equation}
\label{eq:ij}
i_j := \arg \max_{i\in s_j \cup R_{1, \ldots, w-1}} g(i, S_{1, \ldots,w-1} \cup \{i_1,\ldots, i_{j-1}\}).
\end{equation}
}

%Also suppose $\theta(E,S):= \cup_{e\in E} \theta(e,s)$. 

\mcomment{
also we introduce $\zeta(\tau)$ which is counterpart of $\gamma(\tau)$ for the removed elements. 
Define $\zeta(\tau):=\{c_1, \ldots, c_\ell\}$,
where
\begin{equation}
\label{eq:cij}
c_j :=  \theta(i_j, S(w-1,\tau,j-1)).
\end{equation}
where,
\begin{equation}
\label{eq:diff}
S{(w-1,\tau, j-1)} := S_{1,\cdots, w-1} \cup \{i_1,\cdots, i_{j-1}\} \setminus \{c_1,\cdots, c_{j-1}\}
\end{equation}
\begin{lemma} \label{Sindep}
$S(w-1,\tau,j)$ is an independent set of $\mathcal{M}$.
\end{lemma}
}

\begin{prop} \label{Sindep}
%For each window $w\in [W]$, $S_{1,\cdots, w}$ 
In each iteration of the algorithm $SH$ is an independent set of $\mathcal{M}$.
%Similarly, $SH(s)$ is also an independent set $\mathcal{M}$.
\end{prop}

We define history $T(w,s)$ and $Supp(T(w,s))$
%$T_w, T_{1,\cdots, w}, T(w,s)$ and $Supp(T_w), Supp(T_{1,\cdots, w}), Supp(T(w,s))$ and $R_{1,\cdots, w}, R(w,s)$ 
similar to Definition~\ref{def:T}, %\ref{def:Supp}, 
in Section~\ref{sec:cardinality} but we define it based on  $SH$ (instead of sets $H_{\ell}$).
%as defined above in Definition~\ref{def:Hbar}. %To avoid confusion we redefine it here:

\matcomment{
\begin{definition} [History]\label{def:T}
For slot $s$ in (stochastic) window $w$ define 
\begin{align*}
T(w,s):= \{SH(s')|s\succ s'\}  and \\
Supp(T(w,s)):=\bigcup_{ s\succ s'} SH_{\ell}^{s'},
\end{align*}
If the configuration is not clear from the context, we make the notation explicit by $T(w,s)(Y)$ for configuration $Y$ (refer to Lemma~\ref{config}).
We use shorthand $R(w,s)$ to denote $Supp(T(w,s))$.
\end{definition}
}

\matcomment{
\begin{definition} \label{def:Tmatroid}
For window $w\in [W]$, define 
\begin{equation}
T_w:= \{H_{\ell}^s|s\in w, 1\le \ell \le \alpha\},
\end{equation}
%at the end of window $w$.
moreover,  
\[
T_{1,\cdots,w}:= \bigcup_{i=1}^w T_i,
\]
and 
\begin{equation}
    T(w,s):=T_{1,\cdots, w-1} \cup \{ {H_{\ell}^{s'} | s\succ_w s', 1\le \ell \le \alpha }  \}.
\end{equation}
\end{definition}
\begin{definition} \label{def:SuppMatroid}
For slot $s$ in window $w$ define
\[
Supp(T_w):=\bigcup_{1\le \ell \le \alpha, s\in w} H_{\ell}^s,
\] 
also,
\[
Supp(T_{1,\cdots, w}):= \bigcup_{i=1}^w Supp(T_i),
\]
and,
\[
Supp(T(w,s)):= Supp(T_{1,\cdots, w-1}) \cup \bigcup_{1\le \ell \le \alpha; s\succ_w s'} H_{\ell}^{s'}.
\]
(Note that $Supp(T_{1,\cdots, w})=R_{1,\cdots, w}$). Furthermore, define $R(w,s):=Supp(T(w,s)).$
 \mcomment{define $R(w,s)$}
%Furthermore, for slot $s$ in window $w$ define
%for window $w$, and slot $s\in w$, define
\end{definition}
}
\mcomment{
\begin{prop} \label{prop:Tsucc}
For slots $s,s'$ in window $w$, such that $s \succ_w s'$, we have $T(w,s') \subseteq T(w,s)$.
\end{prop}
}

\matcomment{
Moreover, as described in Algorithm~\ref{alg:matroid} define
\begin{equation}
\label{eq:Swbar}
{{S}}_w:=H_{\alpha}^{\alpha \beta},
\end{equation}
and
\begin{equation}
\bar{{S}}_w:=\bar{H}_{\alpha}^{\alpha \beta},
\end{equation}
\toRemove{
\begin{equation}
\label{eq:taustar}
\tau^*:=\arg \max_{\tau: |\tau|=\alpha} f((S_{1,\ldots,w-1} \cup \gamma(\tau))\setminus \zeta(\tau)) - f(S_{1,\ldots,w-1}).
\end{equation}
}
}

\mcomment{
Define the union of all greedy subsequences $\tau$ of length $\alpha$ from the $\alpha \beta$ slots in a window $w$.
%, and $S_w$ to be  the best subsequence among those. 
%That is,
\begin{equation}
\label{eq:Rw}
R_w = \cup_{\tau: |\tau|=\alpha} \gamma(\tau),
\end{equation}
and denote $R_{1,\cdots, w}:=R_1\cup \cdots \cup R_{w}$. 
}
\toRemove{
Moreover, define $S_w$ to be  the best subsequence among those. 
\begin{equation}
\label{eq:Sw}
S_w=\gamma(\tau^*),
\end{equation} also define 
\begin{equation}
\label{eq:Swbar}
\bar{S}_w=\zeta(\tau^*),
\end{equation}
}

\mcomment{
\begin{lemma}
Size of set $R_w$ is a constant. $|R_w|\le \alpha {\alpha \beta \choose \alpha}$.
\end{lemma}
}

\begin{lemma}
The size of the shortlist $R$ that Algorithm~\ref{alg:matroid} uses is at most $4k \beta \log (2/\epsilon)$. % = O(\log (1/\epsilon))$.
\end{lemma}

\begin{proof}
Similar to the proof of Lemma~\ref{lem:Rsize}.
\end{proof}

\toRemove{
also define
\begin{equation}
\label{eq:Swhat}
\hat{S}_w=\{c_{j_1},\cdots, c_{j_t}\}, \text{where } (\gamma(\tau^*)\cap A)=  \{i_{j_1},\cdots, i_{j_t}\}.
\end{equation}
}

%Note that $i_j$ (refer to \eqref{eq:ij}) can be set as  either an item in slot $s_j$ or an item {\it from a previous greedy subsequence} in $R_1\cup \cdots \cup R_{w-1}$. The significance of the latter relaxation will become clear in the analysis.
%Here, $s_j \cup R_1 \cup \cdots \cup R_{w-1}$ denotes the union of items in slot $s_j$ and 

\mcomment{
At the end of the algorithm, $A^*$ will be elements of $\bar{S}_w$ that are selected by the Algorithm~\ref{alg:matroidmax} as part of the shortlist.
}
\toRemove{
In other words, $\hat{S}_w$ is counterpart of elements of $S_w\cap A$ that are removed by $g$. 
Also note that in the main Algorithm~\ref{alg:main}, we  remove $\zeta(\tau^*)$ from $S_{1, \cdots, w-1}\cup S_w$ at the end of window $w$ and make $S_{1,\cdots, w}$ an independent set of $\mathcal{M}$.
}

%$S_w$ represents the best subsequence among all the subsequences in $R_w$.
%\mcomment{sa}
%As such, identifying the sets $R_w$ and $S_w$ involves looking forward in a slot $s_j$ to 

%To obtain an online implementation of this procedure, 

\mcomment{
We use online Algorithm \ref{alg:matroidmax} for the \textit{secretary problem with replacement} described earlier, in order to find the item with the maximum $g$ value as in eq.~\eqref{eq:ij}, among all the items in the slot. 
Note that $R_w$ %, $S_w$ 
and $\bar{S}_w$ can be computed {\it exactly at the end} of window $w$. }
%This online procedure will result in selection of a set $H_w$ potentially larger than $R_w$, while ensuring that each element from $R_w$ is part of $H_w$ with a high probability $1-\delta$ at the cost of adding extra $\log(1/\delta)$ items to the shortlist. Note that $R_w$ and $S_w$ can be computed {\it exactly at the end} of window $w$. 
%Algorithm \ref{alg:tmp} summarizes the overall structure of our algorithm. 
%The  following notation is used for brevity. For any collection of sets $V_1, \ldots, V_\ell$, we use $V_{1,\ldots, \ell}$ to denote $V_1 \cup \cdots \cup V_\ell$. 
%In the algorithm, for any item $i$ and set $V $, we define $\Delta_f(i|V):=f(V\cup \{i\})- f(V)$.
%The subroutine to compute $H_w$ for window $w$ uses Algorithm \ref{alg:SIImax} for each subsequence $\tau$ and each slot $s_j$ in that subsequence. Consequently, the parameters for Algorithm \ref{alg:SIImax} for slot $j$ are: number of items $X_j$ which is $n/k\beta$ in expectation, $\delta >0$, and value $a_i$ of an item $i\in [X_j]$ is as given by the expression inside maximizer in \eqref{eq:ij}.
%\scomment{I will come back to make this more precise with parameter specifications, after finishing the subroutine description and analysis.}
\mcomment{
The algorithm returns both the shortlist $A$ which similar to~\cite{us} is of size $O(k)$
%as stated in the following proposition
, as well as the set $A^*$.
Note that we remove $\hat{S}_w$ from $A^*$ at the end of window $w$.}

Similar to previous section, conditioned on $T \mcomment{_{1,\cdots, w-1}}$, different elements of $S^*$ have different probability of appearing in a slot $s$. By subsampling set $Z_s$, make these probabilities even  (Note that $T$, and $Z_s$ are for the purpose of analysis. %The algorithm does not keep track of these variables).

\mcomment{
\begin{lemma}
%\label{lem:Zs}
For any slot $s$ in a window $w\in [W]$, 
given  $T(w,s)$, 
all $i, i' \in S^*, i\ne i'$ will appear in $Z_s$ independently with probability $\frac{1}{k\beta}$, i.e.,  the random variables  $\mathbf{1}(i\in Z_s | T(w,s))$ are i.i.d. for all $i \in S^*$, and 
\[
\Pr(i \in Z_s | T(w,s)) = \Pr(i'\in Z_s | T(w,s)) = \frac{1}{k\beta} \ .
\]
\end{lemma}
}
\mcomment{
%\begin{restatable*}{lemma}{lemmazs} 
\begin{lemma}\label{lemma:zs}
For all $i, i' \in S^*\backslash \{Z_{s_1} \cup \ldots \cup Z_{s_{j-1}}\}$,
\begin{align*} 
\label{eq:pk}
&\Pr(i\in Z_{s_j}| Z_{s_1}, \ldots , Z_{s_{j-1}}, T_{1,\cdots, w-1})
= \Pr(i'\in Z_{s_j}| Z_{s_1}  \ldots  Z_{s_{j-1}},T_{1,\cdots, w-1})  
\ge \frac{1}{k} \ .
\end{align*}
\end{lemma}
}
%\end{restatable*}{lemma}
\toRemove{
\begin{proof}
The proof is similar to Lemma ? in~\cite{us}, and it is based on Lemma \ref{lem:Zs},
\end{proof}
}

%We will use Lemma~\ref{lem:Zs}.
%which also holds true for the new definition of $T$.

%In the following lemma, we lower bound the marginal gain of a randomly picked element of optimal solution in slot $s$ with respect to previously selected items. 

\mcomment{
\begin{lemma}
\label{lem:margij}
For slot $s$ in window $w$, and $1\le \ell \le \alpha'=\min(t,\alpha)$, {\small
$$\Ex[\Delta(m_{\ell}^s|S_{1,\ldots, w-1} \cup H_{\ell-1}^{s-1} %\{i_1, \ldots, i_{j-1}\}
)|T(w,s), Z_s\ne \emptyset ] \ge \frac{1}{k}\left(f(S^*)-f(S_{1,\ldots, w-1} \cup H_{\ell-1}^{s-1}
%\{i_1, \ldots, i_{j-1}\}
)\right)\ .$$}
\end{lemma}
}

\begin{lemma}
%\label{lem:margijMatroid}
\label{lem:asGoodas}
For slot $s$, and a randomly selected element $a$ in $Z_s$, %in a window $w$, and $1\le \ell \le \alpha'=\min(t,\alpha)$, 
{\small
$$\Ex[\Delta(a|SH(s)
)|T(w,s), Z_s\ne \emptyset ] \ge \frac{1}{k}\left(f(S^*)-f(SH(s)
%\{i_1, \ldots, i_{j-1}\}
)\right)\ .$$}
\end{lemma}
\begin{proof}
The proof is similar to Lemma~\ref{lem:margij}, for a randomly selected item $a\in Z_s$.
\end{proof}

\matcomment{
\begin{proof}
From Definition~\ref{def:ml},
%of $\gamma(\cdot)$ (refer to \eqref{eq:gamma}), 
$m_{\ell}^s$ is chosen greedily to maximize the increment
\[
\underset{x\in s\cup R(w,s)}{\arg\max} \Delta(x|SH(w,\ell,s)),
\]
\mcomment{$R$ should be $R(w,s_j)$}
%$\Delta_f(i|S_{1,\ldots, w-1} \cup i_{1,\ldots, s-1})$ over all 
So $m_{\ell}^s$ belongs to $s \cup R(w,s)
%Supp(T_{1,\ldots, w-1}) 
\supseteq Z_{s}$. 
Therefore, we can lower bound the marginal gain of $m_{\ell}^s$ w.r.t. previously selected items $SH(w,\ell,s)$ by the marginal gain of a randomly picked item $i$ from $Z_{s}$ as follows.
\begin{eqnarray*} 
& & \Ex[\Delta(m_{\ell}^s|SH(w,\ell,s))|T(w,s), Z_s\ne \emptyset] \\
(\text{using }\eqref{eq:pk} \mcomment{and \ref{}}) & \ge & \frac{1}{k} \sum_{i\in S^*} \Delta(i|SH(w,\ell,s)) \\
(\text{using  Lemma~\ref{marginalsum}, monotonicity of $f$} ) & \ge & \frac{1}{k} \left( f(S^*)-f(SH(w,\ell,s)) \right) \\
\end{eqnarray*}
\end{proof}
}
\mcomment{
We can lower bound the increment assuming $a$ is randomly picked item from $Z_{s_j}\cap S^*$:
\begin{eqnarray*} 
& & \Ex[\Delta_f(a,S(w-1,\tilde{\tau}_w,j-1)) |T_{1,\ldots, w-1}=T, i_1, \ldots, i_{j-1},a\in S^*\cap Z_{s_j}] \\
& \ge & \frac{1}{k} \Ex[\sum_{a\in S^*\backslash \{Z_1,\ldots Z_{s_{j-1}}\}} \Ex[\Delta_f(a,S(w-1,\tilde{\tau}_w,j-1))|T, i_1, \ldots, i_{j-1}]] \\
%(\text{using  Lemma~\ref{marginalsum}, monotonicity of $f$} ) 
& \ge & \frac{1}{k} \Ex[ (f(S^*\backslash  \{Z_1,\ldots Z_{s_{j-1}}\})-f(S(w-1,\tilde{\tau}_w,j-1))) | T]\\
%(\text{using  monotonicity of }f)
& \ge & \frac{1}{k}  \Ex[(f(S^*\backslash  \cup_{s'\in w} Z_{s'})-f(S(w-1,\tilde{\tau}_w,j-1))| T]\\
& \geq & \frac{1}{k} ((1-\frac{\alpha}{k})f(S^*) -f(S(w-1,\tilde{\tau}_w, j-1)).
\end{eqnarray*}
%Here, we used the submodularity of $f$ and 
The last inequality uses 
Lemma \ref{sample} for submodular function $f$,
and the observation from Lemma \ref{lem:Zs} that given $T$, every $i\in S^*$ appears in $\cup_{s'\in w} Z_{s'}$ independently with probability $\alpha/k$, so that every $i\in S^*$ appears in $S^*\backslash\cup_{s'\in w} Z_{s'}$ independently with probability $1-\frac{\alpha}{k}$; %along with Lemma \ref{sample} for submodular function $f$. 
%so that $\Ex[|\cup_{s'\in w} Z_{s'}|] \le {\alpha}$.
%\end{proof}
}

Now we use the Brualdi lemma (refer to~\ref{lem:Brualdi}), to create a bijection $\pi$ between a base of matroid containing the current solution %$S(w-1,\tilde{\tau},j-1)$ 
$SH(s)$
and the optimal solution. Then for a random element $a$ of the optimal solution in  %$Z_{s_j}$, 
$Z_s$,
if we remove its corresponding element $\pi(a)$ from the current solution, we can still lower bound the value of the remaining set. 

%Then we lower bound the value of the set remaining after we remove an element of optimal solution in slot $Z_{s_j}$, 

\begin{lemma}\label{replaced}
Let $S'$ be the extension of %$S(w-1,\tilde{\tau}_w,j-1)$
$SH(s)$
to a base of $\mathcal{M}$ (refer to Lemma~\ref{extension}).
Let $\pi$ be the bijection from Brualdi lemma (refer to Lemma~\ref{lem:Brualdi}) from $S^*$ to $S'$.
Then, for a randomly selected element $a$ in $Z_s$
%for all $j=1,\ldots, \alpha'$, \\
{\small \begin{eqnarray*}
\Ex[f(SH(s) - \pi(a))|T(w,s), Z_s\ne \emptyset]
 \ge (1-\frac{1}{k})f(SH(s))\ .
\end{eqnarray*}}
\end{lemma}
\begin{proof}
Since $\pi$ is a bijection from $S^*$ to $S'$, from Brualdi's lemma (lemma~\ref{lem:Brualdi}), 
$SH(s)- \pi(a) +a \in \mathcal{I}$, for all $a \in S^*$. 
Recall the definition of $Z_{s}$.
Suppose $a$ is a randomly picked item from $ Z_{s}$.
Since $Z_{s}\ne \emptyset$,  using Lemma~\ref{lem:Zs} conditioned on $T(w,s)$, the element $a$ can be equally any element of $S^*$ with probability
$1/k$. Therefore, $\pi(a)$ would be any of $SH(s)$ with probability at most
$1/k$, i.e., 
%(It is because $\pi'$ might map some elements of $S^*$ to $S'\setminus SH()$).
\[
\Pr(\pi(a)=e | T(w,s)) \le 1/k, \ \text{for } e \in  SH( s), 
\]
Now the lemma follows from the definition of $\pi$ and lemma~\ref{sample}.
%the lemma follows.
\end{proof}

\mcomment{
\begin{proof}
Since $\pi$ is a bijection from $S^*$ to $S'$, from Brualdi's lemma (lemma~\ref{lem:Brualdi}), 
$SH(w,\ell,s)- \pi'(a) +a \in \mathcal{I}$, for all $a \in S^*$. Further, $\pi'(a)=\pi(a)$ if $\pi(a)\in  S(w,\ell,s)$ and $\pi'(a)=\emptyset$ otherwise.
Recall the definition of $Z_{s_j}$.
Suppose $a$ is a randomly picked item from $S^*\cap Z_{s_j}$.
Since $Z_{s_j}\ne \emptyset$,  using Lemma~\ref{lem:Zs} conditioned on $T(w,s)$, the element $a$ can be equally any element of
$S^*$
with probability
$1/k$.
Therefore, $\pi'(a)$ would be any of $SH(w,\ell,s)$ with probability at most $1/k$
(It is because $\pi'$ might map some elements of $S^*$ to the $\emptyset$).
Now based on the definition of $\pi$ and lemma~\ref{sample} the lemma follows.
\end{proof}
}

\mcomment{
Next, we can lower bound the expected marginal gain of the next element selected in by $\gamma$ in slot $s_j$ of $\tilde{\tau}$. 
}

\toRemove{
\begin{lemma}\label{exg}
%Suppose function $g$ is as defined in equation~\eqref{eq:g}.
%\mcomment{ , and $\rho:=\frac{k}{k-\alpha}$}.
For all $j=1,\ldots, \alpha'$, \\
\mcomment{
\begin{align}  \label{recursion}
\Ex[f({SC}(w,j) ) 
- f(SC(w,j-1) ) | T(w,s_j)] \nonumber 
\ge
 \frac{1}{k}(f(S^*)-2 f({SC}(w,j-1))  \ .
\end{align}
}
\mcomment{ba expted}
\begin{align}  
\Ex[f({SC}(w,j) ) 
- f(SC(w,j-1) ) | T(w,s_j)] \nonumber 
\ge
 \frac{1}{k}\Ex[f(S^*)-2 f({SC}(w,j-1) | T(w,s_{j-1})]  \ . 
\end{align}

\mcomment{
{\small \begin{eqnarray*}
&\Ex[g(i_j,SH(w,\ell,s)) | T(w,s), s=s_j] \ge 
 %(1-\frac{1}{k-\alpha})
 \frac{1}{k}
 (f(S^*)-
 2 f(SH(w,\ell,s)))\ .
\end{eqnarray*}}
}
\end{lemma}
}

\begin{lemma}\label{exg}
For slots $s$,
\begin{align}  
\Ex[f({SH}(s) ) 
- f(SH(s-1) ) | T(w,s),Z_s\ne \emptyset] \nonumber 
\ge
 \frac{1}{k}\Ex[f(S^*)-2 f({SH}(s-1) | T(w,s)]  \ . 
\end{align}

\end{lemma}

\begin{proof}
\matcomment{
In the Algorithm~\ref{alg:matroid}, at the end of window $w$, we set 
$S=S + S_w -\bar{S}_w$.
}
%Suppose $a\in s_j\cap S^*$.
Suppose $a\in Z_s$, 
and, let $S'$ be the extension of %$S(w-1,\tilde{\tau}_w,j-1)$
%$SH(w,\ell,s)$ 
$SH(s)$ 
to a base of $\mathcal{M}$, and $\pi$ be the bijection from Brualdi's Lemma (refer to Lemma~\ref{lem:Brualdi}) from $S^*$ to $S'$.
Thus the expected value of the function $g$ on the element selected by the algorithm in slot $s$ (the element with maximum $g$ in the slot $s$) is as follows. 
%\begin{eqnarray*}
\begin{align*}
\Ex[f(SH( {s})|T(w,s)]   \ge& 
\Ex[f( SH(s-1) + a - \pi(a)) 
 |T(w,s), a\in Z_{s}] 
\\ \ge &
\Ex[f(SH(s-1) - \pi(a) )
|T(w,s), a\in  Z_{s}]+
  \Ex[\Delta(a|SH(j-1)- \pi(a) )
|T(w,s), a\in  Z_{s}] 
\\ \ge &
\Ex[f(SH(s-1) - \pi(a) )
|T(w,s), a\in Z_{s}]
+\Ex[\Delta(a|SH(s-1) )
|T(w,s), a\in Z_{s}].
\end{align*}
%\end{eqnarray*}

The first inequality is from the definition of function $g$ as it is defined in equation~\ref{eq:g}. The last inequality is from submodularity of $f$.
Now from the last inequality  and lemma~\ref{replaced} we have
\begin{align*}
\Ex[f({SH}(s))  |T(w,s), Z_s \ne \emptyset]\ge &
 (1-\frac{1}{k} ) f(SH(s-1) )
 + \Ex[ \Delta(a|{SH}(s-1) \})
 |T(w,s), a\in  Z_{s}]. 
\end{align*}
Now from lemma~\ref{lem:asGoodas} and the above inequality we can show
\begin{align*}
\Ex[f({SH}(s)  |T(w,s), Z_s\ne \emptyset]&\ge
(1-\frac{1}{k} ) f(SH(s-1) )
+ \frac{1}{k} (f(S^*)-f({SH}(s-1)) )\ . 
\end{align*}
Thus,
\begin{align}  %\label{recursion}
\Ex[f({SH}(s) ) 
- f(SH(s-1) ) | T(w,s), Z_s \ne \emptyset] \nonumber 
\ge
 \frac{1}{k}(f(S^*)-2 f({SH}(s-1))  \ . 
 %\label{recursion}
\end{align}

\matcomment{
Hence, by taking expectation on $T(w,s)$, and by Proposition~\ref{prop:Tsucc},
\begin{align}  
\Ex[f({SH}(j) ) 
- f(SH(j-1) ) | T(w,s_j)] \nonumber 
\ge
 \frac{1}{k}\Ex[f(S^*)-2 f({SH}(j-1) | T(w,s_{j-1})]  \ . 
\end{align}
}

\mcomment{
Hence,
\begin{align*}
&\Ex[g(i_j,SH(j-1)) | T(w,s_j)] \ge 
 %(1-\frac{1}{k-\alpha})
 \frac{1}{k}
 (f(S^*)-2 f(S(w,j-1))) \ .
\end{align*}
}

\toRemove{
\begin{align}  %\label{recursion}
f({S}(w-1,\tilde{\tau}_w,j) ) 
- f(S(w-1,\tilde{\tau}_w,j-1) ) %\nonumber 
\ge
 \frac{1}{k-\alpha}((1-\frac{\alpha}{k})f(S^*)-2f({S}(w-1,\tilde{\tau}_w,j-1))  \ . 
 %\label{recursion}
\end{align}
}

\end{proof}

\mcomment{
\begin{proof}
Let $\pi_0= (1-\frac{\alpha}{k}) f(S^*) - 2\rho\Ex[f(S_{1,\ldots, w-1}) | T_{1,\ldots, w-1}=T]$, and for $j \ge 1$,
\begin{align}
\pi_j:=(1-\frac{\alpha}{k}) f(S^*) - 2\rho\Ex[f(S(w-1,\tilde{\tau}_w,j)) | T_{1,\ldots, w-1}=T, i_1,\ldots, i_{j-1}],
\end{align}
Then, subtracting and adding $\frac{1}{2}(1-\frac{\alpha}{k}) f(S^*)$ from the left hand side of lemma~\ref{exg}, and taking expectation conditional on $T_{1,\ldots, w-1}=T, i_1, \ldots, i_{j-2}$, we get
\[-\frac{1}{2} ( \Ex[\pi_{j} | T, i_1, \ldots, i_{j-2}] + \pi_{j-1} ) \ge \frac{\rho}{k} \pi_{j-1}\]
which implies
\[\Ex[\pi_j|T, i_1, \ldots, i_{j-2}] \le \left(1-\frac{2\rho}{k}\right) \pi_{j-1} \le \left(1-\frac{2\rho}{k}\right)^j \pi_0\ .\]
By martingale stopping theorem, this implies:
\[\Ex[\pi_t|T] \le \Ex\left[\left(1-\frac{2\rho}{k}\right)^t \left| T\right. \right] \pi_0 \le \Ex\left[e^{-2t/(k-\alpha)}| T\right] \pi_0\ .\]
where stopping time $t=|\tilde{\tau}_w|\le \alpha \beta$ is bounded. 
Therefore, by martingale stopping theorem we can replace $t$ with $|\tilde{\tau}_w|$.
%($t=|\tilde \tau_w| \le \alpha\beta$ is bounded, therefore, martingale stopping theorem can be applied).
\end{proof}
}

\mcomment{
\begin{lemma}
\label{cor:asGoodas}
$$\Ex\left[\left(1-\frac{\alpha}{k}\right) f(S^*) - 2\rho f(S(w-1,\tilde{\tau}_w,|\tilde{\tau}_w|)) | T\right]
\le \Ex\left[e^{-\frac{2|\tilde \tau_w|}{k-\alpha}} \left|\right. T\right] \left(\left(1-\frac{\alpha}{k}\right) f(S^*)-2\rho f(S_{1,\ldots, w-1})\right).$$
\end{lemma}
}

\toRemove{
\begin{proof}
In the appendix.
\end{proof}
}

\toRemove{
\begin{proof}
Let $\pi_0= (1-\frac{\alpha}{k}) f(S^*) - 2\Ex[f(S_{1,\ldots, w-1}) | T_{1,\ldots, w-1}=T]$, and for $j \ge 1$,
$$\pi_j:=(1-\frac{\alpha}{k}) f(S^*) - 2\Ex[f(S_{1,\ldots, w-1}\cup \{i_1, \ldots, i_j\}$$
$$\setminus\{c_1,\cdots,c_j\}) | T_{1,\ldots, w-1}=T, i_1,\ldots, i_{j-1}],$$
Then, subtracting and adding $\frac{1}{2}(1-\frac{\alpha}{k}) f(S^*)$ from the left hand side of the recursion~\ref{recursion} in lemma~\ref{exg}, and taking expectation conditional on $T_{1,\ldots, w-1}=T, i_1, \ldots, i_{j-2}$, we get
$$-\frac{1}{2} ( \Ex[\pi_{j} | T, i_1, \ldots, i_{j-2}] + \pi_{j-1} ) \ge \frac{1}{k-\alpha} \pi_{j-1}$$
which implies
$$\Ex[\pi_j|T, i_1, \ldots, i_{j-2}] \le \left(1-\frac{2}{k-\alpha}\right) \pi_{j-1} \le \left(1-\frac{2}{k-\alpha}\right)^j \pi_0\ .$$
By martingale stopping theorem, this implies:
$$\Ex[\pi_t|T] \le \Ex\left[\left(1-\frac{2}{k-\alpha}\right)^t \left| T\right. \right] \pi_0 \le \Ex\left[e^{-2t/(k-\alpha)}| T\right] \pi_0\ .$$
where stopping time $t=|\tilde{\tau}_w|$. ($t=|\tilde \tau_w| \le \alpha\beta$ is bounded, therefore, martingale stopping theorem can be applied).

\end{proof}
}

\mcomment{
Next, we compare $\gamma(\tilde \tau_w)$ to $S_w=\gamma(\tau^*)$ . Here, $\tau^*$ was defined has the `best' greedy subsequence of length $\alpha$ (refer to %\eqref{eq:Sw} and
\eqref{eq:taustar}). 
To compare it with $\tilde \tau_w$,}

Similar to Defintion~\ref{def:B} in the previous section define:

\begin{definition}
For a fixed slot $s$, define $$B_s:=\Ex \left[f\left(SH({s})\right) \right] .$$
\end{definition}
Then, we get the following equation,
\begin{align*}\label{eq:wrec}
&B_{s} \ge q \times \left(B_{s-1} + \frac{1}{k}(f(S^*)-2B_{s-1}) \right) + (1-q) \times B_{s-1}, & \forall 1\le s \le k \beta \nonumber \\
&B_0=0.
\end{align*}
where $f(S) \ge B_{k \beta}$.

From the above recursion we get
$$
\Ex[f(S)]/OPT \ge \frac{1}{2} \left(1-(1-2q/k)^{k\beta}  \right) \ge \frac{1}{2}(1-e^{-2q\beta }) \ge \frac{1}{2}(1-e^{-2}-2\epsilon).
$$

\begin{remark}
Note that similar to~\cite{us}, we can show that $\Ex[f(S\cap R)] \ge (1-\epsilon) \OPT$.
\end{remark}

\matroidThm

\begin{cor} \label{thm:preemption}
For the matroid secretary problem in the %preemption model, and 
matroid secretary problem 
that uses shortlist of size at most $\eta(k)=k$,
there is an algorithm 
%the algorithm~\ref{alg:main} 
%(with $\alpha=\beta=1$) 
that achieves a constant competitive ratio. %$\frac{1}{2} (1-1/e) (1-1/e^2-\epsilon)$.
\end{cor}

%% file: mmatroids.tex
In this section, we present  algorithms
for monotone submodular function maximization subject to $p$-matchoid constraints.
These constraints generalize many basic combinatorial constraints such as the cardinality constraint, the intersection of $p$ matroids, and matchings in graphs. 
A formal definition of
a $p$-matchoid is in~\cite{Chekuri} and in the appendix.
Throughout this section,  $k$ would refer to the size of the largest feasible set.

\toRemove{
\begin{definition} \label{def:matchoid}
(\textbf{Matchoids}). Let $\mathcal{M}_1 = (\mathcal{N}_1, \mathcal{I}_1), \cdots ,\mathcal{M}_q = (\mathcal{N}_q, \mathcal{I}_q)$ be $q$ matroids over overlapping groundsets. Let $\mathcal{N} = \mathcal{N}_1\cup  \cdots \cup \mathcal{N}_q$ and $\mathcal{I} = \{S\subseteq \mathcal{N} : S\cap \mathcal{N} \in \mathcal{I}_{\ell},
 \forall \ell\}$. The finite set system
$\mathcal{M}_p = (\mathcal{N} , \mathcal{I})$ is a $p$-matchoid if for every element $e \in \mathcal{N}$ , $e$ is a member of $\mathcal{N}_{\ell}$ for at most $p$ indices $\ell \in [q]$.
\end{definition}
}
%p-matchoids generalizes matchings and intersections of matroids, among
%others

\mcomment{
There are some subtle differences in the algorithm as well as in the analysis.
}
We make some modifications in the algorithm in Section~\ref{sec:matroid}, and the analysis provided there. 
The main difference in the algorithm is that we update functions $g$ and $\theta$ defined in Definition~\ref{def:g}.
Here, function $\theta$, instead of one item, might remove up to $p$ items form the current independent set $S$. Each removed item corresponds to different ground set $\mathcal{N}_i$, in which the new item lies (based on the definition of $p$-matchoid constraints, %Definition~\ref{def:matchoid},
there are at most $p$ such elements).

\begin{definition} \label{def:gmatchoid}
For each matroid $M_{\ell}=(\mathcal{N}_{\ell},\mathcal{I}_{\ell})$, and $\ell\in [q]$  define:
\begin{equation*}
\Omega_{\ell}(e,S):= \{e'\in S| S+e-e' \in \mathcal{I}_{\ell} \}.
\end{equation*}
For an element $e$ in the input, suppose $e\in N_{\ell_i}$, for $i=1,\cdots, t\le p$.
Define 
\begin{equation*}
\lambda(e,S):= \prod_{i=1}^{t} {\Omega_{\ell_i}(e,S)}.
\end{equation*}
For a combination vector 
$r=(r_1,\cdots, r_p)\in \lambda(e,S)$, where
$r_i \in \Omega_{\ell_i}(e,S)$,
define the union of all the components of $r$ as:
\begin{equation*}
\mu(r):=  \{r_1,\cdots, r_p\}.
\end{equation*}
\begin{equation*}
g_r(e,S):= f(S+e-\mu(r))-f(S).
\end{equation*}
Also define:
\begin{equation*}
\theta(e,S):= \mu( \arg\max_{r\in \lambda(e,S) } g_r(e,S)).
\end{equation*}
%Thus, $\theta(e,S) \in \lambda(e,S)$.
Furthermore define,
\begin{equation*}
\label{eq:newg}
g(e,S):= \max_{r\in \lambda(e,S) } g_r(e,S).
\end{equation*}

\end{definition} 

\toRemove{
For each index $\ell\in [q]$  define the following functions as before:
%$\theta$ as: 
$$ \theta_{\ell}(e,S) := \arg\max_{e'\in S} \{ f(S+e-e')| S+e-e' \in \mathcal{I}_{\ell} \} $$
Also define $g_{\ell}$ as 
% $$g(e):=\max_{e': S+e-e' \in \mathcal{I}} F(S+e-e') - F(S)$$
 $$g_{\ell}(e,S):= f(S+e-\theta_{\ell}(e,S)) - f(S)$$
and
$$g(e,S):= f(S+e-\bigcup_{i}\theta_i(e,S)) - f(S)$$
 }

Now using the new definition of $g$, we employ the \textit{oneline max algorithm}, %Algorithm 1, in~\cite{us} 
to find:
\[
m_{\ell} \leftarrow\underset{x\in s\cup R}{\arg\max }\  g(x,S).
\]
Accordingly we will update line 5 of Algorithm~\ref{alg:matroidconst}, by this new definition of $g$ in eq.~\eqref{eq:newg}.
\mcomment{ 
As in the online subroutine for the main algorithm, we run Algorithm~\ref{alg:matroidmax} with the new function $g$ defined in eq.~\eqref{eq:newg}.}
It returns  element  $e$ with maximum  $g(e,S)$, and it achieves a  $1-\delta$ competitive ratio   with shortlists of size logarithmic in $1/\delta$.
Here, the output of $\theta$ in Algorithm~\ref{alg:matroidconst} is a set instead of only one item:
\[
 o_{\ell} :=  \theta(m_{\ell}, SH(w,s)).
\]
%Additionally, we  make some changes in Algorithm~\ref{alg:matroidconst}. 
%We define $H_{\ell}^s$ and $\bar{H}_{\ell}^s$ defined in Definition~\ref{def:Hbar}, using the new definition of $g$ in eq.~\eqref{eq:newg}.
Note that in each update a set would be removed from $SH$, whereas for the matroid constraints it was only one item.
\mcomment{
\begin{definition} \label{def:Hbar}
 For a slot $s \in \{1,\cdots, \alpha \beta\}$ in window $w$.
Let's denote by $H_{\ell}^s$, the set $H_{\ell}$ as defined in the Algorithm~\ref{alg:card} at the end of slot $s$.  Similarly, define $\bar{H}_{\ell}^s$ to be the set $\bar{H}_{\ell}$ at the end of slot $s$. (also initialize  $H_{\ell}^0=\bar{H}_{\ell}^0=\emptyset$)
\end{definition}
}
Furthermore, we define %$C_j, \bar{C}_j$, 
$SH(w,s)$ %, SC(w,j)$, $S_w$ and $\bar{S}_w$ 
similar to their definition in Section~\ref{sec:matroid}, using new definition of $g$ and $\theta$.

\mcomment{
\begin{definition}
For slot $s$ in window $w$, and $1 \le \ell \le \alpha$, define \mcomment{ $s\ge 1$ define for $s=0$}
\begin{equation*}
SH{(w, \ell, s)} :=( S_{1,\cdots, w-1} \cup H_{\ell-1}^{s-1} ) \setminus \bar{H}_{\ell-1}^{s-1}\ .
\end{equation*}
\end{definition}
\begin{definition} \label{def:mlmatroid}
Define $m_{\ell}^s$ to be $m_{\ell}$ as defined in Algorithm~\ref{alg:cardconst}, for slot $s$, which is
\begin{equation*}
m_{\ell}^s:=\underset{x\in s\cup R(w,s)}{\arg\max}{g(x,SH(w,\ell,s))}, 
\end{equation*}
\begin{equation*}
r_{\ell}^s:=\theta(m_{\ell},SH(w,\ell,s))\ .
\end{equation*}
Also for the sequence $\tilde{\tau}_w=(s_1,\cdots, s_t)$ defined in Definition~\ref{def:tauw}, define sequence $\mcomment{\mu_w}=(i_1,\cdots, i_{\alpha'})$, and $\nu_w:=(q_1,\cdots, q_{\alpha'})$
for $\alpha'=\min(t,\alpha)$,
where
\begin{equation*}
i_j:=m_{j}^{s_j},
\end{equation*}
and
\begin{equation*}
q_j:=r_{j}^{s_j},
\end{equation*}
Moreover, for $1\le j \le \alpha'$ define 
\begin{equation*}
C_j:=H_{j}^{s_{j+1}-1}
\end{equation*}
and
\begin{equation*}
\bar{C}_j:=\bar{H}_{j}^{s_{j+1}-1}
\end{equation*}
If $j+1> \alpha'$, set $s_{j+1}:=\alpha \beta+1$. 
We also use the notation $i_{1,\cdots, j}=(i_1,\cdots, i_j)$, for $1\le j\le \alpha'$.
\end{definition}
\begin{definition}
\begin{equation*}
SC{(w,j)} := (S_{1,\cdots, w-1} \cup C_{j-1}) \setminus \bar{C}_{j-1}\ .
\end{equation*}
\end{definition}
}

\mcomment{
In particular, we define $\gamma$ similar to~\eqref{eq:gamma} but using the new definition of $g$ in equation~\eqref{eq:newg}.
%$\zeta(\tau):=\{C_1, \ldots, C_\ell\}$
Moreover, for a subsequence $\tau=(s_1,\ldots, s_\ell)$
define 
\begin{equation}
\zeta(\tau):=\bigcup_{j=1}^{\ell} C_j
\end{equation}
where each $C_j$ is a set defined as
\begin{equation}
\label{eq:cij}
C_j :=  
\theta(i_j, S(w-1,\tau,j-1)).
\end{equation}
\toRemove{
\begin{equation}
\label{eq:cij}
C_j :=  
\theta(i_j, S_{1, \ldots,w-1} \cup \{i_1,\ldots, i_{j-1}\}).
\end{equation}
}
Note that in contrast with the definition of $\zeta(\tau)$ for the matroid constraints~\eqref{eq:cij}, in which $c_j$ is only one item, now each $C_j$ is a subset of the current independent set $S$. Further, the definition of $S$ and $\bar{S}_w$, in~\eqref{eq:diff} and \eqref{eq:Swbar}, will be updated accordingly using the new definition of $\zeta(\tau)$.
}

\matchoidThm

\begin{proof}
The proof is based on the recursion we get in Lemma \ref{lem:recursionMatchoid} in the Appendix. It is similar to proof of %Lemma~\ref{cor:asGoodas} for
the matroid constraints.
%which omitted.
\end{proof}

%% file: Supp/CardSupp.tex
\toRemove{
\begin{lemma} \label{lem:tw}
For any real $\delta \in (0,1)$, and $k\ge \alpha \beta$, then given $T$,
\[
|\tilde{\tau}_w| \ge (1-\delta)(1-1/\beta) \alpha
\]
with probability $1-exp(-\frac{\delta^2 \alpha}{3})$.
\end{lemma}
\begin{proof}
By definition 
$$
|\tilde{\tau}_w| = |s\in w: Z_s\ne \emptyset|
$$
%Similar to eq.~(12) in~\cite{us}, 
then from Lemma~\ref{lem:Zs},
given $T(w,s)$, for $i\in S^*$, we have $\Pr(i\in Z_s)=\frac{1}{k\beta}$.
%By new construction of $T$ and
Suppose $A:=\cup_{s\in w} Z_s$. Now using Chernoff bound we can show that 
\[
\Pr(|A|<(1-\delta) \alpha) < exp(-\delta^2 \alpha/3)
\]
We also have $i\in Z_s|T$ and $i'\in Z_s|T$ are independent for $i,i'\in S^*$.
Now given that $|A|\le \alpha$ (otherwise ignore some elements), for element $e\in \cup_{s\in w} Z_s$, the probability that $e$ lies in the same slot as some other element of  $\cup_{s\in w} Z_s$ is at most $\frac{\alpha}{\alpha \beta}=\frac{1}{\beta}$. Thus $(1-1/\beta)$ fraction of $A$ will be in different slots. Thus
\[
\Pr(|\tilde{\tau}_w|<(1-\delta)(1-1/\beta) \alpha) < exp(-\delta^2 \alpha/3)
\]
\end{proof}
}

\subsection{Some useful properties of $(\alpha, \beta)$ windows}

We revisit the properties proven for $(\alpha,\beta)$-windows in~\cite{us}. Because of some changes made in the algorithm we need to provide new proofs for some of these properties.

%We first prove some useful properties of $(\alpha, \beta)$ windows, defined in Definition~\ref{def:windows} and used in Algorithm \ref{alg:main}. 

The first observation is that every item will appear uniformly at random in one of the $k\beta$ slots in $(\alpha,\beta)$ windows. 

\rcomment{
\begin{definition}
For each item $e\in I$, define $Y_e\in [k\beta]$ as the random variable indicating the slot in which $e$ appears. We call vector $Y\in [k\beta]^n$ a \textit{configuration}.
\end{definition}
}
%Since the length of each slot is coming from $B(n,\frac{1}{k\beta})$, random variables $\{Y_e\}_{e\in I}$ are i.i.d. with uniform distribution on all $k\beta$ slots. 

\begin{lemma} \label{lem:indep}
For each configuration $Y$, random variables $\{Y_e\}_{e\in I}$ are i.i.d. with uniform distribution on all $k\beta$ slots. 
\end{lemma}

This follows from the uniform random order of arrivals, and the use of the balls in bins process to determine the number of items in a slot during the construction of $(\alpha,\beta)$ windows. 
The proof can be found in~\cite{us}.
%A proof is provided in Appendix \ref{app:windows}.

%Further, we define the following new quantities to aid the analysis. 
%\scomment{Earlier $T_w$ was defined as $T_w:= \bigcup_{\tau} (\tau, \gamma(\tau))$. I am not sure how one would define  union over such things. }

\rcomment{
\begin{definition}
Let's denote by $H_{\ell}^s$, the set $H_{\ell}$ as defined in the Algorithm~\ref{alg:card} at the end of slot $s$ in window $w$. 
%Define $T_w:=\bigcup_{i=1}^{\alpha} H_i$.
\mcomment{
More precisely, define $H_{\ell}^s$ recursively as:
\[
H_{0}^{s} = \emptyset
\]
\[
H_{\ell}^{s} := H_{\ell-1}^{s} + \underset{}{}
\]
}
\end{definition}
}

Next, we make important observations about the probability of assignment of items in $S^*$ in the slots in a window $w$, given the $T(\bar w,s)$.
%sets $R_{1,\ldots, w-1}, S_{1,\ldots, w-1}$ \mcomment{(refer to \eqref{eq:Rw}, \eqref{eq:Sw} for definition of these sets)}.  For the purpose of analysis, we define the following new random variable $T_w$ that will track all the useful information from a window $w$.  

\rcomment{
\begin{definition} \label{def:T}
For window $w\in [W]$, define 
\begin{equation}
T_w:= \{H_{\ell}^s|s\in w, 1\le \ell \le \alpha\},
\end{equation}
%at the end of window $w$.
moreover,  
\[
T_{1,\cdots,w}:= \bigcup_{i=1}^w T_i,
\]
and 
\begin{equation}
    T(w,s):=T_{1,\cdots, w-1} \cup \{ {H_{\ell}^{s'} | s\succ_w s', 1\le \ell \le \alpha }  \}.
\end{equation}
We denote by $s_0$ the first slot in window $w$. Note that $T(w,s_0)= T_{1,\cdots, w-1}$.
\end{definition}
\begin{definition} \label{def:Supp}
For slot $s$ in window $w$ define
\[
Supp(T_w):=\bigcup_{1\le \ell \le \alpha, s\in w} H_{\ell}^s,
\] 
also,
\[
Supp(T_{1,\cdots, w}):= \bigcup_{i=1}^w Supp(T_i),
\]
and,
\[
Supp(T(w,s)):= Supp(T_{1,\cdots, w-1}) \cup \bigcup_{1\le \ell \le \alpha; s\succ_w s'} H_{\ell}^{s'}.
\]
(Note that $Supp(T_{1,\cdots, w})=R_{1,\cdots, w}$). %\mcomment{define $R(w,s)$} 
Furthermore, define $R(w,s):=Supp(T(w,s)).$
%Furthermore, for slot $s$ in window $w$ define
%for window $w$, and slot $s\in w$, define
\end{definition}
}

\rcomment{
\begin{prop} \label{prop:Tsucc}
For slots $s,s'$ in window $w$, such that $s \succ_w s'$, we have $T(w,s') \subseteq T(w,s)$.
\end{prop}
}

\mcomment{
\begin{definition}
Define $T_w:= \{(\tau, \gamma(\tau))\}_\tau$, for  all $\alpha$-length subsequences $\tau=(s_1,\ldots, s_\alpha)$ of the $\alpha\beta$ slots in window $w$. Here, $\gamma(\tau)$ is a sequence of items as defined in \eqref{eq:gamma}. Also define  $Supp(T_{1,\cdots ,w}) :=\{e| e\in\gamma(\tau) \text{ for some } (\tau, \gamma(\tau))\in T_{1,\cdots, w} \} $ (Note that $Supp(T_{1,\cdots, w})=R_{1,\ldots, w}$).
%take the union of $(\gamma(\tau), \tau)$
%$T_{1,\cdots, w} := \bigcup_{\omega \le w, \tau \in w} (\tau, \gamma(\tau)) $
\end{definition}
}

%\begin{obs}
%For $1\le w\le W$,  $T_{1,\ldots, w}$ and $S_{1, \ldots, w}$  are independent of ordering of elements within a slot.
%\end{obs}
\begin{lemma}
\label{config}
For any window $w\in [W]$, and slot $s$ in $w$,  $T_{1,\ldots, w}$, $T(w,s)$ and $S_{1, \ldots, w}$ 
\toRemove{are uniquely defined for each configuration $Y$.}
are independent of the ordering of elements within any slot, and 
are determined by the configuration $Y$.
\end{lemma}
\begin{proof}
Given the assignment of items to each slot, it follows from line 5 and 8 of Algorithm~\ref{alg:cardconst} \mcomment{ it follows from the definition of $\gamma(\tau)$ and $S_w$ (refer to \eqref{eq:gamma} and \eqref{eq:Sw}) } that $T_{1,\ldots, w}$, $T(w,s)$ and $S_{1, \ldots, w}$ are independent of the ordering of items within a slot. Since each $\arg\max$ in line 5 is independent of ordering elements in a slot.
Now, since  the assignment of items to slots are determined by the configuration $Y$, we obtain the desired lemma statement.
%and the position of each reassignments that are determined by $(Y,Z)$.
\end{proof}
Following the above lemma, given a configuration $Y$, we will some times use the notation  $T_{1,\ldots, w}(Y)$, $S_{1, \ldots, w}(Y)$, $T(w,s)(Y)$ and $H_{\ell}^s(Y)$ to make this mapping explicit.

The following lemma works for both stochastic and regular windows.

\begin{lemma}
\label{lem:pijBound}
For any item $i\in S^*$, (stochastic) window $w$,
%\in \{1,\ldots, W\}$
and slot $s$ in window $w$, define
\begin{equation}
\label{eq:pijBound}
p_{is}:=\Pr(i \in s \cup Supp(T(w,s)) | T(w,s)).
\end{equation}
Then, %for any pair of slots $s',s''$ after slot $s$, i.e. $s' \succeq s$, $s'' \succeq s$,
%in windows $w, w+1, \ldots, W$,
\begin{equation}
p_{is}\ge \frac{1}{k\beta} \ .
\end{equation}
\end{lemma}

\mcomment{
\begin{lemma}
\label{lem:pijBound}
For any item $i\in S^*$, window $w \in \{1,\ldots, W\}$, and slot $s$ in window $w$, define
\begin{equation}
\label{eq:pijBound}
p_{is}:=\Pr(i \in s \cup Supp(T) | T_{1,\ldots, w-1}=T).
\end{equation}
Then, for any pair of slots $s',s''$ in windows $w, w+1, \ldots, W$,
\begin{equation}
p_{is'}=p_{is''} \ge \frac{1}{k\beta} \ .
\end{equation}
\end{lemma}
}
\begin{proof}
%Let $s$ be a slot in the first $w-1$ windows and $s'$ be a slot in window $w$.
If $i\in Supp(T(w,s))$ then the statement of the lemma is trivial, so consider $i\notin Supp(T(w,s))$. For such $i$, we have $p_{is}=\Pr (Y_i=s  | T(w,s)=T)$. 
%$p_{is}=\Pr (Y_i=s  | T_{1,\ldots, w-1}=T)$. 
%Now, consider any pair of slots $s,s'$ with $s$ in windows $1,\ldots, w-1$ and $s'$ in window $w$.

%begin{lemma}
%label{beforeafterw}
%label{afterw}
%or $w\in [W]$, consider any $i\in S^*$ and $i\notin Supp(T_{1,\ldots, w-1})$. 
We show that \mcomment{for any pair of slots $s,s'$, where $s$ is a slot in first $w-1$ windows and $s'$ is a slot in window $w$, }
for any slot $s'$, where $s'$ appears before slot $s$, i.e., $s \succ s'$, \mcomment{modify definition of $\succ$}

\begin{equation}\label{eq1}
\Pr(T(w,s)=T|Y_i=s') \le \Pr(T(w,s)=T|Y_i=s) \ .
\end{equation}

\mcomment{
\begin{equation}\label{eq1}
\Pr(T_{1,\ldots, w-1}=T|Y_i=s) \le \Pr(T_{1,\ldots, w-1}=T|Y_i=s') \ .
\end{equation}
}
{
And, for any %Similarly we can show the following:
%\begin{lemma} \label{afterw}
pair of slots $s', s''$ 
%in windows $\{w, w+1 ,\cdots, W\}$, 
on or after slot $s$, i.e., $s' \succeq s$ and $s'' \succeq s$,
\begin{equation}
\label{eq:afterw}
\Pr ( T(w,s)=T | Y_i=s' ) =\Pr ( T(w,s)=T | Y_i=s'').
\end{equation}
}
%end{lemma}
%begin{proof}
To see \eqref{eq1}, suppose for a configuration $Y$
\rcomment{for stoch window}
(corresponding to slot $s$ for the case of stochastic window)
we have $Y_i=s'$ and $T(w,s)(Y)=T $. 
Since $i\notin Supp(T(w,s))$, then by definition of $T(w,s)$, we have
that $i\notin H_{\ell}^{s'}$ for slot $s'$ and any index $1\le \ell \le L$.

Therefore, if we remove $i$ from slots before slot $s$, i.e., $\{s':s\succ s'\}$, (i.e., consider another configuration where $Y_i$ is in slot $s$ or after $s$, i.e. in $\{s':s' \succeq s\}$ ), then $T(w,s)$ would not change.
This is because either $i$ is not the output of $\arg\max$ in the definition of $H_{\ell}^s$ (refer to \eqref{eq:maxmi}, \eqref{eq:Hi})
for slot $s$ and $1\le \ell \le L$, and therefore its removal will not change the output of $\arg\max$ and  $H_{\ell}$; or $i$ is the output of $\arg\max$ for slot $s$, and some index $1\le \ell \le L$, but $\Delta(H_{\ell}|S) \ge \Delta(H_{\ell-1}+ i|S)$. In that case, removing $i$ will not change $H_{\ell}$ either. Thus, removing $i$ will not change $T(w,s)$.

Also by adding $i$ to slot $s$, $T(w,s)$ will not change (since $T(w,s)$ does not cover $s$)
%since $s'$ is not in window $1,\cdots, w-1$.
Suppose configuration $Y'$ is a new configuration obtained from $Y$ by changing $Y_i$ from $s'$ to $s$. 
Therefore $T(w,s)(Y') = T$. 

Also remember that from Lemma~\ref{lem:indep}, %lemma~\ref{eqprob}, $\Pr (Y) = \Pr(Y')$.
%Now consider this mapping that take a configuration $Y$ with $Y_i=s$ and map it to configuration $Y'$ with $Y'=s'$.
%This mapping is 
This mapping shows that $\Pr(T(w,s)=T|Y_i=s') \le \Pr(T(w,s)=T|Y_i=s)$. 
The proof for \eqref{eq:afterw} is similar.
The rest of the proof is by applying Bayes rule and it is similar to Lemma 7 in~\cite{us}.

\mcomment{
To see \eqref{eq1}, suppose for a configuration $Y$ we have $Y_i=s$ and $T_{1,\cdots, w-1}(Y)=T $. 
Since $i\notin Supp(T)$, then by definition of $T_{1,\ldots, w-1}$, we have
that $i\notin H_{\ell}^s$ for slot $s$ and any index $1\le \ell \le L$.
}
\mcomment{ in any of the  windows $1,\cdots, w-1$.}
\mcomment{
we have that $i\notin \gamma(\tau)$ for any  $\alpha$ length subsequence $\tau$ of slots in any of the windows $1,\ldots, w-1$.
}
\mcomment{
Therefore, if we remove $i$ from windows ${1,\cdots, w-1}$ \mcomment{(i.e., consider another configuration where $Y_i$ is in windows $\{w, \ldots, W\}$)}, then $T_{1,\cdots, w-1}$ would not change. This is because either $i$ is not the output of $\arg\max$ in the definition of $H_{\ell}^s$ (refer to \eqref{eq:maxmi}, \eqref{eq:Hi})
\mcomment{$\gamma(\tau)$ (refer to  \eqref{eq:gamma})} 
for slot $s$ and $1\le \ell \le L$ \mcomment{i=1 or 0}, and therefore its removal will not change the output of $\arg\max$ and  $H_{\ell}$; or $i$ is the output of $\arg\max$ for slot $s$, and some index $0\le \ell-1 < \alpha$, but $\Delta(H_{\ell} \ge H_{\ell-1}+ i)$. In that case, removing $i$ will not change $H_{\ell}$ either. Thus, removing $i$ will not change $T_{1,\cdots, w-1}$.
\mcomment{
for any 
$\tau$, so that its removal will not change the output of argmax}. 
}
\mcomment{
Also by adding $i$ to slot $s'$, $T_{1,\cdots, w-1}$ will not change since $s'$ is not in window $1,\cdots, w-1$.
Suppose configuration $Y'$ is a new configuration obtained from $Y$ by changing $Y_i$ from $s$ to $s'$. 
Therefore $T_{1,\cdots ,w-1}(Y') = T$. }

\mcomment{
Also remember that from Lemma~\ref{lem:indep}, %lemma~\ref{eqprob}, $\Pr (Y) = \Pr(Y')$.
%Now consider this mapping that take a configuration $Y$ with $Y_i=s$ and map it to configuration $Y'$ with $Y'=s'$.
%This mapping is 
This mapping shows that $\Pr(T_{1,\ldots, w-1}=T|Y_i=s) \le \Pr(T_{1,\ldots, w-1}=T|Y_i=s')$. 
The proof for \eqref{eq:afterw} is similar. The rest of the proof is by applying Bayes rule and it is similar to Lemma 7 in~\cite{us}.
}
%end{proof}

\mcomment{
%Now, consider any slot $s$ in window $1,\ldots, w-1$.
By applying Bayes' rule to \eqref{eq1} we have 
\[
\Pr (Y_i=s  | T_{1,\ldots, w-1}=T) \frac{ \Pr(T_{1,\ldots, w-1}=T) }{\Pr(Y_i=s) } 
\le \Pr (Y_i=s'  | T_{1,\ldots, w-1}=T) \frac{ \Pr(T_{1,\ldots, w-1}=T) }{\Pr(Y_i=s') } \ .
\]
Also from Lemma~\ref{lem:indep}, $\Pr(Y_i=s)  = \Pr(Y_i=s')$ thus 
\[
\Pr (Y_i=s  | T_{1,\ldots, w-1}=T) 
\le \Pr (Y_i=s'| T_{1,\ldots, w-1}=T)  \ .
\]
Now, for any pair of slots $s',s''$ in  windows $w, w+1, \cdots, W$, by applying Bayes' rule to the equation \eqref{eq:afterw},
we have $p_{is'}=\Pr (Y_i=s'  | T_{1,\ldots, w-1}=T) =\Pr (Y_i=s''  | T_{1,\ldots, w-1}=T)=p_{is''}$.
That is, $i$ has as much probability to appear in $s'$ or $s''$ as any of the other (at most $k\beta$) slots in windows $w, w+1, \ldots, W$. 
As a result $p_{is''}=p_{is'} \ge \frac{1}{k\beta}$.
}
\end{proof}

\begin{lemma} \label{addtoslot}
Fix a slot $s'$, $T$, and $j\notin Supp(T)$. Suppose that there exists  some configuration  $Y'$ such that $T(w,s') (Y') =T$ and $Y_j'=s'$. Then, given any configuration $Y''$ with $T(w,s')(Y'')=T$, we can replace  $Y''_j$ with $s'$ to obtain a new configuration $\bar Y$ that also satisfies $T(w,s')(\bar Y)=T$.
\end{lemma}

\mcomment{
\begin{lemma} \label{addtoslot}
Fix a slot $s'$, $T$, and $j\notin Supp(T)$. Suppose that there exists  some configuration  $Y'$ such that $T_{1,\cdots, w-1} (Y') =T$ and $Y_j'=s'$. Then, given any configuration $Y''$ with $T_{1,\ldots, w-1}(Y'')=T$, we can replace  $Y''_j$ with $s'$ to obtain a new configuration $\bar Y$ that also satisfies $T_{1,\ldots, w-1}(\bar Y)=T$.
\toRemove{
Suppose there exists at least one configuration $\bar{Y}$ such that   $T_{1,\cdots, w-1} (\bar{Y}) =T$ and $\bar{Y}_j=s$ for $j\notin Supp(T)$.
Then for any configuration $Y$ with  $T_{1,\cdots, w-1} (Y) =T$, by setting $Y_j=s$, we get a new configuration $Y'$
such that $T_{1,\cdots, w-1} (Y') =T$.}
\end{lemma}
}
\begin{proof}
%Suppose $Y$ is a configuration such that $T_{1,\cdots, w-1} (Y) =T$. 
Suppose the slot $s'$ lies in window $w'$. If $s'\succeq s$ then the statement is trivial. So suppose $s \succ s'$.
\mcomment{If $w' \ge w$ then the statement is trivial. So suppose $w' < w$.}
Create an intermediate configuration by { removing} the item $j$ from $Y''$, call it $Y^-$. Since $j\notin Supp(T(w,s')(Y'')) = Supp(T)$ we have $T(w,s') (Y^-) =T$. 
In fact, for every slot $s$ and $1\le \ell \le L$, the set $H_{\ell}^s$ for $Y''$ will be the same as that for $Y^-$, i.e., $H_{\ell}^{s'}(Y'')=H_{\ell}^{s'}(Y^-)$.
\mcomment{
In fact, for every subsequence $\tau$, the greedy subsequence for $Y''$, will be same as that for $Y^-$, i.e.,   $\gamma_{Y''}(\tau) = \gamma_{Y^-}(\tau)$.
}
Now add item $j$ to slot $s'$ in $Y^-$, to obtain configuration $\bar Y$. We claim $T(w,s') (\bar Y) =T$.
%We only need to show that $j$ will not get selected in slot $s$ in configuration $Y'$.
%Suppose not, then there is a subsequence $\tau$  in the algorithm ending in slot $s$ such that $\gamma_{Y'} (\tau) =\{i_1, \cdots, i_{t-1}, j\}$
By construction of $T_{1,\ldots, w}$, we only need to show that $j$ will not be in $H_{\ell}^{s'}(\bar{Y})$ for slot $s'$ and any $1\le \ell \le L$.

\mcomment{we only need to show that $j$ will not be part of the greedy subsequence $\gamma_{\bar Y}(\tau)$ for any subsequence $\tau, |\tau| = \alpha$ containing the slot $s'$ when the input is in configuration $\bar Y$. }

To prove by contradiction, suppose that $j\in H_{\ell}^{s'}(\bar{Y})$, for some $1\le \ell \le L$.

Note that since the slots before $s'$ are the same for $\bar{Y}$ and $Y^-$, we have 
\[
H_{\ell-1}^{s'-1} (\bar{Y}) = H_{\ell-1}^{s'-1} (Y^-)= H_{\ell-1}^{s'-1}(Y'),
\] and
\[
H_{\ell}^{s'-1} (\bar{Y}) = H_{\ell}^{s'-1} (Y^-)= H_{\ell}^{s'-1}(Y').
\]
Suppose $j$ gets selected in slot $s'$ for some index $1\le \ell \le L$, i.e. $j\in H_{\ell}^{s'} (\bar{Y})$. Thus, 
\[
j=\underset{x\in s'\cup R(w',s')}{\arg \max} {\Delta(x|S_{1,\cdots, w'-1}\cup H_{\ell-1}^{s'-1}(\bar{Y}))},
\] and 
\[\Delta( H_{\ell-1}^{s'-1}(\bar{Y})+j |S_{1,\cdots, w'-1}) > \Delta(H_{\ell}^{s'-1}(\bar{Y}) | S_{1,\cdots, w'-1}).\]
Hence, 
\[
j=\underset{x\in s'\cup R(w',s')}{\arg\max} {\Delta(x|S_{1,\cdots, w'-1} \cup H_{\ell-1}^{s'-1}(Y'))},
\] and 
\[\Delta( H_{\ell-1}^{s'-1}(Y')+j |S_{1,\cdots, w'-1}) > \Delta(H_{\ell}^{s'-1}(Y') | S_{1,\cdots, w'-1}).\]
Thus $j\in H_{\ell}^{s'}(Y')$. In other words $j\in Supp(T)$ which is a contradiction.

\mcomment{
To prove by contradiction, suppose that $j$ is part of greedy subsequence for some $\tau$ ending in the slot $s'$. }
\toRemove{We only need to show that $j$ will not get selected in slot $s$ in configuration $Y'$.}
\mcomment{
Suppose $j$ gets selected in slot $s$ for some subsequence $\tau$ ending in slot $s$.
%Thus $\gamma_{Y'}(\tau) = \{i_1, \cdots, i_{t-1}, j\}$. Note that since slots before $s$ are the same as $Y^-$ and $T_{1,\cdots, w-1} (Y^-) =T$, $\gamma_{Y^-}(\tau) = \{i_1, \cdots, i_{t-1}, i_t\}$.
\toRemove{Suppose that $\gamma_{Y^-}(\tau)  := \{i_1, \cdots, i_{t-1}, i_t\} = \gamma_{Y''}(\tau) $.} 
For this $\tau$, let $\gamma_{Y^-}(\tau)  := \{i_1, \cdots, i_{\alpha-1}, i_\alpha\} = \gamma_{Y''}(\tau) $. Note that since the items in the  slots  before $s'$ are identical for $\bar Y$ and $Y^-$, \toRemove{and $T_{1,\cdots, w-1} (Y^-) =T$,} we must have that $\gamma_{\bar Y}(\tau) = \{i_1, \cdots, i_{\alpha-1}, j\}$, i.e.,
$\Delta_f(j | S_{1,\ldots,w'-1} \cup \{i_1,\ldots, i_{\alpha-1}\} ) \ge \Delta_f(i_\alpha | S_{1,\ldots,w'-1} \cup \{i_1,\ldots, i_{\alpha-1}\} ) $.
%$j=\arg \max_{i\in j\cup U_s \cup T_{1, \ldots, w-1}} \Delta_f(i | S_{1,\ldots,w-1} \cup \{i_1,\ldots, i_{t-1}\} \cup \{i\}) $
%$i_t=\arg \max_{i\in U_s \cup T_{1, \ldots, w-1}} \Delta_f(i | S_{1,\ldots,w-1} \cup \{i_1,\ldots, i_{t-1}\} \cup \{i\}) $
On the other hand, since $T_{1,\cdots, w'-1} (Y') = T_{1, \cdots, w'-1}(Y'') = T (\text{restricted to $w'-1$ windows})$,  we have that $\gamma_{Y'} (\tau) =\{i_1, \cdots, i_\alpha\}$.
However, $Y'_j=s'$. Therefore $j$ was not part of the greedy subsequence $\gamma_{Y'}(\tau)$ even though it was in the last slot in $\tau$, implying $\Delta_f(j | S_{1,\ldots,w'-1} \cup \{i_1,\ldots, i_{t-1}\} ) < \Delta_f(i_t | S_{1,\ldots,w'-1} \cup \{i_1,\ldots, i_{t-1}\} ) $. This  contradicts the earlier observation.
}
\end{proof}

\begin{lemma}
\label{lem:ijindep}
For any window $w$,  $i,j\in S^*, i\ne j$ and $s\in w$, the random variables $\mathbf{1}(Y_i=s|T(w,s))$ and $ \mathbf{1}(Y_j=s|T(w,s))$ are independent. That is, given $T(w,s)$,  items $i,j\in S^*, i\ne j$ appear in slot $s$ in $w$ independently.
\end{lemma}

\mcomment{
\begin{lemma}
\label{lem:ijindep}
For any window $w$,  $i,j\in S^*, i\ne j$ and $s,s'\in w$, the random variables $\mathbf{1}(Y_i=s|T_{1,\cdots, w-1}=T)$ and $ \mathbf{1}(Y_j=s'|T_{1,\cdots, w-1}=T)$ are independent. That is, given $T_{1,\cdots, w-1}=T$,  items $i,j\in S^*, i\ne j$ appear in any slot $s$ in $w$ independently.
\end{lemma}
}

\begin{proof}
Proof is similar to Lemma 8 in~\cite{us} and it is based on the previous Lemma.
\mcomment{
To prove this, we show that $\Pr(Y_i=s|T_{1,\cdots, w-1}=T)=\Pr(Y_i=s|T_{1,\cdots, w-1}=T \text{ and } Y_j =s')$. Suppose $Y'$ is a configuration such that $Y'_i=s$ and $Y'_j=s'$, and $T_{1,\cdots, w-1}(Y')=T$. Assume there exists another feasible slot assignment of $j$, i.e., there is another configuration $Y''$ such that $T_{1,\cdots, w-1} (Y'') =T$ and $Y''_j=s''$ where $s''\ne s'$. (If no such configuration $Y''$ exists, then $\mathbf{1}(Y_j=s')|T$ is always $1$, and the desired lemma statement is trivially true.) Then, we prove the desired independence by showing that there exists a feasible configuration where slot assignment of $i$ is $s$, and $j$ is $s''$. This is obtained by changing $Y_j$ from $s'$ to $s''$ in $Y'$, to obtain another configuration $\bar{Y}$. In Lemma~\ref{addtoslot}, we show that this change will not effect $T_{1,\cdots, w-1}$, i.e., $T_{1,\cdots, w-1} (\bar Y)=T $. 
Thus configuration $\bar Y$ satisfies the desired statement.
}
\end{proof}

%% file: Supp/Suppmmatroids.tex
Now we can generalize Lemma~\ref{replaced} to $p$-matchoid constraints.

\mcomment{
\begin{definition} \label{def:Hbar}
 For a slot $s \in \{1,\cdots, \alpha \beta\}$ in window $w$.
Let's denote by $H_{\ell}^s$, the set $H_{\ell}$ as defined in the Algorithm~\ref{alg:card} at the end of slot $s$.  Similarly, define $\bar{H}_{\ell}^s$ to be the set $\bar{H}_{\ell}$ at the end of slot $s$. (also initialize  $H_{\ell}^0=\bar{H}_{\ell}^0=\emptyset$)
\end{definition}
}

\matcomment{
\begin{definition}
For slot $s$ in window $w$, and $1 \le \ell \le \alpha$, define \mcomment{ $s\ge 1$ define for $s=0$}
\begin{equation}
SH{(w, \ell, s)} :=( S_{1,\cdots, w-1} \cup H_{\ell-1}^{s-1} ) \setminus \bar{H}_{\ell-1}^{s-1}\ .
\end{equation}
\end{definition}
\begin{definition} \label{def:mlmatroid}
Define $m_{\ell}^s$ to be $m_{\ell}$ as defined in Algorithm~\ref{alg:cardconst}, for slot $s$, which is
\begin{equation}
m_{\ell}^s:=\underset{x\in s\cup R(w,s)}{\arg\max}{g(x,SH(w,s))}, 
\end{equation}
\begin{equation}
r_{\ell}^s:=\theta(m_{\ell},SH(w,s))\ .
\end{equation}
Also for the sequence $\tilde{\tau}_w=(s_1,\cdots, s_t)$ defined in Definition~\ref{def:tauw}, define sequence $\mcomment{\mu_w}=(i_1,\cdots, i_{\alpha'})$, and $\nu_w:=(q_1,\cdots, q_{\alpha'})$
for $\alpha'=\min(t,\alpha)$,
where
\begin{equation}
i_j:=m_{j}^{s_j},
\end{equation}
and
\begin{equation}
q_j:=r_{j}^{s_j},
\end{equation}
Moreover, for $1\le j \le \alpha'$ define 
\begin{equation}
C_j:=H_{j}^{s_{j+1}-1}
\end{equation}
and
\begin{equation}
\bar{C}_j:=\bar{H}_{j}^{s_{j+1}-1}
\end{equation}
If $j+1> \alpha'$, set $s_{j+1}:=\alpha \beta+1$. 
We also use the notation $i_{1,\cdots, j}=(i_1,\cdots, i_j)$, for $1\le j\le \alpha'$.
\end{definition}
\begin{definition}
\begin{equation}
SH{(w,j)} := (S_{1,\cdots, w-1} \cup C_{j-1}) \setminus \bar{C}_{j-1}\ .
\end{equation}
\end{definition}
}

\mcomment{
Let $\gamma(\tilde \tau_s)=(i_1,\ldots, i_t)$ defined as follows 
%with $\gamma(\cdot)$ as defined in \eqref{eq:gamma}. 
, based on function $g$ defined in~\eqref{eq:newg}, for $p$-matchoid constraints.
\begin{equation}
\label{eq:gamma}
\gamma(\tau):=\{i_1, \ldots, i_t\},
\end{equation}
where
\begin{equation}
\label{eq:ij}
i_j := \underset{i\in s_j \cup R_{1, \ldots, w-1}}{\arg \max}  g(i, S{(w-1,\tau, j-1)}).
\end{equation}
Also let
\begin{equation}
\label{eq:diff}
S{(w-1,\tau, j-1)} := S_{1,\cdots, w-1} \cup \{i_1,\cdots, i_{j-1}\} \setminus \bigcup_{r\le j-1} C_r.
\end{equation}
where $C_r$ is defined in~\eqref{eq:cij}.
}

For any slot $s$ in window $w$, and element $b\in \mathcal{N}_{i}$, 
let $S'_{i}$ be the extension of $SH(w,s)$
to a base of $\mathcal{M}_{i}$ (refer to Lemma~\ref{extension}), and $\pi_{i}$ be the bijection from Brualdi lemma (refer to Lemma~\ref{lem:Brualdi}) from $S^*$ to $S'_{i}$. 
Further, let's denote
\begin{equation} \label{eq:piMatroid}
\pi(b):=\{\pi_{i}(b)|b \in \mathcal{N}_{i}\}.
\end{equation}

\mcomment{
For any $j\in \{1,\ldots, t\}$, and element $b\in \mathcal{N}_{\ell}$, 
let $S'_{\ell}$ be the extension of %$S_{1,\cdots, w-1}\cup \{i_1, \cdots, i_{j-1}\}\setminus\bigcup_{r\le j-1} C_r$ 
%$S(w-1,\tilde{\tau}_w,j-1)$
$SH(w,s)$
to a base of $\mathcal{M}_{\ell}$ (refer to Lemma~\ref{extension}), and $\pi_{\ell}$ be the bijection from Brualdi lemma (refer to Lemma~\ref{lem:Brualdi}) from $S^*$ to $S'_{\ell}$. 
Further, let's denote
\begin{equation}
\pi(b):=\{\pi_{\ell}(b)|b \in \mathcal{N}_{\ell}\}.
\end{equation}
}

\toRemove{
\begin{lemma}\label{replacedMatchoid}
Suppose the sequence $\tilde \tau_w=(s_1, \ldots, s_t)$ defined as in Definition \ref{def:tauw}, let $\gamma(\tilde \tau_s)=(i_1,\ldots, i_t)$, with $\gamma(\cdot)$ as defined in \eqref{eq:gamma}. 
For any $j\in \{1,\ldots, t\}$, and element $b\in \mathcal{N}_{\ell}$, 
let $S'_{\ell}$ be the extension of $S_{1,\cdots, w-1}\cup \{i_1, \cdots, i_{j-1}\}\setminus\bigcup_{r\le j-1} C_r$ to a base of $\mathcal{M}_{\ell}$ (refer to Lemma~\ref{extension}), and $\pi_{\ell}$ be the bijection from Brualdi lemma (refer to Lemma~\ref{lem:Brualdi}) from $S^*$ to $S'_{\ell}$. 
Further, let's denote by $\pi(b):=\{\pi_{\ell}(b)|b \in \mathcal{N}_{\ell}\}$, then
{\small \begin{eqnarray*}
\Ex[f(S_{1,\ldots, w-1} \cup \{i_1, \ldots, i_{j-1}\}\setminus (\bigcup_{r\le j-1} C_r \cup \pi(a)|T_{1,\ldots, w-1}, i_{1,\ldots, j-1}, a\in S^*\cap Z_{s_j}]\\
 \ge (1-\frac{p}{k})f(S_{1,\ldots, w-1} \cup \{i_1, \ldots, i_{j-1}\}\setminus\{c_1,\cdots,c_{j-1}\})\
\end{eqnarray*}}
\end{lemma}
}

\mcomment{
\begin{lemma}\label{replaced}
Let $S'$ be the extension of %$S(w-1,\tilde{\tau}_w,j-1)$
$SH(w,s)$
to a base of $\mathcal{M}$ (refer to Lemma~\ref{extension}).
Let $\pi$ be the bijection from Brualdi lemma (refer to Lemma~\ref{lem:Brualdi}) from $S^*$ to $S'$.
Then, for all $j=1,\ldots, \alpha'$, \\
{\small \begin{eqnarray*}
\Ex[f(SH(w,s) - \pi(a))|T(w,s), a\in S^*\cap Z_{s}]
 \ge (1-\frac{1}{k})f(SH(w,s))\ .
\end{eqnarray*}}
\end{lemma}
\begin{proof}
Since $\pi$ is a bijection from $S^*$ to $S'$, from Brualdi's lemma (lemma~\ref{lem:Brualdi}), 
$SH(w,s)- \pi(a) +a \in \mathcal{I}$, for all $a \in S^*$. 
Recall the definition of $Z_{s_j}$.
Suppose $a$ is a randomly picked item from $S^*\cap Z_{s_j}$.
Since $Z_{s_j}\ne \emptyset$,  using Lemma~\ref{lem:Zs} conditioned on $T(w,s)$, the element $a$ can be equally any element of $S^*$ with probability
$1/k$. Therefore, $\pi(a)$ would be any of $SH(w,s)$ with probability at most
$1/k$, i.e., 
%(It is because $\pi'$ might map some elements of $S^*$ to $S'\setminus SH()$).
\[
\Pr(\pi(a)=e | T(w,s)) \le 1/k, \ \text{for } e \in  SH(w,\ell, s), 
\]
Now based on the definition of $\pi$ and lemma~\ref{sample} the lemma follows.
\end{proof}
}

\begin{lemma}\label{replacedMatchoid}
For slot $s$ in window $w$, and $\pi$ as defined in eq.~\eqref{eq:pi},
{\small \begin{eqnarray*}
\Ex[f(SH(w,s) - \pi(a))|T(w,s), a\in S^*\cap Z_{s}]
 \ge (1-\frac{p}{k})f(SH(w,s))\ .
\end{eqnarray*}}

\mcomment{
{\small \begin{eqnarray*}
\Ex[f(S(w-1,\tilde{\tau}_w,j-1)\setminus \pi(a)|T_{1,\ldots, w-1}, i_{1,\ldots, j-1}, a\in S^*\cap Z_{s_j}]
 \ge (1-\frac{p}{k-\alpha})f(S(w-1,\tilde{\tau}_w,j-1)).\
\end{eqnarray*}}
}
\end{lemma}

\begin{proof}
%It follows from Brualdi lemma and lemma~\ref{sample}.

%where each element has probability at least $p$ to appear in $A$ (not necessarily independently)
The proof is similar to the proof of Lemma~\ref{replaced}. 
For $\ell \in [q]$,
since $\pi_{\ell}$ is a bijection from $S^*\cap \mathcal{N}_{\ell}$ to $S'_{\ell}$, 
\mcomment{
by Brualdi's lemma (lemma~\ref{lem:Brualdi}), there is an onto mapping $\pi'_{\ell}$ from $S^*\cap \mathcal{N}_{\ell}$ %$OPT$ 
 to $S(w-1,\tilde{\tau}_w,j-1)\setminus (\pi(a))\cup \{\emptyset\})$,
such that 
}
%$S_{1,\cdots, w-1}\cup \{i_1, \ldots, i_{j-1}\} - \pi(a) +a \in M$ for all $a \in S^*$.
we have $S(w,s)-\pi_i(a)+a \in M_{i}$, for all $a\in S^*$.

\mcomment{
$S(w-1,\tilde{\tau}_w,j-1)\setminus  \pi(a)- \pi'_{\ell}(a) +a \in M_{\ell}$, for all $a \in S^*$. Further, $\pi'_{\ell}(a)=\pi_{\ell}(a)$ if 
$\pi_{\ell}(a)\in  S(w-1,\tilde{\tau}_w,j-1)$ 
%$\pi_{\ell}(a)\in  S_{1,\cdots, w-1}\cup \{i_1, \cdots, i_{j-1}\}\setminus \bigcup_{r\le j-1} C_r$ 
and $\pi'_{\ell}(a)=\emptyset$ otherwise.
}

Recall the definition of $Z_{s_j}$.
Suppose $a$ is a randomly picked item from $S^*\cap Z_{s}$.
Since $Z_{s_j}\ne \emptyset$,  using Lemma~\ref{lem:Zs} conditioned on $T(w,s)$, the element $a$ can be equally any element of $S^*$ with probability
$1/k$. Therefore, $\pi_i(a)$ would be any element of $SH(w,s)$ with probability at most
$1/k$, i.e., 
\[
\Pr(\pi_i(a)=e | T(w,s)) \le 1/k, \ \text{for } e \in  SH(w, s), i\in[q] 
\]

\mcomment{
Recall the definition of $Z_{s_j}$ (refer to definition~\ref{def:tauw}).
Suppose $a$ is a randomly picked item from $S^*\cap Z_{s_j}$.
Note that from Lemma~\ref{lem:Zs}, conditioned on $T_{1,\cdots, w-1}$ and $i_1,\cdots,i_{j-1}$, the element $a$ can be equally any element of %$S^*$ 
%$S^*\backslash  \{Z_1,\ldots Z_{s_{j-1}}\}$
$S^*\backslash  \{i_1,\ldots i_{{j-1}}\}$
with probability at least $1/(k-\alpha)$.
Therefore, $\pi'_{\ell}(a)$ would be any of $S(w-1,\tilde{\tau}_w,j-1)$ with probability at most $1/(k-\alpha)$ 
(since $\pi'_{\ell}$ might map some elements of $S^*$ to the empty set).
}

For element $e\in SH(w,s)$, let $\mathcal{N}(e)$ be the set of indices $i$ such that $e\in \mathcal{N}_{i}$. Because of the $p$-matchoid constraint, we have $|\mathcal{N}(e)|\le p$.
Define 
\[
\pi^{-1}(e):=\{t | t\in \mathcal{N}_{i}, \text{ for some }  i \in \mathcal{N}(e) \text{ and }  \pi_{i}(t)=e \}. 
\]
we have also $|\pi^{-1}(e)|\le p$. Thus, each element $e \in SH(w,s)$ belongs to %$\bigcup_{\ell} \pi_{\ell}(a)$
$\pi(a)$ with probability at most $p/k$:
\[
\Pr(e\in \pi(a) |T(w,s), a\in  Z_{s} ) = \Pr(a\in  Z_{s}\cap \pi^{-1}(e) | T(w,s)) \le \frac{p}{k}. 
\]
 Now we apply Lemma~\ref{sample}.
It is crucial to note that in Lemma~\ref{sample}  each element do not necessarily need to be selected  independently.
Definition of $\pi$ and lemma~\ref{sample} imply the lemma.
\end{proof}

\begin{cor}\label{cor:replaced}
For slot $s$ in window $w$, and $\pi$ as defined in eq.~\eqref{eq:pi}.
Then, for all slot $s$, %$j=1,\ldots, \alpha'$, 
{\small \begin{eqnarray*}
\Ex[f(SH(w,s)- \pi(a))|T(w,s), a\in  Z_{s}]
 \ge (1-\frac{p}{k})f(SH(w,s))\ .
\end{eqnarray*}}
\end{cor}

%\subsection{Proof of Lemma 13}
%Furthermore the main difference in the analysis is that instead of recursion~\ref{recursion}, we get the following new recursion:

\begin{lemma} \label{lem:recursionMatchoid}
For all slots $s$ %$j=1,\ldots, \alpha'$, 
\begin{align}  
\Ex[f({SH}(w,s) ) 
- f(SH(w,s-1) ) | T(w,s)] \nonumber 
\ge
 \frac{1}{k}\Ex[f(S^*)-(p+1) f({SH}(w,s-1) | T(w,s-1)]  \ . 
\end{align}
\end{lemma}

\begin{proof}
In the Algorithm~\ref{alg:matroid}, %at the end of window $w$, we set %$S_{1,\cdots, w-1} = {S}_{1,\cdots, w-1} \setminus \bar{S}_w$.
%$S_{1,\cdots, w} = S_{1,\cdots, w-1}\cup S_w \setminus \bar{S}_w$.
%$S=S + S_w -\bar{S}_w$.
Suppose $a\in s\cap S^*$. 
\mcomment{
Moreover, let $S'$ be the extension of $SH(w,j-1)$ 
to a base of $\mathcal{M}$, and $\pi$ be the bijection from Brualdi's Lemma (refer to Lemma~\ref{lem:Brualdi}) from $S^*$ to $S'$.
}
Moreover, let $S'_{\ell}$ be the extension of $SH(w,s-1)$ to an independent set in $\mathcal{M}_{\ell}$, and $\pi_{\ell}$ be the bijection in Brualdi lemma (refer to Lemma~\ref{lem:Brualdi}) from $S^*_{\ell}$ to $S'_{\ell}$.
Further, let's denote
\begin{equation} \label{eq:piMatroid}
\pi(b):=\{\pi_{i}(b)|b \in \mathcal{N}_{i}\}.
\end{equation}
Then, the expected value of the function $g$ on the element selected by the algorithm in slot $s$ (the element with maximum $g$ in the slot $s$) is as follows. 
%\begin{eqnarray*}
\begin{align*}
&\Ex[f(SH(w, s)|T(w,s)]  
\\ \ge& 
\Ex[f( SH(w,s-1) + a - \pi(a)) 
 |T(w,s), a\in Z_{s}] 
\\ \ge &
\Ex[f(SH(w,s-1) - \pi(a) )
|T(w,s), a\in  Z_{s}]+
\\ & \Ex[\Delta(a|SH(w,s-1)- \pi(a) )
|T(w,s), a\in  Z_{s}] 
\\ \ge &
\Ex[f(SH(w,s-1) - \pi(a) )
|T(w,s), a\in  Z_{s}]
\\+&\Ex[\Delta(a|SH(w,s-1) )
|T(w,s), a\in  Z_{s}].
\end{align*}
%\end{eqnarray*}

The first inequality is from the definition of function $g$ as it is defined in equation~\ref{eq:g}. The last inequality is from submodularity of $f$.
Now from the last inequality  and lemma~\ref{replacedMatchoid} we have
\begin{align*}
\Ex[f({SH}(w,s))  |T(w,s)]\ge &
 (1-\frac{p}{k} ) f(SH(w,s-1) )
 + \Ex[ \Delta(a|{SH}(w,s-1) \})
 |T(w,s), a\in Z_{s}]. 
\end{align*}
Now from lemma~\ref{lem:asGoodas} and the above inequality we can show
\begin{align*}
\Ex[f({SH}(w,j)  |T(w,s)]&\ge
(1-\frac{p}{k} ) f(SH(w,s-1) )
+ \frac{1}{k} (f(S^*)-f({SH}(w,s-1)) )\ . 
\end{align*}
Thus,
\begin{align}  %\label{recursion}
\Ex[f({SH}(w,s) ) 
- f(SH(w,s-1) ) | T(w,s)] \nonumber 
\ge
 \frac{1}{k}(f(S^*)-(p+1) f({SH}(w,s-1))  \ . 
 %\label{recursion}
\end{align}
Hence, by taking expectation %on $T(w,s_{j-1})$, and by Proposition~\ref{prop:Tsucc},
\begin{align}  
\Ex[f({SH}(w,s) ) 
- f(SH(w,s-1) ) | T(w,s)] \nonumber 
\ge
 \frac{1}{k}\Ex[f(S^*)-(p+1) f({SH}(w,s-1) | T(w,s)]  \ . 
\end{align}

\mcomment{
Hence,
\begin{align*}
&\Ex[g(i_j,SH(w,j-1)) | T(w,s)] \ge 
 %(1-\frac{1}{k-\alpha})
 \frac{1}{k}
 (f(S^*)-2 f(S(w,j-1))) \ .
\end{align*}
}

\toRemove{
\begin{align}  %\label{recursion}
f({S}(w-1,\tilde{\tau}_w,j) ) 
- f(S(w-1,\tilde{\tau}_w,j-1) ) %\nonumber 
\ge
 \frac{1}{k-\alpha}((1-\frac{\alpha}{k})f(S^*)-2f({S}(w-1,\tilde{\tau}_w,j-1))  \ . 
 %\label{recursion}
\end{align}
}

\end{proof}

\mcomment{
\begin{lemma}\label{exgmatchoid}
Suppose function $g$ is as defined in Definition~\ref{def:gmatchoid}, 
%and $\rho:=\frac{k}{k-\alpha}$.
%For the sequence $\tilde \tau_w=(s_1, \ldots, s_t)$, and $\gamma(\tilde \tau_s)=(i_1,\ldots, i_t)$. Then, 
For all $j=1,\ldots, t$, \\
{\small \begin{eqnarray*}
\Ex \left[g(i_j,S(w-1,\tilde{\tau}_w,j-1))|T_{1,\ldots, w-1}, i_{1,\ldots, j-1}\right]
 \ge 
 %(1-\frac{1}{k-\alpha})
 \frac{1}{k}
 \left(
 (1-\frac{\alpha}{k})f(S^*)-\rho (p+1) f(S(w-1,\tilde{\tau}_w,j-1))\right)\
\end{eqnarray*}}
\end{lemma}
\begin{proof}
The proof is similar to the proof of Lemma~\ref{exg} with some changes regarding matchoid constraints.
In the algorithm~\ref{alg:main}, at the end of window $w$, we set %$S_{1,\cdots, w-1} = {S}_{1,\cdots, w-1} \setminus \bar{S}_w$.
$S_{1,\cdots, w}= \bar{S}_w$.
%$S_{1,\cdots, w} = S_{1,\cdots, w-1}\cup S_w \setminus \bar{S}_w$.
Suppose $a\in s\cap S^*$. 
Moreover, let $S'_{\ell}$ be the extension of $S_{1,\cdots, w-1}\cup \{i_1, \cdots, i_{j-1}\}\setminus \bigcup_{r\le j-1} C_r$ to an independent set in $\mathcal{M}_{\ell}$, and $\pi_{\ell}$ be the bijection in Brualdi lemma (refer to Lemma~\ref{lem:Brualdi}) from $S^*_{\ell}$ to $S'_{\ell}$.
Thus the expected value of the function $g$ on the element selected by the algorithm in slot $s$ (the element with maximum $g$ in the slot $s$) would be
\begin{eqnarray*}
&\Ex[f({S}(w-1,\tilde{\tau}_w,j))
 |T_{1,\ldots, w-1}, i_{1,\ldots, j-1}]  
\\\ge&
\Ex[f({S}(w-1,\tilde{\tau}_w,j-1)\cup\{a\} \setminus \pi(a)  |T_{1,\ldots, w-1}, i_{1,\ldots, j-1}, a\in S^*\cap Z_{s}] 
\\\ge&
\Ex[f({S}(w-1,\tilde{\tau}_w,j-1) \setminus \pi(a) )|T_{1,\ldots, w-1}, i_{1,\ldots, j-1}, a\in S^*\cap Z_{s}]
\\&+\Ex[\Delta_f(a|S(w-1,\tilde{\tau}_w,j-1) \setminus \pi(a) )|T_{1,\ldots, w-1}, i_{1,\ldots, j-1}, a\in S^*\cap Z_{s}] \\
\ge&
\Ex[f({S}(w-1,\tilde{\tau}_w,j-1) \setminus \pi(a) ) |T_{1,\ldots, w-1}, i_{1,\ldots, j-1}, a\in S^*\cap Z_{s}]
\\&+\Ex[\Delta_f(a|S(w-1,\tilde{\tau}_w,j-1) |T_{1,\ldots, w-1}, i_{1,\ldots, j-1}, a\in S^*\cap Z_{s}].
%\end{align}
\end{eqnarray*}
The first inequality is from the definition of function $g$ as it is defined in equation~\ref{eq:newg} and the fact that the algrotihm selects an element in slot $s$ with maximum value of $g$. The second inequality is from submodularity and the last inequality is from monotonicity of $f$.
Now from the last inequality  and Lemma~\ref{exgmatchoid}, we can show,
\begin{align*}
&\Ex[f({S}(w-1,\tilde{\tau}_w,j) |T_{1,\ldots, w-1}, i_{1,\ldots, j-1}]& \\ &\ge
(1-\frac{p}{k-\alpha} ) f(S(w-1,\tilde{\tau}_w,j-1))&\\
&+ \Ex[ \Delta_f(a|{S}_{1,\ldots, w-1} \cup \{i_1, \ldots, i_{j-1}\} \setminus\bigcup_{r\le j-1} C_r )|T_{1,\ldots, w-1},  i_{1,\ldots, j-1}, a\in S^*\cap Z_{s}]. &
\end{align*}
Now from lemma~\ref{lem:asGoodas} and the above inequality we can show
\begin{align*}
&\Ex[f({S}(w-1,\tilde{\tau}_w,j)) |T_{1,\ldots, w-1}, i_{1,\ldots, j-1}]& \\  &\ge
(1-\frac{p}{k-\alpha} ) f(S(w-1,\tilde{\tau}_w,j))
+ { \frac{1}{k}\left((1-\frac{\alpha}{k})f(S^*)-f({S}(w-1,\tilde{\tau}_w,j-1)) \right)\ . }&
\end{align*}
Thus,
\begin{align}\label{newrecursion}
&\Ex[f({S}(w-1,\tilde{\tau}_w,j))  
\nonumber 
- f(S(w-1,\tilde{\tau}_w,j-1))] \nonumber
\ge
{ \frac{1}{k}\left((1-\frac{\alpha}{k})f(S^*)-\rho (p+1)f({S}(w-1,\tilde{\tau}_w,j-1))\right)\ . }
\end{align}
\end{proof}
}

%we can show the following theorem:

%\matchoidThm
%\matchoidThm

\mcomment{
\textbf{Proof of Theorem~\ref{opttheoremmatchoid}}. 
Similar to Lemma~\ref{lem:Sw}, by solving the recursion in Lemma~\ref{exgmatchoid} and similar to the analysis of matroid constraints, Theorem~\ref{opttheoremmatchoid} follows.
For any real $\delta'\in (0,1)$, if parameters 
$k,\alpha, \beta$ satisfy \settinga, then the set $S_{1,\ldots, W}$ tracked by Algorithm \ref{alg:main} satisfies
\begin{align*}
\mathbb{E}[f(S_{1,\ldots, W})] &\ge (1-\delta')^2 (\frac{1}{\rho(p+1)}(1-1/e^{\rho (p+1)}) ) \OPT \\ 
&\ge  \frac{1}{\rho}(1-\delta')^2 (\frac{1}{p+1}(1-1/e^{p+1}) ) \OPT \\
&\ge (1-\frac{\alpha}{k}) (1-\delta')^2 (\frac{1}{p+1}(1-1/e^{p+1}) ) \OPT 
%\\ & \ge (1-\delta')^3 (\frac{1}{2}(1-1/e^{2}) ) \OPT
\end{align*}
%$$\mathbb{E}[f(S_{1,\ldots, W})] \ge (1-\frac{\alpha}{k}) (1-\delta')^2 (\frac{1}{p+1}(1-1/e^{p+1}) ) \OPT.$$
Similar to Theorem~\ref{opttheorem}, \[\Ex[f(A^*)] := \mathbb{E}[f(S_{1,\cdots, W} \cap A)] \ge (1-\frac{\epsilon}{2})\mathbb{E}[f(S_{1,\cdots, W})].\]
where $A^*:= S_{1,\cdots, W} \cap A$ is the %size $k$ 
subset of shortlist $A$ returned by Algorithm \ref{alg:main}.
Thus we can achieve a $\frac{1}{p+1}(1-1/e^{p+1} - \epsilon -O(k^{-1})) OPT$ lower bound.
The running time will be diSHussed in the streaming section.
}

%% file: Supp/Suppstreaming.tex
\subsection{Proof of Theorem~\ref{opttheoremmatchoid}}
\begin{proof}
Th difference between Algorithm~\ref{alg:main} in this paper and the main Algorithm  in~\cite{us} is that, 
%we remove elements of $\bar{S}_w$  from $S$ at the end of each window $w$. 
at the end of each window $w$,  besides adding some new elements to the current solution $S$, we remove some items from $S$ to make $S$ an independent set in each matroid.
Similar to the proof of Theorem 2 in~\cite{us}, we use a slightly modified  implementation of the algorithm such that $\gamma(\tau)$ can be comutted in a memory efficient manner ( refer to  Theorem 2 in~\cite{us}). Therefore we are able to store $\gamma(\tau)$ for all $\alpha$-subsequence $\tau$ in $O(1)$ space.
The same implemenation will allow us to
we keep track of all parameters in Algorithm~\ref{alg:main} including $\bar{S}_w, S_w, R_w, \hat{S}_w$ in a memory efficient way using memory $O(k)$. 
The other difference between the two algorithms is in the subroutine~\ref{alg:matroidmax} that finds the element with maximum $g$ in a slot. 
In~\cite{us}, $g(e,S)$ can be computed using only one oracle access, whereas in the new definition of $g$ in equation~\ref{eq:newg}, we need access to independence oracle of $p$ matroids that $e$ belongs to, in order to check the independence of $S+e-e'$ for each $e' \in S$. At most $p \kappa$ elements of $S$ are eligible (they are in the ground set of a matroid that $e$ also belongs to). Hence, 
in order to create $\Omega_{\ell}(e,S)$, 
 for each arriving element $e$ in the input, we need $O(p\kappa)$ access to the \textit{independence oracle}. Similarly the total access to the value oracle is $O(p\kappa)$. 
 In order to compute 
$\lambda(e,S)$, we need to consider all $\kappa^p$ combinations and query value oracle. Therefore the  number of access to the value oracle is $O(p\kappa+\kappa^p)$ per element. 
But, since the first element $a_0$ is computed in the beginning of each slot
for each $\tau$, we would have in average an additional $O(k^2/n)$ function evaluation per element.
Thus the number of objective  function evaluations for each item, amortized over $n$ items, is $O(p\kappa+\kappa^p+\frac{k^2}{n})$.
%where $\kappa= \max_{i\in[q]} rk(\mathcal{M}_i)$.
\end{proof}

\mcomment{
\subsection{Experiment: {Uniform Matroid}}
The simplest constraint that we can impose is the uniform matroid or equivalently the cardinality constraint.
In the simplest form our algorithm is similar to~\cite{us}.
We compare our algorithm to the state of the art algorithm in the streaming setting~\cite{kazemi2019submodular}.
%As we established an upper bound on the constant factor $\eta_{\epsilon}(k)$ in theorem~\ref{thm:streaming}, 
The performance of our algorithm crucially depends on the choice of 
$\alpha$ and $\beta$. The running time also is a function of $\alpha$ and $\beta$, and it grows rapidly as we increase $\alpha$ and $\beta$. 
Surprisingly, our algorithm outperforms~\cite{kazemi2019submodular}
%substantially 
even with relatively small choices of $\alpha=6$ and $\beta=2$.
We also observe that the utility of the output returned by our algorithm can be very  close to  what the optimal offline algorithm, namely the Greedy algorithm achieves. In Figure~\ref{fig:uniform}, we have plotted the performance of all three algorithms on the YouTube dataset.
Note that in our experiment we use a simplistic version of our algorithm in which we subsample from the shortlist in beginning of each window.
%and only use that subsample rather than the entire shortlist.
Furthermore we observe that our algorithm is slower than~\cite{kazemi2019submodular}, but the interesting fact about our algorithm 
%as stated in Theorem~\ref{thm:streaming} 
is that it is highly parallel, thus it has the potential to become $\eta_{\epsilon}(k)$  times faster.
%using parallelization. %(here roughly 1000 times).  
\begin{figure}[ht!]
\centering
\includegraphics[width=80mm]{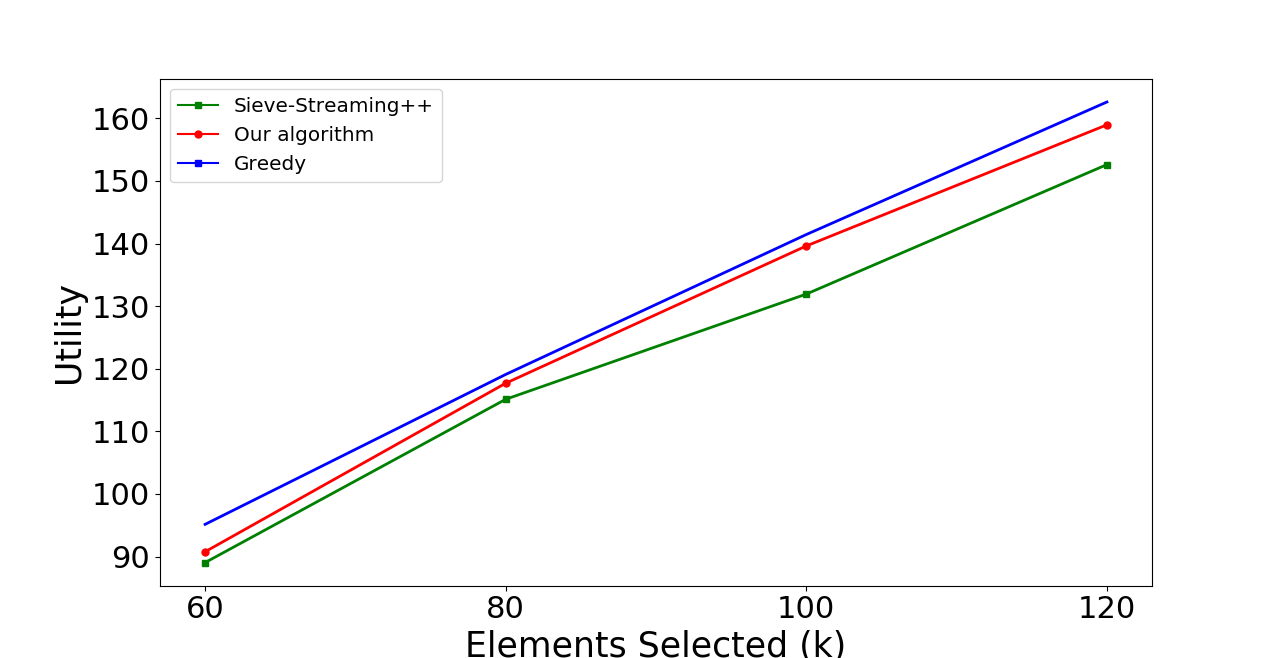}
\caption{The plot is for unifrom matroid, $\alpha =6$ and $\beta=2$}
\label{fig:uniform}
\end{figure}
}